\documentclass[a4paper,11pt]{article}
\pdfoutput=1
\usepackage[normalem]{ulem}
\usepackage[utf8]{inputenc}
\usepackage{cite}
\usepackage[left=2cm,right=2cm,top=2cm,bottom=3cm]{geometry}
\usepackage{xcolor}
\usepackage{float}

\usepackage{epsf}
\usepackage{graphicx}

\usepackage{amsmath}
\usepackage{amssymb}
\usepackage{mathrsfs}
\usepackage{slashed}
\usepackage{mathtools}

\usepackage{amsfonts}
\usepackage{xfrac} 
\usepackage{booktabs} 
\usepackage{soul} 

\usepackage{color}
\definecolor{cred}{RGB}{180,50,40}
\definecolor{purple}{RGB}{180,90,180}
\definecolor{darkgreen}{RGB}{0, 100, 0}

\usepackage[
colorlinks=true
,urlcolor=black
,anchorcolor=black
,citecolor=black
,filecolor=black
,linkcolor=black
,menucolor=black
,linktocpage=true
,pdfproducer=medialab
,pdfa=true
]{hyperref}

\usepackage{braket}

\begin{document}

\allowdisplaybreaks
\setcounter{footnote}{0}
\vspace*{-1.5cm}
\begin{flushright}
TUM-HEP-1303/20 \\

\vspace*{2mm}
\end{flushright}
\begin{center}
\vspace*{1mm}
\vspace{1cm}

\vspace*{.5cm}
{\Large\bf The fate of \pmb{$V_1$} vector leptoquarks: the impact of future flavour data}

\vspace*{0.8cm}

{\bf C.~Hati $^{a}$, J.~Kriewald $^{b}$, J. Orloff $^{b}$ and A.~M.~Teixeira $^{b}$}

\vspace*{.5cm}
$^{a}$ Physik Department T70, Technische Universit\"at M\"unchen,\\
James-Franck-Stra{\ss}e 1, D-85748 Garching, Germany

\vspace*{.2cm}
$^{b}$ Laboratoire de Physique de Clermont (UMR 6533), CNRS/IN2P3,\\
Univ. Clermont Auvergne, 4 Av. Blaise Pascal, F-63178 Aubi\`ere Cedex,
France
\end{center}

\vspace*{2mm}

\begin{abstract}
\noindent
Motivated by the recent experimental progress on the $B$-meson decay anomalies (in particular the angular observables in $B\to K^\ast\mu\mu$),
we rely on a simplified-model approach to study the prospects of vector leptoquarks in what concerns numerous flavour observables, identifying several promising decay modes which would allow to (indirectly) probe such an extension. Our findings suggest that the confirmation of the $B$-meson decay anomalies, in parallel with
positive signals (at Belle II or LHCb) for $\tau\to \phi \mu$, $B_{(s)}$-meson decays to $\tau^+ \tau^-$ and $\tau^+ \mu^-$ ($\tau^+ e^-$) final states, as well as an observation of certain charged lepton flavour violation decays (at COMET or Mu2e), would contribute to strengthen the case for this scenario. We also illustrate how the evolution of the experimental determination of $R_{D^{(*)}}$ could be instrumental in 
falsifying an explanation of the anomalous $B$-meson decay data via a vector $V_1$ leptoquark.

\end{abstract}


\section{Introduction}
One of the key predictions of the Standard Model (SM) is the universality of interactions for the charged leptons of different generations. Extensive experimental observations confirm that this is indeed the case for several electroweak precision observables, as for example for $Z\to \ell \ell$ decays~\cite{Tanabashi:2018oca, ALEPH:2005ab}. However, certain recent experimental measurements suggest that hints for the violation of lepton flavour universality (LFUV) might be present in a number of observables, which would thus unambiguously point towards the presence of New Physics (NP). The LFUV observables concern the flavour changing neutral current (FCNC) quark transitions $b\to s \ell^+\ell^-$, 
and the charged current quark transitions $b\to c \ell^- \nu$: the former are loop-suppressed within the SM, thus providing a high sensitivity to probe NP effects; the latter can occur at the tree-level and are only subject to Cabibbo-Kobayashi-Maskawa (CKM) suppression within the SM. 
Among these observables, ratios of potentially LFU violating $B$-meson decays are of particular interest, since they are free of the theoretical hadronic uncertainties arising from the form factors, as these cancel out in the ratios. The most relevant LFUV ratios for our study are  $R_{D^{(*)}}$ (corresponding to the charged current transition $b\to c \ell^- \nu$) and  $R_{K^{(\ast)}}$ (corresponding to the neutral current transition $b\to s \ell^+\ell^-$), respectively defined as
\begin{equation}\label{eq:RDRK:def}
  R_{D^{(*)}} \,= \,\frac{\text{BR}(B \to D^{(*)} \,\tau^- \,\bar\nu)}{
    \text{BR}(B \to  D^{(*)}\, \ell^- \,\bar\nu)}\, , \quad R_{K^{(\ast)}} \,= \,
  \frac{\text{BR}(B \to K^{(*)}\, \mu^+\,\mu^-)}{\text{BR}(B \to
  K^{(*)}\, e^+\,e^-)}\,,
\end{equation}
where $\ell=e,\,\mu$. 
A number of experimental measurements~\cite{Lees:2012xj,Lees:2013uzd,Amhis:2019ckw,Huschle:2015rga,Adachi:2009qg, Bozek:2010xy,Aaij:2015yra,Hirose:2016wfn, Abdesselam:2019dgh, Aaij:2019wad, Aaij:2017vbb, Abdesselam:2019wac, Aaij:2015esa, Wehle:2016yoi} shows deviations from the theoretical SM predictions~\cite{Ligeti:2016npd,Crivellin:2016ejn, Amhis:2019ckw,Bigi:2016mdz,Bigi:2017jbd,Bordone:2016gaq,Capdevila:2017bsm,Iguro:2020cpg}. In particular, the current measurements of $R_D$~\cite{Amhis:2019ckw, Abdesselam:2019dgh} and
$R_{D^\ast}$~\cite{Aaij:2015yra,Hirose:2016wfn,Amhis:2019ckw,Abdesselam:2019dgh}
respectively reveal $1.4\sigma$ and $2.5\sigma$ deviations with respect to their SM predictions~\cite{Bigi:2016mdz,Bigi:2017jbd,Iguro:2020cpg} and, when combined, this amounts to a deviation of $3.1\sigma$ from the SM expectation~\cite{Ligeti:2016npd,Crivellin:2016ejn,Amhis:2019ckw}. In the neutral current $b\to s \ell^+\ell^-$ transitions,
the measurements of $R_K$~\cite{Aaij:2019wad,Aaij:2014ora} in the dilepton invariant mass squared bin $[1.1,6]~\text{GeV}^{2}$ show a deviation from the corresponding SM prediction~\cite{Bordone:2016gaq,Capdevila:2017bsm} at the level of $2.5\sigma$; for $R_{K^*}$, the measurements 
in the dilepton invariant mass squared bins $q^2\in[1.1,6]~\text{GeV}^{2}$ and $q^2\in[0.045,1.1]~\text{GeV}^{2}$~\cite{Aaij:2017vbb} reveal tensions with the SM expectations~\cite{Bordone:2016gaq,Capdevila:2017bsm}
with significances of $2.5\sigma$ and $2.4\sigma$,
respectively.
The recent Belle collaboration results for $R_{K^{\ast}}$ in the analogous bins~\cite{Abdesselam:2019wac} are consistent with both the SM and the LHCb measurements~\cite{Aaij:2017vbb}.
Furthermore, the LHCb measurement of $R_K$ in the bin $[1.1,6]~\text{GeV}^{2}$ has been recently updated~\cite{LHCb:2021trn}, now exhibiting a $3.1\,\sigma$ tension with respect to the SM prediction.

In addition to the LFUV ratios, further discrepancies with respect to the SM have also been identified in a small number of lepton flavour specific observables in $b\to s \ell^+\ell^-$ neutral current transitions - this is the case of several angular observables in both charged and neutral $B^{0,+}\to K^\ast \mu^+\mu^-$ decays (as recently reported by the LHCb collaboration~\cite{Aaij:2020nrf, Aaij:2020ruw}), for which tensions between observation and SM expectations lie around the $3\sigma$ level.
Very recent measurements~\cite{LHCb:2021zwz} of the differential branching fraction of $B_s\to\phi\mu^+\mu^-$ decays further corroborate the picture.
Moreover, LHCb recently updated~\cite{LHCb:2021qbv,LHCb:2021awg} their analysis of $B_{(s)}\to \mu^+\mu^-$ decays leading to an improved measurement of the $B_{(s)}\to \mu^+\mu^-$ branching fractions.

Many of the initial attempts to address the $B$-meson decay anomalies in terms of beyond the standard model (BSM) scenarios
have relied upon Effective Field Theory (EFT) approaches (see e.g.~\cite{Ligeti:2016npd,Bigi:2017jbd,Capdevila:2017bsm,Alguero:2019ptt,Aebischer:2019mlg,Ciuchini:2019usw,Datta:2019zca,Arbey:2019duh,Shi:2019gxi,Bardhan:2019ljo,Alok:2019ufo,Bhattacharya:2019eji,Alok:2017qsi,Ghosh:2014awa,Glashow:2014iga,Bhattacharya:2014wla,Freytsis:2015qca,Ciuchini:2017mik,Jaiswal:2017rve,Jaiswal:2020wer,Bhattacharya:2019dot,Biswas:2020uaq,Bhattacharya:2018kig} for some relevant studies). Extensive efforts have also been devoted to explain the anomalies - either separately or combined - in terms of specific NP constructions: 
among the most minimal scenarios
studied, heavy $Z^\prime$ mediators were identified as possible solutions (see for example ~\cite{Altmannshofer:2014cfa,Crivellin:2015mga,Crivellin:2015lwa,Sierra:2015fma,Crivellin:2015era,Celis:2015ara,Bhatia:2017tgo,Kamenik:2017tnu,Chen:2017usq,Camargo-Molina:2018cwu,Darme:2018hqg,Baek:2018aru,Biswas:2019twf,Allanach:2019iiy,Crivellin:2020oup}); likewise, numerous studies addressed the scalar and the vector leptoquark hypotheses (e.g.~\cite{Hiller:2014yaa,Gripaios:2014tna,Sahoo:2015wya,Varzielas:2015iva,Alonso:2015sja,Bauer:2015knc,Hati:2015awg,Fajfer:2015ycq,Das:2016vkr,Becirevic:2016yqi,Sahoo:2016pet,Cox:2016epl,Crivellin:2017zlb,Becirevic:2017jtw,Cai:2017wry,Dorsner:2017ufx,Buttazzo:2017ixm,Greljo:2018tuh,Sahoo:2018ffv,Becirevic:2018afm,Hati:2018fzc,Fornal:2018dqn,deMedeirosVarzielas:2018bcy,Aebischer:2018acj,Aydemir:2019ynb,Mandal:2018kau,deMedeirosVarzielas:2019okf,Yan:2019hpm,Bigaran:2019bqv,Popov:2019tyc,Hati:2019ufv,Crivellin:2019dwb,Saad:2020ihm,Dev:2020qet,Saad:2020ucl,Balaji:2019kwe,Cornella:2019hct,Mandal:2019gff,Babu:2020hun,Martynov:2020cjd,Fuentes-Martin:2020bnh,Guadagnoli:2020tlx}); further examples include
$R-$parity violating supersymmetric models (see for instance~\cite{Deshpand:2016cpw,Altmannshofer:2017poe,Das:2017kfo,Earl:2018snx,Trifinopoulos:2018rna,Trifinopoulos:2019lyo,Cohen:2019cge,Earl:2019adq,Hu:2019ahp,Hu:2020yvs,Altmannshofer:2020axr}, as well as other interesting
constructions~\cite{Greljo:2015mma,Arnan:2017lxi,Geng:2017svp,Choudhury:2017qyt,Choudhury:2017ijp,Grinstein:2018fgb,Cerdeno:2019vpd,Crivellin:2019dun,Arnan:2019uhr,Gomez:2019xfw}). 

Despite the large number of alternatives, only a select few scenarios can successfully put forward a simultaneous explanation for both charged and neutral current $B$-meson decay anomalies. Standard Model extensions relying on a $V_1$ vector leptoquark transforming as $(\mathbf{3},\mathbf{1},2/3)$ under the SM gauge group have received considerable attention in the literature~\cite{Assad:2017iib,Buttazzo:2017ixm,Calibbi:2017qbu,Bordone:2017bld,Blanke:2018sro,Bordone:2018nbg,Kumar:2018kmr,Angelescu:2018tyl,Balaji:2018zna,Fornal:2018dqn,Baker:2019sli,Cornella:2019hct,DaRold:2019fiw,Hati:2019ufv, Fuentes-Martin:2019ign, Fuentes-Martin:2020luw, Fuentes-Martin:2020hvc}, being currently the only single-leptoquark solution capable of simultaneously addressing both charged and neutral current anomalies.

\bigskip
In this work, our goal is to evaluate the prospects of 
a vector leptoquark transforming as $(\mathbf{3},\mathbf{1},2/3)$ as a viable hypothesis to address the current LFUV hints, fitting in a simplfied-model approach the $V_1$ couplings to SM fermions, further considering how current and upcoming experimental data may strengthen or disfavour such an hypothesis. The viability of the vector leptoquark $V_1$ as a solution of the current LFUV hints has been explored in detail in the existing literature:
in most of the existing studies, driven by an explanation of the anomalous data, only selected vector leptoquark couplings to 
specific quark and lepton flavours (while setting others to vanishing values) have been analysed;
other studies adopt an approach driven by an ultra-violet (UV) complete model (often assisted by additional flavour symmetries) to achieve the pattern of vector leptoquark couplings preferred by the anomalous data. 
In contrast, we pursue a distinct avenue in this study in what concerns the underlying analysis. We rather adopt a completely data-driven approach, in which we start from a "democratic" matrix for the vector leptoquark couplings to SM quarks and leptons, without any pre-bias towards particular flavours; the existing constraints from various flavour observables, together with the anomalous LFUV data, then determine the phenomenologically allowed ranges for the vector leptoquark couplings to different flavours of quarks and leptons. 
We thus present a fit of the full $3\times 3$  matrix of the vector leptoquark couplings taking into account all the relevant (and most stringent) measurements of various flavour observables as well as the anomalous LFUV data. As we proceed to discuss, our findings suggest that searches for a number of rare decays and transitions - conducted at Belle II and coming charged lepton flavour violation (cLFV) dedicated experiments - may help strengthening the case for $V_1$ models, or then contribute to exclude them as single-mediator explanations to the $R_{K^{(\ast)}}$ and $R_{D^{(\ast)}}$ anomalies.

Starting from a general effective theory framework, we first perform global fits taking into account the current experimental status of several observables associated with the anomalous $B$-meson decays, in complementary channels and kinematically interesting regions. We focus on the impact of the latest $b\to s\ell\ell$ data from LHCb~\cite{LHCb:2021zwz,LHCb:2021qbv,LHCb:2021awg,LHCb:2021trn,Aaij:2020nrf, Aaij:2020ruw} and how the new global fits of current flavour data favour specific classes of NP realisations. 

The bulk of our work is then devoted to the study of $V_1$.
Numerous ultra-violet (UV) complete models 
for vector leptoquarks with $\mathcal{O}$(TeV) mass, capable of addressing both charged and neutral current $B$-decay anomalies (while being consistent with the constraints from the charged lepton flavour violating decays $K_L\rightarrow \mu e$ and $K\rightarrow \pi \mu e$) have been proposed in the literature~\cite{Assad:2017iib,Buttazzo:2017ixm,Calibbi:2017qbu,Bordone:2017bld,Blanke:2018sro,Bordone:2018nbg,Kumar:2018kmr,Angelescu:2018tyl,Fornal:2018dqn,Baker:2019sli,Greljo:2018tuh,Cornella:2019hct,DaRold:2019fiw,Hati:2019ufv},
accompanied by extensive studies regarding 
LHC signatures and other phenomenological aspects. 
In this work, we pursue a distinct approach, 
relying on a simplified-model parametrisation of the vector leptoquark interactions with SM fermion fields, and study the impact of this scenario for a large set of observables -- various leptonic and semileptonic meson decays and cLFV observables (in addition to the ``anomalous'' $B$-meson  observables). 
The cLFV observables are very relevant not only given the current stringent constraints on the model, but also in view of the excellent (near future) projected experimental sensitivities.

Following 
an extensive global fit for the $V_1$ couplings in view of the experimental data on anomalous $B$-meson decay observables 
and data on other relevant (semileptonic) processes
(meson decays and cLFV observables offering stringent constraints), we identify 
several tauonic modes and cLFV transitions which are likely to play a key r\^ole in testing the vector leptoquark scenario as a unified explanation to the $B$-decay anomalies. 
As we emphasise in this work, $\tau\to \phi \mu$ decays emerge as one of the ``golden channels'' for probing the $V_1$ hypothesis at Belle II; if the $B$-meson decay anomalies persist at their current level, sizeable contributions - within future experimental reach -, are also expected for $B_s \rightarrow \tau^+ \mu^-$, $B^+ \rightarrow K^+ \tau^+ e^-$, $B^+ \rightarrow K^+ \tau^+ \mu^-$ and $B_s \rightarrow \phi \tau^+ \mu^-$ decays, as also noticed in~\cite{Glashow:2014iga,Angelescu:2018tyl}. Owing to their impressive expected future sensitivity, the upcoming cLFV experiments dedicated to searching for neutrinoless $\mu - e$ conversion in Aluminium nuclei, Mu2e and COMET, will probe a large part of the preferred $V_1$ parameter space via 
$\mu$ and $e$ couplings to all quark generations.
Furthermore, and as a direct consequence of accommodating the charged current anomalies in $R_{D^{(*)}}$, our study reveals that a number of $b\to s\tau\tau$ branching fractions are expected to be enhanced with respect to the SM (by one to two orders of magnitude), as first pointed out in~\cite{Capdevila:2017iqn}.

Our study is organised as follows: following an EFT-based 
global fit in Section~\ref{sec:WilsonCfits} (allowing to identify the currently best favoured NP classes of models), Section~\ref{VLQmodel} is devoted to vector leptoquark realisations: in particular, and after introducing the simplified approach to this class of NP models, we perform a global fit of the $V_1$ couplings to SM fermions, taking into account a thorough set  of flavour observables. Finally, Section~\ref{sec:futureimpact} contains a discussion of the prospects for probing the vector leptoquark hypothesis as a solution to the anomalous $B$-meson decay data through several mesonic and leptonic decays to be searched for at Belle II and future cLFV-dedicated facilities. 
Complementary and/or detailed information relevant to the study is collected in several appendices. 
\section{Semileptonic $\pmb{B}$-meson decays: the impact of new LHCb data}\label{sec:WilsonCfits}
In this section we will address the $B$-meson decay
observables currently pointing towards a violation of LFU. 
Our goal is to evaluate how new recent data~\cite{LHCb:2021zwz,LHCb:2021qbv,LHCb:2021awg,LHCb:2021trn,Aaij:2020nrf, Aaij:2020ruw} has impacted the global fits carried out for an EFT approach to NP models 
(i.e. for generic BSM realisations), in terms of the relevant semileptonic Wilson coefficients $C^{q q^\prime; \ell \ell^\prime }$. We aim at investigating how well-motivated scenarios for (sets of) $C^{q q^\prime; \ell \ell^\prime }$ 
- resulting from different effective operators - remain viable or have become disfavoured.

We thus begin by briefly commenting on charged current  $b\to c\ell\nu$ decay data; we then proceed to 
discuss the experimental status of several observables associated with neutral current semileptonic $B$-meson decays, focusing our attention on global fits of 
$b\to s\ell\ell$ transitions, and how the status of the latter has evolved in recent months.

\subsection{\bf{$\pmb{b \to c \tau \nu}$ data and new-physics interpretations}}

A number of reported results from several experimental collaborations have
suggested a possible violation of lepton flavour
universality in the charged current decay mode $B \to D^{(*)} \ell \nu$, parametrised by the $R_{D}$ and $R_{D^{(*)}}$ ratios (see Eq.~(\ref{eq:RDRK:def})). 
The latest average values of these observables, given by the HFLAV collaboration~\cite{Amhis:2019ckw}, are 
\begin{eqnarray}\label{eq:RD*:expSMsigma}
 R_{D}\, =\, 0.340\,\pm\, 0.027\,\pm\,0.013\,,\quad
& R_{D}^{\text{SM}}\, =\, 0.299\, \pm\, 0.003 \, \quad
& (1.4 \sigma) \,;
 \nonumber\\
R_{D^*}\, =\, 0.295\,\pm\, 0.011\,\pm\,0.008\,,\quad
& R_{D^*}^{\text{SM}}\, =\, 0.258\, \pm\, 0.005 \, \quad
& (2.5 \sigma) \,.
 \end{eqnarray}

\noindent
The relevant effective Lagrangian for the charged current
transitions $d_k\to u_j\bar{\nu}\ell^{-}$ can be expressed as
\begin{eqnarray}\label{eq:Heff:RD}
\mathcal L_\text{eff} = -{4\,G_F \over \sqrt2} \,V_{jk} &\times&\left[
	(1+C_{V_L}^{jk;\ell i}) (\bar{u}_j \,\gamma_\mu\, P_L\, d_k) (\bar{\ell}\, \gamma^\mu\, P_L \,\nu^i) +
	C_{V_R}^{jk;\ell i} (\bar{u}_j \,\gamma_\mu \,P_R \,d_k) (\bar{\ell} \,\gamma^\mu\, P_L \,\nu^i) \right. \nonumber\\
	&+& C_{S_L}^{jk;\ell i} (\bar{u}_j \,P_L \,d_k) (\bar{\ell} \,P_L\, \nu^i)
	  + C_{S_R}^{jk;\ell i} (\bar{u}_j \,P_R \,d_k) (\bar{\ell}\, P_L\, \nu^i) \nonumber\\
	&+& \left. C_{T_L}^{jk;\ell i} (\bar{u}_j \sigma_{\mu\nu} \,P_L\, d_k) (\bar{\ell} \sigma^{\mu\nu}\, P_L\, \nu^i) \right] + \mathrm{H.c.}\,,
\end{eqnarray}
in which we have assumed the neutrinos to be left-handed and where, for the SM, we have $C_i=0$, $\forall i\in \{S_L,S_R,V_L,V_R,T_L\}$. For the convenient
double ratios $R_{D}/R_{D}^\text{SM}$ and $R_{D^\ast}/R_{D^\ast}^\text{SM}$ (which combine the current experimental averages with the SM predictions), the current data can be summarised
as $R_{D}/R_D^\text{SM} \,= \,1.14 \pm 0.10\, , \;
R_{D^{\ast}}/R_{D^{\ast}}^\text{SM} 
\,= \,1.14 \pm 0.06$,
where the statistical and systematical errors have been added in quadrature. 

To perform a numerical analysis of the transition $B \to D^{(*)} \tau \nu$ (and fit the above double ratios) one further requires knowledge of the hadronic form factors which parameterise the vector, scalar and tensor current matrix elements. However, under the simplifying assumption of a non-vanishing single type of NP operator at a time - i.e. $C_i\neq 0$, $i\in \{S_L,S_R,V_L,V_R,T_L\}$ -, it is possible to draw some qualitative conclusions from the approximate numerical forms for the double ratios using a heavy quark effective theory (HQET) formalism~\cite{Iguro:2018vqb,Sakaki:2012ft,Tanaka:2012nw,Hati:2015awg,Neubert:1991td,Hagiwara:1989cu,Caprini:1997mu}.

In particular, and if one assumes that all the relevant Wilson coefficients are real, then the following qualitative observations can be readily made. The operator corresponding to $C_{V_L}$ contains the same Lorentz structure as the SM contribution and the NP amplitude adds to the SM one, thus leading to similar enhancements to both $R_{D}$ and $R_{D^\ast}$, which are proportional to $(1+C_{V_L})^2$. In turn, this leads to similar fractional enhancements to $R_{D}/R_{D}^\text{SM}$ and $R_{D^\ast}/R_{D^\ast}^\text{SM}$. Therefore, $C_{V_L}$ is one of the most favoured choices for explaining the anomalous $R_{D}$ and $R_{D^\ast}$ data. On the other hand, if the new physics contribution is purely 
a right-handed vector current
($C_{V_R}$ type), then for a real $C_{V_R}$, $R_{D}$ is proportional to $(1+C_{V_R})^2$ while $R_{D^\ast}$ is roughly proportional to $(1-C_{V_R})^2$. Under such circumstances, it is then not possible to simultaneously explain both $R_{D}$ and $R_{D^\ast}$ data. 
However, and as discussed in~\cite{Iguro:2018vqb}, this conclusion is no longer valid for a complex $C_{V_R}$. The scalar operators corresponding to $C_{S_L}$ and $C_{S_R}$ contain the pseudoscalar Dirac bilinear and therefore are not subject to helicity suppressions, leading to stringent constraints from the (relatively large) branching ratios of $B_c\rightarrow \tau \nu$. The tensor operator, corresponding to $C_{T_L}$, is subject to tensions from the recent measurement of the $D^{*}$ longitudinal polarisation  $f_L^{D^*}$, which is currently about $1.6\, \sigma$ higher than the SM prediction  and has a discriminatory power between the scalar and tensor solutions~\cite{Alok:2017qsi,Blanke:2019qrx,Shi:2019gxi}. Choices based on pure right-handed operators seem to be disfavoured by LHC data~\cite{Shi:2019gxi,Cornella:2019hct}. Finally, scenarios that only present scalar contributions are in conflict with both LHC and $B_c\rightarrow \tau \nu$ data.

\subsection{Neutral current $\pmb{b\to s\ell\ell}$ decays}
\label{sec:bsll}
A number of anomalies reported in $b\to s\ell\ell$ observables currently stand as promising hints of NP, among them those parametrised by the 
$R_{K^{(*)}}$ ratios, defined in Eq.~(\ref{eq:RDRK:def}).
The latest averages of the reported anomalous experimental data, together with the SM predictions can be expressed as~\cite{Aaij:2019wad,Aaij:2017vbb,Abdesselam:2019wac}
	\begin{eqnarray}\label{eq:RK*:expSMsigma}
	  R_{K [1.1,6]}^{\text{LHCb}}\, &=&\, 0.846\,\pm^{0.042}_{0.039}\,
	  \pm^{0.013}_{0.012}\,, \quad
	 R_{K}^{\text{SM}}\, =\, 1.0003\, \pm\, 0.0001
	  \,,
	 \nonumber\\
	 R_{K^*[0.045, 1.1]}^{\text{LHCb}}\, &=&\, 0.66 ^{+0.11}_{-0.07} \,\pm\,
	 0.03\,, \quad
	 R_{K^*[0.045, 1.1]}^{\text{Belle}}\, =\, 0.52 ^{+0.36}_{-0.26} \,\pm\,
	 0.05\,, \quad
	R_{K^*[0.045, 1.1]}^{\text{SM}}\,  \sim\,  0.93 \,,
	 \nonumber\\
	 R_{K^* [1.1, 6]}^{\text{LHCb}} \,&=& \,0.69 ^{+0.11}_{-0.07}\, \pm
	0.05\,, \quad
	 R_{K^* [1.1, 6]}^{\text{Belle}} \,= \,0.96 ^{+0.45}_{-0.29}\, \pm
	0.11\,, \quad
	 R_{K^* [1.1, 6]}^{\text{SM}} \,\sim \, 0.99\, ,
	\end{eqnarray}
where the dilepton invariant mass squared bin (in $\text{GeV}^{2}$) is identified by the associated subscripts. 
Further anomalies have also been reported in the neutral current decay modes of $B$-mesons for semileptonic final states including muon pairs\footnote{Notice that here we refer to the neutral and {\it charged} $B$-meson decays, i.e. 
$B^{0,+} \to K^{*} \mu \mu$ decays.}.
Among them, one concerns the observable $\Phi \equiv d {\rm BR}(B_s\to\phi\mu\mu)/ dq^2$ in the bin
$q^2\in[1,6]\,{\rm GeV}^2$~\cite{Aaij:2015esa}, presently exhibiting a tension with the SM prediction around $3\sigma$. Further discrepancies with respect to the SM, typically at the $3\sigma$ level, have also emerged in relation to the angular observables. In particular, this is the case of
$P_5^{\prime}$ in $B \to K^\ast \ell^+ \ell^-$	processes: 
the results from the LHCb collaboration for $P_5^{\prime}$ regarding muon final states ($B \to K^\ast	\mu^+ \mu^-$ decays) reveal a discrepancy with respect to the SM~\cite{Aaij:2013qta,Aaij:2015oid}; the Belle collaboration~\cite{Abdesselam:2016llu,Wehle:2016yoi} reported that $P_5^{\prime}$ results for electrons show a better agreement with theoretical SM expectations than those for muons. 
More recently, similar measurements have also been reported by the ATLAS~\cite{Aaboud:2018krd} and CMS~\cite{Sirunyan:2017dhj} collaborations. 
The 2015 LHCb results~\cite{Aaij:2015oid} and the ATLAS result~\cite{Aaboud:2018krd} for $P_5^{\prime}$ in the low dimuon invariant mass-squared
range, $q^2\in[4,6]\,{\rm GeV}^2$, indicate a $\approx 3.3\sigma$ discrepancy with respect to the
SM prediction~\cite{Aebischer:2018iyb}. Belle
results corroborate the latter findings, showing a deviation of $2.6\sigma$ from the SM expectation in the bin $q^2\in[4,8]\,{\rm GeV}^2$~\cite{Wehle:2016yoi}. 
The reported CMS measurement (possibly as a consequence of insufficient statistics) is still consistent with the SM expectation within $1\sigma$~\cite{Sirunyan:2017dhj}. 
Among the angular observables it is important to stress that $F_L$, $P_4^{\prime}$ , $P_5^{\prime}$ and $P_8^{\prime}$ have been a driving force in the evolution of the global fits.
Very recently, the LHCb collaboration has updated the results for the angular observables relying on 4.7 fb$^{-1}$ of data~\cite{Aaij:2020nrf, Aaij:2020ruw}: local discrepancies of $2.5\sigma$ and $2.9\sigma$, respectively in the bins $q^2\in[4,6]\,{\rm GeV}^2$ and $q^2\in[6,8]\,{\rm GeV}^2$ GeV$^2$, were reported.
While these lepton flavour dependent observables are also sensitive to the presence of NP~\cite{Altmannshofer:2008dz,Bobeth:2011nj,Matias:2012xw,DescotesGenon:2012zf,Matias:2014jua}, they are nevertheless subject to hadronic uncertainties (for example form factors, power corrections and charm resonances~\cite{Khodjamirian:2010vf,Khodjamirian:2012rm,Lyon:2014hpa,Descotes-Genon:2014uoa,Capdevila:2017ert,Blake:2017fyh,Jager:2012uw,Jager:2014rwa,Ciuchini:2015qxb,Ciuchini:2016weo,Bobeth:2017vxj,Gubernari:2020eft}) contrary to the LFUV ratios, which are in general free of the latter sources of uncertainty.

\bigskip
A way to consistently analyse the aforementioned anomalous experimental data is to adopt the ``effective approach'', in which all possible short-distance NP effects are encoded in the Wilson coefficients related to a complete EFT basis. 
Within a weak effective theory (WET), the effective Lagrangian for a general $d_j \rightarrow d_i \ell^- \ell^{\prime +}$ transition can be expressed as~\cite{Buchalla:1995vs,Bobeth:1999mk,Ali:2002jg,Hiller:2003js,Bobeth:2007dw,Bobeth:2010wg}
\begin{equation}\label{eqn:effL}
\mathcal L_\text{eff} =
 \frac{4 G_F}{\sqrt{2}} V_{3j}\,V_{3i}^{\ast}\Big[ \sum_{
    \begin{array}{c}
      k=7,9,\\
      10,S,P
    \end{array}
  } \hspace{-5mm}\left(C_k (\mu) \,\mathcal{O}_k(\mu) + 
  C_k^{'} (\mu)\,\mathcal{O}_k^{'} (\mu)\right) + 
  C_T (\mu) \,\mathcal{O}_T(\mu) +
  C_{T_5} (\mu) \,\mathcal{O}_{T_5}(\mu)\Big]\, ,
\end{equation}
with $V_{ij}$ denoting the CKM matrix and in which the relevant operators are defined as
\begin{align}\label{eqn:operators}
\mathcal O_7^{ij} &= \,\frac{e \,m_{d_j}}{(4\pi)^{2}}
(\bar d_i\,
\sigma_{\mu\nu}\,P_R\, d_j)\,F^{\mu\nu}\:\text, \quad\quad\quad
\mathcal{O}_9^{ij;\ell \ell^{\prime}} =\, \frac{e^{2}}{(4\pi)^2}
(\bar
d_i \,\gamma^{\mu}\,P_L\, d_j)(\bar \ell \,\gamma_\mu \,\ell^\prime)\:\text,  \nonumber \\
\mathcal{O}_{10}^{ij;\ell \ell^{\prime}} &=\,
\frac{e^{2}}{(4\pi)^2}(\bar d_i \,\gamma^{\mu}\,P_L d_j)(\bar \ell\,
\gamma_\mu \,\gamma_5 \,\ell^\prime)\:\text, \quad
\mathcal{O}_S^{ij;\ell \ell^{\prime}} =\, \frac{e^{2}}{(4\pi)^2}(\bar
d_i \,P_R \,d_j)(\bar\ell \,\ell^\prime)\:\text, \nonumber \\
\mathcal{O}_P^{ij;\ell \ell^{\prime}} &=\, \frac{e^{2}}{(4\pi)^2}(\bar
d_i \,P_R \,d_j)(\bar\ell\,\gamma_5 \,\ell^\prime)\:\text, \quad\quad\quad
\mathcal O_T^{ij;\ell \ell^{\prime}} =\, \frac{e^{2}}{(4\pi)^{2}}(\bar
d_i \sigma_{\mu\nu} \,d_j)(\bar \ell \sigma^{\mu\nu}\,
\ell^\prime)\:\text, \nonumber \\
\mathcal O_{T5}^{ij;\ell \ell^{\prime}} &=\,
\frac{e^{2}}{(4\pi)^{2}}(\bar d_i \sigma_{\mu\nu}\, d_j)(\bar \ell
\sigma^{\mu\nu}\,\gamma_5 \,\ell^\prime)\:\text,
\end{align}
where the primed operators $\mathcal O_{7, 9, 10, S, P}^{\prime}$
correspond to the exchange $P_L \leftrightarrow P_R$. Given the above WET parametrisation,
the first question to address concerns the set(s) of Wilson coefficients seemingly
preferred by the anomalous experimental data, which then leads to the identification of possible phenomenological candidates, and ultimately to the construction of UV complete extensions of the SM. 

\bigskip
Let us then first proceed to obtain model-independent fits for different possible new physics scenarios (in terms of non-vanishing contributions to one or several Wilson coefficients in the FCNC $b\to s\ell\ell$ transitions), with a particular emphasis on the impact of the recent data from the LHCb collaboration~\cite{LHCb:2021zwz,LHCb:2021qbv,LHCb:2021awg,LHCb:2021trn,Aaij:2020nrf, Aaij:2020ruw}. 

\subsubsection{\bf Fits of \pmb{$b\to s\ell\ell$} data: before 2020}

We thus begin by performing a fit of the data on angular distributions, differential branching fractions and the LFUV ratios $R_{K^{(\ast)}}$ - excluding the new measurements of LHCb~\cite{LHCb:2021zwz,LHCb:2021qbv,LHCb:2021awg,LHCb:2021trn,Aaij:2020nrf, Aaij:2020ruw} -  to establish a baseline to study the impact of the new measurements from LHCb. 
The underlying methodology for the fit as well as the details of the statistical methods are described in Appendix~\ref{app:stats}; the specific bins of observables and datasets used for the fit are presented in Appendix~\ref{app:bsll}.

Using these data sets, one can already infer a qualitative behaviour of the fits in terms of the Wilson coefficients (allowing to identify favoured NP ``scenarios'').
While NP contributions exclusively to $C_9^{bs\mu\mu}$ already give a very good fit when compared to the SM~\cite{Descotes-Genon:2015uva,Altmannshofer:2017fio,Alok:2017sui,Altmannshofer:2017yso,Geng:2017svp,Ciuchini:2017mik,Capdevila:2017bsm,Alguero:2019ptt,Alok:2019ufo,Ciuchini:2019usw,Datta:2019zca,Aebischer:2019mlg,Kowalska:2019ley,Arbey:2019duh}, most realistic NP models considered to explain the tensions also generate non-zero contributions to other Wilson coefficients, either by construction or then through operator mixings which occur when renormalisation group (RG) running effects (from the NP mass scale to the observable scale) are taken into account. 
In particular, the $SU(2)_L$ conserving scenario $\Delta C_9^{bs\mu\mu} = - \Delta C_{10}^{bs\mu\mu}$ has received a considerable attention in recent years, as it provides a very good fit to the data. However, following the improvement in the measurement of $R_K$ in 2019~\cite{Aaij:2019wad}, with relatively smaller experimental uncertainties, the preference has been slightly shifted to more involved scenarios, calling upon a larger number of non-vanishing Wilson coefficients. This becomes manifest through tensions between the individual fits  for the LFUV ratios, and for the lepton flavour dependent observables ($\Phi \equiv d {\rm BR}(B_s\to\phi\mu\mu)/ dq^2$ and the angular observable $P_5^{\prime}$ in $B \to K^\ast \ell^+ \ell^-$), under the hypotheses of $\Delta C_9^{bs\mu\mu} = - \Delta C_{10}^{bs\mu\mu}$ (and $C_9^{bs\mu\mu}$) NP ``scenario(s)''. 
Therefore, if the anomalous data for the lepton flavour dependent observables is not due to statistical fluctuations or to long-distance effects, then such tensions suggest non-trivial NP contributions in other lepton flavours, or in distinct Wilson coefficients. For example, in Ref.~\cite{Datta:2019zca}, it was reported that LFUV contributions in $b\to s e^+e^-$, in addition to a minimal ``scenario'' (i.e. $\Delta C_9^{bs\mu\mu} = - \Delta C_{10}^{bs\mu\mu}$ and $C_9^{bs\mu\mu}$) can ease such tensions, further improving the overall fit to data 
with respect to the SM. On the other hand, in~\cite{Alguero:2018nvb} it was observed that if one considers a LFUV scenario which only 
affects muons in conjunction with a non-vanishing LFU NP contribution (i.e. with equal contributions to $e$, $\mu$, and $\tau$), then the anomalous LFUV ratio data can be explained by the LFUV in the muon sector (with sub-leading interferences with LFU NP contributions), while the lepton flavour dependent observables can be fitted combining LFUV and LFU NP, with improved agreement with respect to the overall data. 

Therefore, in our global fit scenarios, we include the above two interesting possibilities, comparing individual and combined effects. 

\begin{table}[h!]
\begin{center}
\begin{tabular}{|c|c|c|c|c|}
\hline
NP ``scenario'' & best-fit & $1\sigma$ range & $\text{pull}_\text{SM}$ & $p$-value\\
\hline
\hline
$\Delta C_9^{bs\mu\mu}$ & $-0.91$ & $ [-1.19, -0.73] $ & $5.13$ & $51.1\%$\\
\hline
$\Delta C_9^{bs\mu\mu} = -\Delta C_{10}^{bs\mu\mu}$ & $-0.45$ & $[-0.54, -0.36]$ & $5.21$ & $53.4 \%$\\
\hline
\hline
$\Delta C_9^{bs\mu\mu}$ & $-0.76$ & $ [-0.97, -0.54]$ & $ 5.08$ & $57.0\% $\\
$\Delta C_{10}^{bs\mu\mu}$ & $0.25$ & $[0.11, 0.40]$ & &\\
\hline
$\Delta C_9^{bs\mu\mu} = -\Delta C_{10}^{bs\mu\mu}$ & $-0.65$ & $[-0.82, -0.49]$ & $5.08$ & $57.1\%$\\
$\Delta C_9^{bsee} = -\Delta C_{10}^{bsee}$ & $-0.34$ & $[-0.55, -0.13]$ & & \\
\hline
$\Delta C_9^{bs\mu\mu} = -\Delta C_{10}^{bs\mu\mu}$ & $-0.35$ & $[-0.45, -0.26] $ & $5.43$ & $66.5\%$\\
$\Delta C_9^\text{univ.}$ & $ -0.68$ & $[-0.92, -0.42]$ & &\\
\hline
\hline
$\Delta C_9^{bs\mu\mu} = -\Delta C_{10}^{bs\mu\mu}$ & $-0.45$ & $[-0.64, -0.27]$ & $5.18$ & $65.0\%$\\
$\Delta C_9^{bsee} = -\Delta C_{10}^{bsee}$ & $-0.14$ & $[-0.37, 0.09]$ & & \\
$\Delta C_9^\text{univ.}$ & $-0.61$ & $[-0.88, -0.32]$ & & \\
\hline
\end{tabular}
\caption{Well-motivated NP ``scenarios'' (sets of Wilson coefficients) and corresponding fits to the data on angular distributions, differential branching fractions and the LFUV ratios $R_{K^{(\ast)}}$ (not including the new measurements of LHCb~\cite{LHCb:2021zwz,LHCb:2021qbv,LHCb:2021awg,LHCb:2021trn,Aaij:2020nrf, Aaij:2020ruw}). The SM $p$-value is found to be $\sim6.8\%$. }
\label{tab:bsll_2019}
\end{center}
\end{table}

As can be seen in Table~\ref{tab:bsll_2019}, 
we find that LFU contributions to $C_9^{bs\mu\mu}$ and $C_9^{bsee}$ ($\Delta C_9^\text{univ.}$, corresponding to the last two blocks in the table) are able to significantly improve the model-independent fits. 
It is interesting to note that these scenarios naturally arise in many simple models attempting a combined explanation of $b\to s\ell\ell$ data together with the anomalous charged current data on $b\to c \tau\nu$. In particular, a sizeable contribution to the charged current Wilson coefficients to explain $b\to c \tau\nu$ calls upon large $\tau$-couplings, which in turn generate a sizeable $C_9^{bs\tau\tau}$. Through RG operator mixing effects~\cite{Crivellin:2018yvo} (evolution from NP scale to the observable scale),  
the $C_9^{bs\tau\tau}$ contribution leads to a LFU contribution for both $C_9^{bs\mu\mu}$ and $C_9^{bsee}$. 
We notice that this LFUV and LFU NP combined ``scenario'' is of relevance for $SU(2)_L$-singlet vector-leptoquark models since, due to the $SU(2)_L$ representation, the charged current couplings in $b\to c\tau \nu$ transitions are identical to the ones appearing in the neutral current $b\to s \ell\ell$ transitions (up to CKM elements in the effective Wilson coefficients).

\subsubsection{\bf Fits of \pmb{$b\to s\ell\ell$} data in 2021}
To estimate the impact of the recent measurements of LHCb~\cite{LHCb:2021zwz,LHCb:2021qbv,LHCb:2021awg,LHCb:2021trn,Aaij:2020nrf, Aaij:2020ruw}, we repeat the Wilson coefficient fit of the previous subsection, but now taking into account the new LHCb data~\cite{LHCb:2021zwz,LHCb:2021qbv,LHCb:2021awg,LHCb:2021trn,Aaij:2020nrf, Aaij:2020ruw}. We further take into account the recently improved limits on $\mathrm{BR}(B^0\to e^+e^-)$ and $\mathrm{BR}(B_s\to e^+e^-)$, ~\cite{Aaij:2020nol}, and a recently improved measurement of the angular observables in $B^0\to K^\ast e^+e^-$ at very low $q^2$, as reported by the LHCb collaboration~\cite{Aaij:2020umj}.

\begin{table}[h!]
\begin{center}
\begin{tabular}{|c|c|c|c|c|}
\hline
NP ``scenario'' & best-fit & $1\sigma$ range & $\text{pull}_\text{SM}$ & $p$-value\\
\hline
\hline
$\Delta C_9^{bs\mu\mu}$ & $-0.92$ & $ [-1.07, -0.77] $ & $6.09$ & $29.2\%$\\
\hline
$\Delta C_9^{bs\mu\mu} = -\Delta C_{10}^{bs\mu\mu}$ & $-0.39$ & $[-0.47, -0.32]$ & $5.51$ & $18.3\%$\\
\hline
\hline
$\Delta C_9^{bs\mu\mu}$ & $-0.86$ & $ [-1.03, -0.66]$ & $ 5.81$ & $28.7\% $\\
$\Delta C_{10}^{bs\mu\mu}$ & $0.10$ & $[-0.02, 0.22]$ & &\\
\hline
$\Delta C_9^{bs\mu\mu} = -\Delta C_{10}^{bs\mu\mu}$ & $-0.62$ & $[-0.79, -0.46]$ & $5.39$ & $20.6\%$\\
$\Delta C_9^{bsee} = -\Delta C_{10}^{bsee}$ & $-0.30$ & $[-0.39, -0.12]$ & & \\
\hline
$\Delta C_9^{bs\mu\mu} = -\Delta C_{10}^{bs\mu\mu}$ & $-0.33$ & $[-0.41, -0.25]$ & $6.35$ & $41.9\%$\\
$\Delta C_9^\text{univ.}$ & $-0.86$ & $[-1.05, -0.66]$ & & \\
\hline
\hline
$\Delta C_9^{bs\mu\mu} = -\Delta C_{10}^{bs\mu\mu}$ & $-0.37$ & $[-0.55, -0.20]$ & $6.08$ & $40.0\%$\\
$\Delta C_9^{bsee} = -\Delta C_{10}^{bsee}$ & $-0.04$ & $[-0.24, 0.15]$ & & \\
$\Delta C_9^\text{univ.}$ & $-0.84$ & $[-1.06, -0.61]$ & & \\
\hline
\end{tabular}
\caption{Well-motivated NP ``scenarios'' and corresponding fits to the data on angular distributions, differential branching fractions and the LFUV ratios $R_{K^{(\ast)}}$ 
(as in Table~\ref{tab:bsll_2019}),  now
including the recent LHCb measurements~\cite{Aaij:2020nrf,Aaij:2020nol,Aaij:2020ruw,Aaij:2020umj,LHCb:2021zwz,LHCb:2021qbv,LHCb:2021awg,LHCb:2021trn} (see Appendix~\ref{app:bsll}). The SM $p$-value is now found to be $\sim1.0\%$. }
\label{tab:bsll_2020}
\end{center}
\end{table}

\begin{figure}[h!]
\centering
\includegraphics[width=0.75\textwidth]{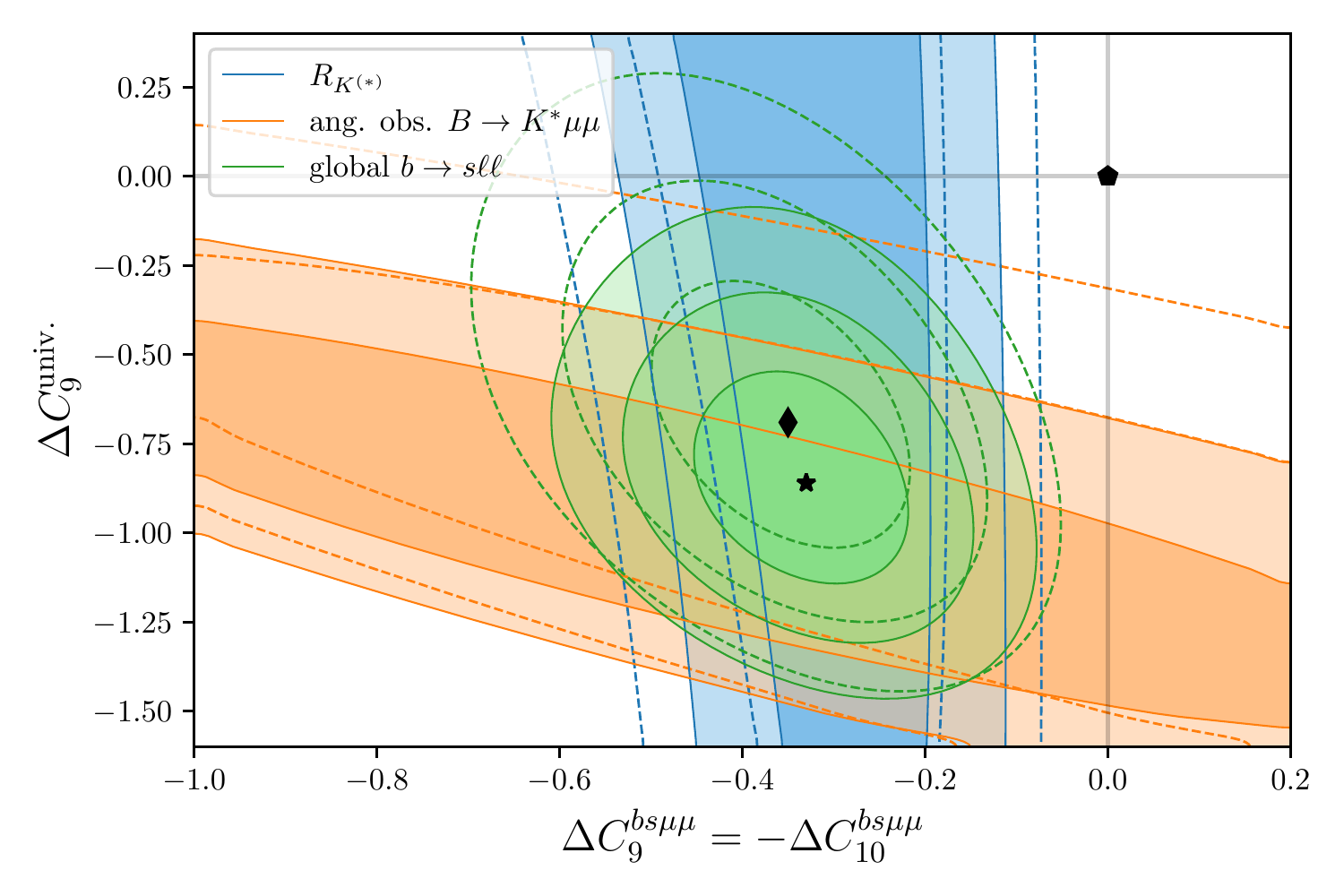}
\caption{Likelihood contours for the $b\to s\ell\ell$-data in the plane spanned by $\Delta C_9^{bs\mu\mu} = -\Delta C_{10}^{bs\mu\mu}$ and $\Delta C_9^\text{univ.}$, corresponding to the scenario with the largest $p$-value (see Tables~\ref{tab:bsll_2019} and~\ref{tab:bsll_2020}). The shaded regions (delimited by full lines) correspond to the $1, 2\,(3) \sigma$ regions around the best-fit point including the recent data; the dashed lines denote the same likelihood contours, without the inclusion of the recent LHCb measurement.
In addition to the angular observables and $R_{K^{(*)}}$, the ``global'' contour (green regions) includes all other $b\to s\ell\ell$ data as listed in Appendix~\ref{app:bsll}.
The black pentagon denotes the SM-value, while the star (diamond) denotes the best-fit point to the current (old) data.
}
\label{fig:wcfit}
\end{figure}

In the fits carried out for the older data sets, the scenario $\Delta C_9^{bs\mu\mu} = -\Delta C_{10}^{bs\mu\mu}$ had a larger $p$-value than the fit which only included NP contributions to $C_9^{bs\mu\mu}$; as can be seen from Table~\ref{tab:bsll_2020}, the situation is now reversed upon inclusion of the new LHCb data. 
Furthermore, $\Delta C_9^{bs\mu\mu} = -\Delta C_{10}^{bs\mu\mu}$ arguably provided an equally good fit (c.f. Table~\ref{tab:bsll_2019}) to the data compared to the hypotheses which included a {\it universal} contribution to $C_9^{bs\ell\ell}$ in addition to the $(V-A)$ contribution, whereas now the hypotheses with a {\it universal} contribution are clearly preferred.
In Fig.~\ref{fig:wcfit} we present the likelihood contours for the ``pre-2020'' and recent data, around the corresponding best-fit points, where it can be seen that a non-vanishing {\it universal} contribution to $C_9$ is now preferred at around $\sim3\sigma$. 
Although the position of the best fit point is only slightly changed (from the former diamond to the current star), the new measurement leads to an improved precision for the model-independent fits. This is manifest  from the comparison of the likelihood contours belonging to either dataset (regions delimited by dashed or solid lines, respectively in association with ``pre-2020'' and full data).

This renders models that attempt a combined explanation of the charged and neutral current $B$-decay anomalies, especially single-mediator scenarios, even more preferable. Following this section's discussion, in the remainder of our study we will focus on such NP realisations, 
in particular extensions of the SM via single 
left-handed vector fields (in our case, a $V_1$ vector leptoquark), which provide the best fit among all the possibilities for single-mediator NP 
scenarios~\cite{Alok:2017qsi,Blanke:2019qrx,Shi:2019gxi}.

\section{Implications for $V_1$ vector leptoquark solutions}\label{VLQmodel}
Among the many possible SM extensions including leptoquarks, in what follows we focus on vector leptoquark
($V_1$) scenarios.
This possibility has received increasing attention in the literature, as it is currently the only single-leptoquark construction that successfully offers a simultaneous solution to both charged and neutral current $B$-meson decay anomalies~\cite{Assad:2017iib,Buttazzo:2017ixm,Calibbi:2017qbu,Bordone:2017bld,Blanke:2018sro,Bordone:2018nbg,Kumar:2018kmr,Angelescu:2018tyl,Balaji:2018zna,Fornal:2018dqn,Baker:2019sli,Cornella:2019hct,DaRold:2019fiw,Hati:2019ufv}. 
As highlighted following the updated global fits carried out in the previous section, the vector leptoquark hypothesis belongs to the class of NP ``scenarios'' most favoured by current data.

However, and in order to account for experimental data, 
$V_1$ should have non-universal couplings to quarks and leptons, and the latter can be realised in a number of ways.
The most minimal possible scenario relies in the assumption that the vector leptoquark is an elementary gauge boson\footnote{There are also models in which the vector leptoquark appears as a composite field, see for instance~\cite{Barbieri:2016las,Cline:2017aed}.}, associated to a non-abelian gauge group extension of the SM, under which the SM fermion generations are universally charged; in the unbroken phase of the underlying extended gauge group, the leptoquark gauge couplings 
also remain universal. Despite its simplicity, 
this scenario is challenged by constraints from the
cLFV decays $K_L\rightarrow \mu e$ and $K\rightarrow \pi \mu e$: current limits force the mass of such a vector leptoquark to be very heavy, $m_V\ge 100$~TeV for $\mathcal O(1)$ couplings~\cite{Hung:1981pd,Valencia:1994cj,Smirnov:2007hv,Carpentier:2010ue,Kuznetsov:2012ai,Smirnov:2018ske}, and thus excessively heavy to account for both the charged and neutral current $B$-meson decay anomalies. 
In order to understand this, notice that while $V_1$ has a  universal coupling to SM fermions in the unbroken phase, after $SU(2)_L$-breaking a potential misalignment of the quark and lepton eigenstates is generated, leading to LFU-violating $V_1$ couplings. Given the constraints from $\tau$ decays, the $c\nu$ coupling generated from $b\tau$ through CKM mixing is not sufficiently large to account for $R_{D^{(\ast)}}$ data~\cite{Feruglio:2017rjo}. On the other hand, for a maximal $c\nu$ coupling (with the neutrino flavour in $c\nu$ different from $\nu_\tau$) generated by $d_i\mu$ and $d_i e$ couplings, important constraints arise from $R_{K^{(*)}}$ data for $i=2,3$, and from kaon decays for $i=1$. Moreover, the $c\nu$ coupling induced by $d\tau$ is heavily CKM-suppressed. Therefore, the only viable possibility is to maximise both $b\tau$ and $s\tau$ couplings, which in turn will induce large couplings between the first two generations of quarks and leptons (given the unitarity of the post-$SU(2)_L$-breaking mixing matrix), thus implying excessive contributions to cLFV.

A possible way to circumvent the above mentioned constraints is to introduce three ``generations'' of vector leptoquarks, belonging to an identical number of copies of the extended gauge group (e.g. Pati-Salam model based on the gauge group $[SU(4)_c]_i \times  [SU(2)_L]_i \times [SU(2)_R]_i$), with each copy acting on a single SM fermion generation (subject to mixing with additional vector-like fermions), with the largest leptoquark-fermion couplings in association with the third family~\cite{Bordone:2017bld}. 
Another possibility to lower the vector leptoquark mass 
relies in an extended gauge group, $SU(4)\times SU(3)' \times SU(2)_L \times U(1)'$ (often referred to as ``4321''-model), with the third fermion family charged under $SU(4) \times SU(2)_L \times U(1)'$, while the lighter families are only charged under $ SU(3)' \times SU(2)_L \times U(1)'$~\cite{Greljo:2018tuh}.
This leads to an approximate $U(2)$ flavour symmetry, which is softly broken by new vector-like fermions, thus allowing to obtain the desired non-universality in the leptoquark couplings. An alternative simplified-model framework, without the need to specify an explicit extended gauge group, was pursued in~\cite{Hati:2019ufv}: working under a single vector leptoquark hypothesis, an effective non-unitary mixing between SM leptons and new vector-like leptons was used to account for the LFUV structure required to 
simultaneously explain both the charged and the neutral current $B$-meson decay anomalies.

\bigskip
Irrespective of the actual NP model including (not excessively heavy) vector leptoquarks, the effects can be understood in terms of contributions to the Wilson coefficients.
Following the discussion of the previous section  (see Tables~\ref{tab:bsll_2019} and~\ref{tab:bsll_2020}),
in order to achieve the preferred contributions for the Wilson coefficients, $C_9^{bs\mu\mu} = -C_{10}^{bs\mu\mu}$ and a universal $\Delta C_9^\text{univ.}$, scenarios in which $V_1$ couples 
at the tree level through a left handed ($V-A$) current 
to muons (as well as to down-type quark flavours $b$ and $s$) appear to be favoured the most by the global fits. A nonvanishing $\Delta C_9^{bsee} = -\Delta C_{10}^{bsee}$ along with $C_9^{bs\mu\mu} = -C_{10}^{bs\mu\mu}$ and a universal $\Delta C_9^\text{univ.}$ also provides a reasonable fit but such hypotheses are subject to stringent constraints from cLFV processes. 
Furthermore, and in order to also address the charged current  data ($R_{D^{(\ast)}}$), sizeable tree-level  $\tau$ couplings to second and third generation quarks must also be present, 
and these induce new contributions to the  $C_{V_L}$ Wilson coefficient. Such large $V_1-\tau$ couplings to second and third generation quarks further lead to a large $C_{9(10)}^{bs\tau\tau}$ which then feeds into the muon and electron counterparts (in a universal way) through
RG running\footnote{We further notice that global fits without the {\it universal} contributions to $C_9^{bs\ell\ell}$ suggest a non-zero {\it tree-level} contribution to the electron coefficients. However, once the universal contribution is added, the direct {\it tree-level} contribution is compatible with $0$ at the $1\sigma$ level. }. 

\bigskip
A simplified-model parametrisation of the vector leptoquark couplings allows not only to perform global fits, but also to understand the phenomenological implications of the relevant flavour structure,
which is paramount to establish the current viability of the model, and its prospects for future testability.
In this section, we thus pursue this approach 
not only regarding the ``anomalous'' $B$-meson observables, 
but also in what concerns the impact of this BSM construction for a large set of observables (various flavour violating meson decays and cLFV modes) - 
relevant in terms of constraints on the model, or then offering excellent prospects of observation in the near future.

\subsection{A simplified-model parametrisation of vector leptoquark $V_1$ couplings}
As mentioned before, in our study we will focus on 
SM extensions via a vector leptoquark $V_1$, arising from an unspecified gauge extension of the SM. The new vector transforms as $(\mathbf 3, \mathbf 1, 2/3)$ under the $SU(3)_c \times SU(2)_L \times U(1)_Y$ gauge group.
%
%
For simplicity, and due to the absence of hints in the data suggesting the presence of right-handed couplings\footnote{In Ref.~\cite{Alguero:2019ptt} a mild drift towards NP contributions in the Wilson coefficients involving right-handed currents ($C_{7,\,9,\,10}^\prime$) was observed; however the results remained compatible with zero at $\sim1\sigma$ level.
Right-handed couplings (corresponding to the Wilson coefficient $C_{V_R}$) are also disfavoured by charged current $R_{D^{(\ast)}}$ data, and by constraints from the LHC.}, we will exclusively focus on {\it left-handed} leptoquark currents. In the mass basis we consider a simplified-model Lagrangian concerning the effective coupling of $V_1$ with the SM fermions, given by
\begin{equation}\label{eq:modelind:L:massbasis}
	\mathcal L \supset \sum_{i,j,k = 1}^{3} V_1^\mu\left(\bar d_L^i \,\gamma_\mu \,K_L^{ik}\, \ell_L^{k} + \bar u_L^j \,V_{ji}^\dagger\, \gamma_\mu\, K_L^{ik} \,U_{kj}^\mathrm{P}\, \nu_L^j\right) + \mathrm{H.c.}\:\text,
\end{equation}
in which $K_L^{ij}$ are effective couplings which are in general complex and non-universal, $V$ denotes the CKM matrix and $U^\mathrm{P} \equiv U_L^{\ell \dagger} U_L^{\nu}$ is the Pontecorvo-Maki-Nakagawa-Sakata (PMNS) leptonic mixing matrix. 
We notice that the above parametrisation is valid irrespective of the 
underlying mechanism responsible for the generation of the effective nonuniversality in the vector leptoquark couplings (see, e.g.~\cite{Bordone:2017bld,Greljo:2018tuh,Balaji:2018zna,Fornal:2018dqn,Hati:2019ufv,Baker:2019sli,Cornella:2019hct,DaRold:2019fiw}). For the sake of simplicity we will further assume that the couplings $V_1-\ell-q$ are real.
The conclusions drawn here should thus hold for generic constructions with real effective couplings and negligible right-handed currents (consistent with zero at the $\sim1\sigma$ level from the EFT fits to the $b\to s\ell\ell$ data~\cite{Alguero:2019ptt}; we recall that this corresponds to negligible $C_{7,\,9,\,10}^\prime$ Wilson coefficients).

\medskip
For the general vector leptoquark scenario under consideration, the most relevant tree-level Wilson coefficients for $b\to s\ell\ell$ transitions and $R_{D^{(\ast)}}$ observable are given by~\cite{Dorsner:2016wpm}
\begin{eqnarray}
C^{ij;\ell \ell^{\prime}}_{9,10} &=& \mp\frac{\pi}{\sqrt{2}G_F\,\alpha_\text{em}\,V_{3j}\,V_{3i}^{\ast} \,m_V^2}\left(K_L^{i
  \ell^\prime} \,K_L^{j\ell\ast} \right)\,,\nonumber \\
  C_{jk,\ell i}^{V_L} &=& \frac{\sqrt{2}}{4\,G_{F}\,m_V^2}\,
  \frac{1}{V_{jk}}\,
  (V\,K_{L}\, U^P)_{ji}\, K_{L}^{k\ell\ast}\,\text.
  \label{eqn:CV}
\end{eqnarray}
Variants of the above coefficients (depending on the flavour indices) are responsible for the leading contributions to most of the $b\to s\ell\ell$ and $R_{D^{(\ast)}}$ observables relevant for the fit. In addition, there are several other observables such as leptonic and semileptonic meson decays, as well as cLFV leptonic decays, which 
are important for the analysis.
The expressions for the branching fractions can be found in Appendix~\ref{app:observables}.
Before proceeding to the description of the global fit, some remarks are in order concerning the evaluation of the latter observables.
We first notice that the vector leptoquark coupling parameters are matched with the Wilson coefficients 
at the leptoquark mass scale, $m_V$; the latter are subsequently 
run down to the $b$-quark mass scale, or to the scale of any other process (observable) considered in the analysis. Therefore, and even though some of the relevant Wilson coefficients are vanishing at the scale of the matching of the EFT to our simplified effective leptoquark model (i.e. at $m_V$), they can be generated from operator mixing during RG running to the scale of a given observable. In particular, in our fits we take into account all running effects using the \texttt{wilson} package~\cite{Aebischer:2018bkb} in association with the \texttt{flavio} package~\cite{Straub:2018kue}.

We recall that the 
non-trivial effective $V_1$ couplings can potentially induce new contributions to cLFV
observables such as radiative decays $\ell_i \rightarrow \ell_j \gamma$ and 3-body decays
$\ell_i \rightarrow 3 \ell_j$ at loop level, and neutrinoless
$\mu - e$ conversion in nuclei (at tree-level). In view of the very good current experimental sensitivity, these observables will provide some of the most stringent constraints on the $V_1$ couplings to SM fermions; as already mentioned, the expected improvements on the future sensitivities offer the possibility to further probe the vector leptoquark couplings.  
Another important point worth noting is that although the cLFV radiative decays occur at loop level, the associated anapole contributions to vector operators can lead to sizable contributions to neutrinoless $\mu - e$ conversion and $\mu\rightarrow 3e$, with a magnitude comparable to the tree level contributions or, in some cases, even accounting for the dominant contribution. In addition, we find that the dipole operators also significantly contribute  to radiative decays and to neutrinoless $\mu -e$ conversion.

We emphasise that the one-loop dipole and anapole contributions 
from the exchange of a vector
boson generically diverge, and a UV complete framework (with a consistent gauge symmetry breaking pattern) is thus necessary to obtain a finite
result in a gauge independent way. 
Therefore, to reliably evaluate such observables in the context of vector leptoquark exchanges we have chosen to work in the Feynman gauge, 
including the necessary contributions from the Goldstone modes. 
We have thus made the minimal working assumption that the vector leptoquark originates from the breaking of a gauge extension of the SM, which gives rise to a would-be Goldstone boson degree of freedom, subsequently absorbed by the massive vector leptoquark. We include this Goldstone mode (degenerate in mass with $V_1$) to obtain the gauge invariant (finite) form factors for the relevant dipole and anapole contributions. Furthermore, to keep our results as general as possible, we do not include any effects due to an extended scalar sector (possibly necessary to implement the breaking of the extended gauge symmetry) nor from new gauge bosons (which might arise due to the breaking of an extended gauge symmetry); we work under the assumption that, should these states be present, they only have flavour conserving couplings \footnote{It has been noted~\cite{DiLuzio:2018zxy} that the neutral gauge boson associated with $V_1$  will have a mass similar to $V_1$ in most of the minimal UV complete models, therefore making it hard to decouple the new neutral states from the EFT.}. 

Finally, and should the model encompass additional vector-like fermions, as is the case for several  
vector leptoquark realisations~\cite{Bordone:2017bld,Greljo:2018tuh,Hati:2019ufv}, one must also consider the impact of such new states for electroweak precision observables, as for instance the constraints on the $Z$ boson
LFU ratios and cLFV decay modes, as emphasised 
in~\cite{Hati:2019ufv}.

\subsection{Towards a global fit of the vector leptoquark $V_1$ flavour structure}
\label{sec:lqfit}
We are now ready to carry out a comprehensive fit of the relevant couplings of the vector leptoquark to the different generations of SM fermions. Relying on the above simplified-model parametrisation, our goal is thus to constrain the entries of the matrix $K_L$ 
(see Eq.~(\ref{eq:modelind:L:massbasis})).
Under the assumption that the relevant couplings are real, a total of nine free parameters will thus be subject to a large number of constraints stemming from data on several SM-allowed leptonic and semileptonic meson decays, SM-forbidden cLFV transitions and decays, as well as from an explanation of the (anomalous) observables in the $b \to s \ell \ell$ and $b \to c \tau\nu$ systems.

\noindent
\paragraph{Data relevant for the global fit}
In particular, we take into account the data from $b\to s\ell\ell$ decays as listed in Appendix~\ref{app:bsll}.
This includes the binned data of the angular observables in the optimised basis~\cite{Descotes-Genon:2013vna} (Table~\ref{tab:ang_data}), the differential branching ratios (Table~\ref{tab:br_bsll}), and the binned LFUV observables (Table~\ref{tab:lfuv_bsll}).
Other than the binned data, we also include the unbinned data of branching ratios in $B_{(s)}\to \ell\ell$~\cite{Chatrchyan:2013bka, Aaij:2017vad, Aaboud:2018mst, Sirunyan:2019xdu,Aaij:2020nol} and inclusive and exclusive branching ratio measurements of $b\to s\gamma$~\cite{Amhis:2014hma,Misiak:2017bgg,Dutta:2014sxo, Aaij:2012ita}.

\noindent
For the charged current $b\to c\ell\nu$ processes (see Appendix~\ref{app:BFCCC}) we include in addition to the LFUV ratios $R_{D^{(\ast)}}$~\cite{Abdesselam:2017kjf, Abdesselam:2018nnh,Aaij:2015yra,Aaij:2017uff,Huschle:2015rga,Hirose:2016wfn,Abdesselam:2019dgh,Lees:2013uzd} the binned branching fractions of $B\to D^{(\ast)} \ell\nu$ decays~\cite{Aubert:2008yv, Aubert:2007qs,Urquijo:2006wd, Aubert:2009qda}, as listed in Table~\ref{tab:binned_bcellnu}.

Other than studying the contributions of the vector leptoquark in the ``anomalous'' channels, we aim to estimate the favoured ranges of all of its couplings to SM fermions. Consequently, we include a large number of additional observables into the likelihoods.
Since most processes only constrain a product of at least two distinct leptoquark couplings, a successful strategy is to include an extensive set of processes, thus allowing to constrain distinct combinations of couplings (as many as possible).

In addition to the $b\to c\ell\nu$ transitions, we also include certain $b\to u\ell\nu$ decays such as $B^0\to\pi\tau\nu$, $B^+\to\tau\nu$ and $B^+\to\mu\nu$, which are listed in Table~\ref{tab:buellnu}.
In many leptoquark models $B\to K^{(\ast)}\nu\bar\nu$ decays provide very stringent constraints. However this is not the case for $V_1$ vector leptoquarks, due to the $SU(2)_L$-structure: the relevant operators for $B\to K^{(\ast)}\nu\bar\nu$ transitions are absent at the tree-level, and are only induced at higher order, thus  leading to weaker constraints. 
Due to the leading operator being generated at the loop level, a non-linear combination of leptoquark couplings is constrained by this process.
Thus, despite the loop suppression, we include $B\to K\nu\bar\nu$ in the likelihoods, and use the data obtained by Belle~\cite{Grygier:2017tzo,Lutz:2013ftz} and BaBar~\cite{Lees:2013kla,delAmoSanchez:2010bk}. 

To constrain combinations of first and second generation couplings, we further include a large number of binned and unbinned leptonic and semileptonic charged current $D$ meson decays, charged and neutral current kaon decays and SM allowed $\tau$-lepton decays.
The observables and corresponding data-sets can be found in Appendix~\ref{app:sctau} and are listed in Tables~\ref{tab:binned_charm} through \ref{tab:fcnc_strange}.

Finally, cLFV processes impose severe constraints on the parameter space of vector leptoquark couplings; in particular neutrinoless $\mu-e$ conversion in nuclei and the decay $K_L\to e^\pm\mu^\mp$ provide some of the most stringent constraints for vector leptoquark couplings to the first two generations of leptons~\cite{Hati:2019ufv}. In Table~\ref{tab:important_LFV} we present the current experimental bounds and future sensitivities for various cLFV observables yielding relevant constraints to our analysis.
Depending on the fit set-up, either only a few, or then all of these observables are included in the global likelihood, as explicitly mentioned in the following paragraphs.

\begin{table}[h!]
	\hspace{-10mm}
	\begin{tabular}{|c|c|c|}
	\hline
	Observable & Current bound & Future sensitivity\\
	\hline
	\hline
	$\text{BR}(\mu\to e \gamma)$	&
 	\quad $<4.2\times 10^{-13}$ \quad (MEG~\cite{TheMEG:2016wtm})	&
 	\quad $<6\times 10^{-14}$ \quad (MEG II~\cite{Baldini:2018nnn}) \\
	$\text{BR}(\tau \to e \gamma)$	&
 	\quad $<3.3\times 10^{-8}$ \quad (BaBar~\cite{Aubert:2009ag})	 &
 	\quad $<3\times10^{-9}$ \quad (Belle II~\cite{Kou:2018nap}) 	 	\\
	$\text{BR}(\tau \to \mu \gamma)$	&
	 \quad $ <4.4\times 10^{-8}$ \quad (BaBar~\cite{Aubert:2009ag})	 &
 	\quad $<10^{-9}$ \quad (Belle II~\cite{Kou:2018nap})		\\
	\hline
	$\text{BR}(\mu \to 3 e)$	&
	 \quad $<1.0\times 10^{-12}$ \quad (SINDRUM~\cite{Bellgardt:1987du}) 	&
	 \quad $<10^{-15(-16)}$ \quad (Mu3e~\cite{Blondel:2013ia})  	\\
	$\text{BR}(\tau \to 3 e)$	&
 	\quad $<2.7\times 10^{-8}$ \quad (Belle~\cite{Hayasaka:2010np})&
 	\quad $<5\times10^{-10}$ \quad (Belle II~\cite{Kou:2018nap})  	\\
	$\text{BR}(\tau \to 3 \mu )$	&
 	\quad $<3.3\times 10^{-8}$ \quad (Belle~\cite{Hayasaka:2010np})	 &
 	\quad $<5\times10^{-10}$ \quad (Belle II~\cite{Kou:2018nap})		\\
	\hline
	$\text{CR}(\mu- e, \text{N})$ &
	 \quad $<7 \times 10^{-13}$ \quad  (Au, SINDRUM~\cite{Bertl:2006up}) &
 	\quad $<10^{-14}$  \quad (SiC, DeeMe~\cite{Nguyen:2015vkk})    \\
	& &  \quad $<2.6\times 10^{-17}$  \quad (Al, COMET~\cite{Krikler:2015msn,KunoESPP19,Adamov:2018vin})  \\
	& &  \quad $<8 \times 10^{-17}$  \quad (Al, Mu2e~\cite{Bartoszek:2014mya})\\
	\hline
	\hline
	$\text{BR}(K_L \to \mu^\pm e^\mp)$ & $< 4.7\times 10^{-12}\quad$~\cite{Tanabashi:2018oca} & ---\\
	\hline
	$\mathrm{BR}(\tau\to\phi\mu)$ & $<8.4\times10^{-8}\quad$~\cite{Tanabashi:2018oca} & $<2\times10^{-9}\quad$ Belle II~\cite{Kou:2018nap}\\
	\hline
	$\mathrm{BR}(B_s\to\mu^\pm\tau^{\mp})$ & $<4.2\times10^{-5}\quad$ LHCb~\cite{Aaij:2019okb} & --- \\
	\hline
	$\mathrm{BR}(B^+\to K^+\tau^+\mu^-)$ & $< 2.8\times 10^{-5}\quad$ BaBar~\cite{Lees:2012zz} & $<3.3\times 10^{-6}\quad$ Belle II~\cite{Kou:2018nap}\\
	\hline
	$\mathrm{BR}(B_s\to\phi\mu^\pm\tau^\mp)$ & $<4.3\times10^{-5}$\cite{Tanabashi:2018oca} & ---\\
	\hline
	\end{tabular}
	\caption{Current experimental bounds and future sensitivities of a selection of the most important cLFV observables which constrain the parameter space of $V_1$ leptoquark models. All upper limits are given at $90\,\%$ confidence level (C.L.).}
	\label{tab:important_LFV}
\end{table}

\paragraph{Results for the simplified-model fit of the $V_1$ couplings}

Firstly, it is important to emphasise that in our analysis we consider all the entries in the $K_L$ coupling matrix as (real) free parameters to be determined by the fit. 
For the leptoquark mass we choose three benchmark-points, $m_{V_1} \in [1.5,\,2.5,\,3.5]\:\mathrm{TeV}$, which allow to illustrate most of the vector leptoquark mass range of interest, while respecting the current bounds from direct searches at colliders~\cite{Khachatryan:2014ura,Aad:2015caa,Aaboud:2016qeg,Aaboud:2019jcc,Aaboud:2019bye,Aad:2020iuy,Sirunyan:2017yrk,Sirunyan:2018vhk,Sirunyan:2018kzh}. 
In particular, notice that masses significantly heavier than a few TeVs preclude a successful explanation of the charged current anomalies, $R_{D^{(*)}}$. For each mass benchmark point we thus obtain best-fit points corresponding to a SM pull around $\sim 6.4\,\sigma$ (with respect to the global likelihood including all lepton flavour conserving observables).

%
\begin{figure}[h]
	\centering
	\includegraphics[width = 0.6\textwidth]{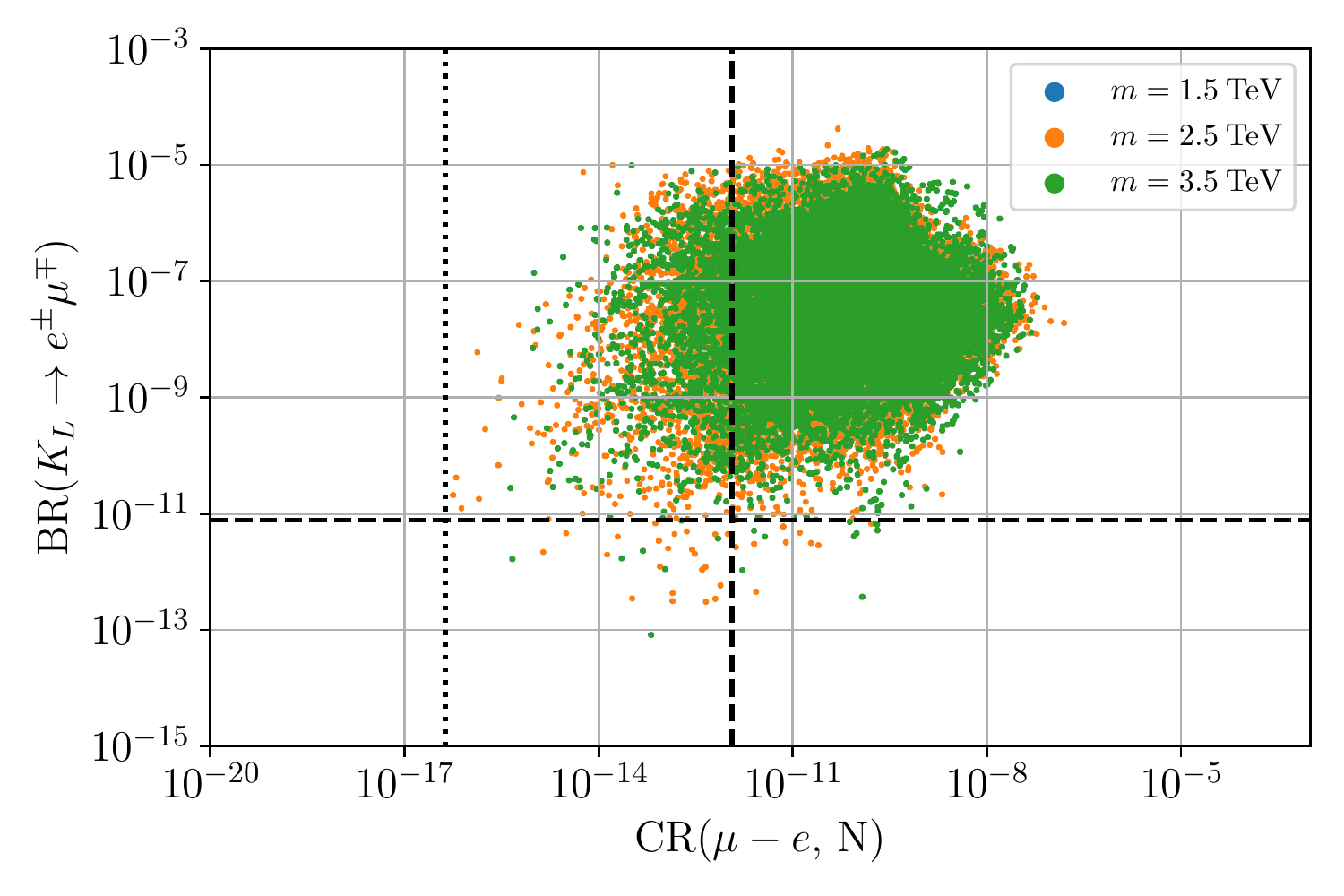}
	\caption{Result of a random scan around the best-fit point (without the inclusion of cLFV bounds on $\mathrm{CR}(\mu - e, \mathrm{Au})$ and $\mathrm{BR}(K_L\to e^\pm\mu^\mp)$ as inputs to the fit). 
	Following a sampling of the global likelihood(s) via MCMC, the sample points shown in the plot are drawn from the posterior distributions of the leptoquark couplings (cf. Appendix~\ref{app:stats}).
	The colour scheme reflects the mass benchmark points: 
	blue, orange and green respectively associated with 
	$m_V$=1.5~TeV, 2.5~TeV and 3.5~TeV.  
	The dashed lines indicate the current bounds at $90\,\%$ C.L., while the dotted line denotes the envisaged future sensitivity of the COMET and Mu2e experiment (for $\mathrm{Al}$ nuclei).}
	\label{fig:woLFV_KL_CR}
\end{figure}
In Fig.~\ref{fig:woLFV_KL_CR}, we present the results of a random scan around the best-fit points for the vector leptoquark scenario here considered, in the plane spanned by two of the most constraining cLFV observables,  $\mathrm{CR}(\mu - e, \mathrm{N})$ and $\mathrm{BR}(K_L\to e^\pm\mu^\mp)$. The sample points are drawn from the posterior frequency distributions of the leptoquark couplings, following Markov Chain Monte Carlo (MCMC) simulations, as described in Appendix~\ref{app:stats}.
It can be easily seen that for the three mass benchmark choices (corresponding to the different colours in the plot) most of the randomly sampled points are excluded by the strong cLFV constraints. 
Although the involved couplings are compatible with $0$, the constraints on first generation couplings derived from lepton flavour conserving low-energy data (as listed in Appendix~\ref{app:Obs}) are considerably weaker than those from LFV processes. This leads to several ``flat directions'' in the likelihood.
The strongest LFV constraints are from $\mathrm{CR}(\mu - e, \mathrm{Au})$ and $\mathrm{BR}(K_L\to e^\pm\mu^\mp)$, while other LFV constraints on second and third generation couplings are weaker, or on par with constraints from lepton flavour conserving low-energy data.
Therefore, we redefine the strategy of the global fit, and now directly include the upper bounds from $\mathrm{CR}(\mu - e, \mathrm{Au})$ and $\mathrm{BR}(K_L\to e^\pm\mu^\mp)$ as {\it inputs} in the fitting procedure for the vector leptoquark couplings.

The inclusion of the current upper limits on the observables $\mathrm{CR}(\mu - e, \mathrm{Au})$ and $\mathrm{BR}(K_L\to e^\pm\mu^\mp)$ as input to the fit will consequently shift the best-fit point towards a lower cLFV prediction, also leading to a slightly lower SM pull. However, we find this to be a good compromise in order to identify regimes in the parameter space not yet disfavoured by the current cLFV data. 
In fact, and since 
$\mathrm{CR}(\mu - e, \mathrm{Au})$ and $\mathrm{BR}(K_L\to e^\pm\mu^\mp)$ are indeed two of the most constraining cLFV observables, once the bounds on the latter observables are respected, most of the sample points will be naturally in agreement with   current bounds on most of other cLFV observables (this is a consequence of correlations with other cLFV $\mu-e$ transitions; processes involving  $\tau$-leptons are comparatively less constraining).

In Table~\ref{tab:fits_wKLCRmue} we present our results for the new fits with their corresponding SM pulls. As can be verified, the SM pull is lower, reduced from $\sim6.4\sigma$ to $\sim5.8\sigma$, of which the contributions to the total $\chi^2$ stemming from the charged current $b\to c\ell\nu$ transitions amounts to $\sim 1.5\sigma$, whereas the contributions from the neutral current $b\to s\ell\ell$ transitions amounts to $\sim 4.3\sigma$. 
Furthermore, we show tentative $90\%$ ranges of the posterior (coupling) distributions, obtained by sampling the global $b\to s\ell\ell$ likelihood using MCMC. The ranges, derived from the histograms of the posterior distributions, are taken as symmetric intervals between the $5^\text{th}$ and $95^\text{th}$ percentiles (cf. Appendix~\ref{app:stats}).
We notice here that the vector leptoquark coupling to the first generation SM fermions are consistent with zero, which is an assumption often invoked in literature for simplified analyses. For second and third generation couplings, the quoted ranges of the corresponding fits are in fair agreement with the (order of magnitude) results for the benchmark ranges of second- and third generation couplings quoted in the literature, e.g.~\cite{Cornella:2021sby,Angelescu:2018tyl}. However, given the differences in the coupling parametrisation choices and underlying statistical treatment, the results are not directly comparable.

\begin{table}[h!]

	\hspace*{-8mm}\begin{tabular}{|c|c|c|c|}
		\hline
				$m_{V_1}$ & $K_L$ best-fit & $K_L$ $90\%$ & $\text{pull}_\text{SM}$\\
				\hline
				\hline
				{\footnotesize$1.5\:\mathrm{TeV}$}&
				{\footnotesize$\begin{pmatrix}
									-5.3\times 10^{-6} & 2.6\times 10^{-3} & -0.079\\
									-9.8\times 10^{-4} & -0.03 & 1.1\\
									-3.4\times 10^{-3} & 0.038 & 0.16
								\end{pmatrix}$}&
				{\footnotesize$\begin{pmatrix}
								(-1.2 \to 1.1)\times 10^{-3} & (-1.5\to 9.1)\times 10^{-3}  & -0.11\to 0.009\\
								-0.034 \to 0.036 & -0.063 \to -0.002 & 0.27\to 1.55\\
								-0.050 \to 0.036 & 1.0\times10^{-3} \to 0.11 & 0.08 \to 0.80
								\end{pmatrix}$}&
				{\footnotesize$5.78$}\\
				\hline
				\hline
				{\footnotesize$2.5\:\mathrm{TeV}$}&
				{\footnotesize$\begin{pmatrix}
					-1.9\times 10^{-5} & 4.3\times 10^{-3} & -0.11\\
					2.1\times 10^{-3} & -0.056 & 1.9\\
					-6.9\times 10^{-3} & 0.063 & 0.27
				\end{pmatrix}$}&
				{\footnotesize$\begin{pmatrix}
					(-1.5 \to 2.3) \times 10^{-3} & (-0.26\to1.1)\times 10^{-2}& -0.17 \to 0.014\\
					-0.059\to 0.068 & -0.13 \to -0.009 & 0.43 \to 2.58 \\
					-0.076 \to 0.072 & 0.009 \to 0.21 & 0.13 \to 1.31
				\end{pmatrix}$}&
				{\footnotesize$5.82$}\\
				\hline
				\hline
				{\footnotesize$3.5\:\mathrm{TeV}$}&
				{\footnotesize$\begin{pmatrix}
					2.9\times 10^{-5} & 5.9\times 10^{-3} & -0.14\\
					3.1\times 10^{-3} & -0.078 & 2.6\\
					0.010 & 0.088 & 0.37
				\end{pmatrix}$}&
				{\footnotesize$\begin{pmatrix}
					(-3.6\to2.9)\times 10^{-3} & (-3.7 \to 14.3)\times 10^{-3} & -0.21 \to 0.017\\
					-0.13\to0.078 & -0.18 \to -0.012 & 0.57 \to 3.23\\
					-0.14 \to 0.11 & 0.023 \to 0.32 & 0.22 \to 1.92
				\end{pmatrix}$}&
				{\footnotesize$5.84$}\\
				\hline
	\end{tabular}
	\caption{Results of the fits including the current experimental bounds on $\mathrm{CR}(\mu - e, \mathrm{Au})$ and $\mathrm{BR}(K_L\to e^\pm\mu^\mp)$ in the likelihood: best fit points and symmetric $90\%$ ranges (see Appendix~\ref{app:stats} for details) of $K_L^{ij}$. The SM pull is reduced from $\sim6.4\sigma$ to $\sim5.8\sigma$.}
	\label{tab:fits_wKLCRmue}
\end{table}

\noindent
Upon inclusion of the current cLFV constraints we find that the shape of the global likelihood consequently enforces small vector leptoquark couplings to the first two generations of charged leptons, leading to predictions consistent with experimental data.
This thus allows to sample the global likelihood (in terms of the leptoquark couplings) via MCMC techniques (as described in Appendix~\ref{app:stats}).
The posterior distributions of the leptoquark couplings are then used to compute predictions for $B$-meson decays into final states containing $\tau$-leptons, and 
several cLFV observables (including tau decays).

\begin{figure}[h]
	\centering
	\includegraphics[width=0.8\textwidth]{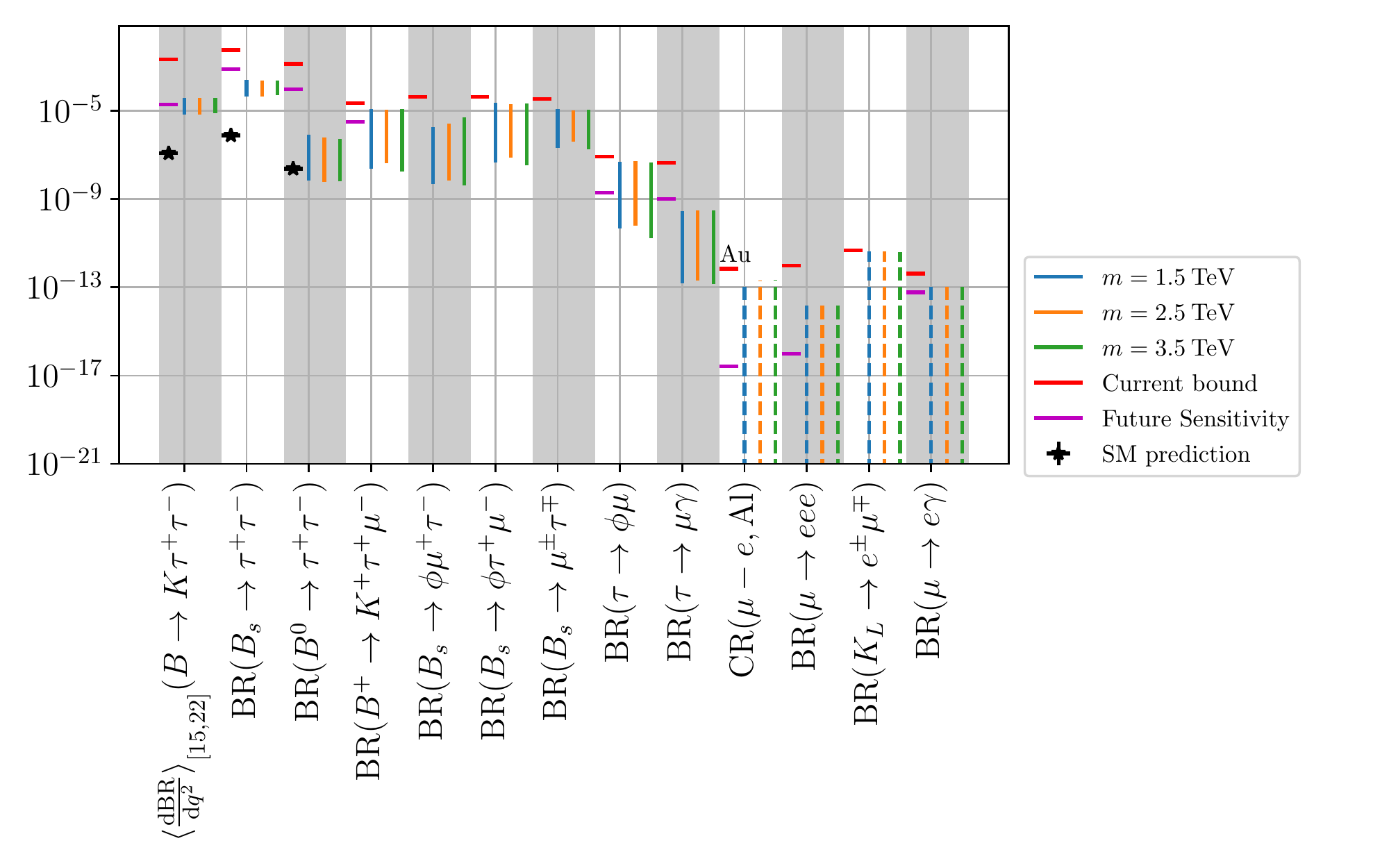}
	\caption{Predicted ranges for several $\tau$-lepton and LFV observables. The blue, orange and green lines respectively denote the $90\%$ range for leptoquark masses of $1.5,\, 2.5\text{ and } 3.5\:\mathrm{TeV}$ while the horizontal red (purple) lines denote the current (future) bound at $90\,\%$ C.L.;  stars denote SM predictions when appropriate.
	The dashed lines correspond to predictions of observables depending only on couplings that are compatible with $0$ and their top edges correspond to $90\%$ upper limits. (The $90\%$ ranges have been obtained as detailed in Appendix~\ref{app:stats}.)
		}
	\label{fig:tau_and_lfv_predictions}
\end{figure}

This is presented in Fig.~\ref{fig:tau_and_lfv_predictions} where, for each observable, we depict the current experimental bounds and future sensitivities, the SM predictions (when relevant), as well as the predictions for the three vector leptoquark mass benchmark points - corresponding to the vertical coloured lines.
The dashed lines describe predictions of observables involving only couplings compatible with vanishing values and thus their top edge corresponds to a $90\%$ upper limit, while no lower limit should be implied.

As can be seen from Fig.~\ref{fig:tau_and_lfv_predictions}, a large part of the 
currently allowed parameter space in the $e \mu$ channel (for the three leptoquark mass benchmark points) will
be probed by the upcoming experiments dedicated to searching for neutrinoless $\mu - e$ conversion in Aluminium nuclei, 
Mu2e and COMET, owing to the expected increase in sensitivity. 
In the case of future non-observation of this process, this will lead to strongly improved constraints on the $V_1$ couplings to first first generation fermions.

Moreover, the sensitivity of the lepton flavour violating process $\tau\to \phi \mu$ is expected to be improved by over an order of magnitude at the Belle II experiment, 
which will allow probing a large region of the parameter space associated with the $\mu\tau$ channel. A priori, 
and as can be seen from Fig.~\ref{fig:tau_and_lfv_predictions},
under the current vector leptoquark hypothesis, $\tau\to \phi \mu$ decays have very strong prospects of being observed at Belle II. Conversely, should such a mode not be observed at Belle II, then the $s-\mu$ and $s-\tau$ couplings of the vector leptoquark will be tightly constrained. As a consequence, it might prove extremely challenging to simultaneously address the anomalous neutral and charged current data within the current model.

\section{Impact of future experiments: 
{\bf Belle II} and cLFV searches}\label{sec:futureimpact}

Following the overview of the vector leptoquark couplings conducted in the previous section, we now proceed to investigate how our working hypothesis can be effectively probed by the coming future experiments, especially  Belle II and cLFV-dedicated facilities. 

Assuming that the above experiments return only negative search results for the most promising modes, we  
then evaluate how the current $V_1$ hypothesis would still stand as a viable explanation for the LFUV $B$-meson decay anomalies.

\subsection{Probing the vector leptoquark $V_1$ at coming experiments}
Concerning the quest for LFUV in  $b\rightarrow s \ell^+\ell^-$ decays,  Belle II is expected to achieve a very high sensitivity for both muon and electron modes, leading to very precise measurements for the ratios $R_K$ and $R_{K^*}$, with the potential to confirm the anomalous LHCb data (if the latter is due to NP effects)~\cite{Kou:2018nap}. 
In what concerns $B$-meson decays to $\tau^+ \tau^-$ final states,
Belle II will also provide the first in-depth experimental exploration of these modes. Notice that the latter remain a comparatively less explored set of observables, with relatively weak bounds on the few modes already being searched for: for example, current bounds on $\text{BR}(B^0 \rightarrow \tau^+ \tau^-)<1.3 \times 10^{-3}$ from LHCb~\cite{DeBruyn:2016tiq} and $\text{BR}(B_s \rightarrow \tau^+ \tau^-)< 2.25 \times 10^{-3}$ from Babar~\cite{TheBaBar:2016xwe} are orders of magnitude weaker than the SM predictions. For the purely leptonic decays, the most recent SM computations now include next-to-leading order (NLO) electroweak corrections and next-to-NLO QCD corrections~\cite{Bobeth:2013uxa,Hermann:2013kca, Bobeth:2013tba},
\begin{eqnarray}
\text{BR}(B_s \rightarrow \tau^+ \tau^-)_{\text{SM}} &=& (7.73 \pm 0.49) \times 10^{-7}\,,\nonumber\\
\text{BR}(B^0 \rightarrow \tau^+ \tau^-)_{\text{SM}} &=& (2.22 \pm 0.19) \times 10^{-7}\, .
\end{eqnarray}

\noindent
Within the SM, the exclusive semileptonic decays of $B$-mesons to $\tau^+ \tau^-$ final states have been studied by several groups: the modes $B\to K^*\tau^+\tau^-$ and $B_s\to \phi\tau^+\tau^-$ have been computed\footnote{The inclusive $B\to X_s\tau^+\tau^-$ process has been addressed in Refs.~\cite{Guetta:1997fw,Bobeth:2011st}, while indirect constraints on $b\to s\tau^+\tau^-$ operators 
were studied in Ref.~\cite{Bobeth:2011st}.}  in~\cite{Hewett:1995dk,Bouchard:2013mia,Kamenik:2017ghi}. 
To avoid contributions from the resonant decays through the narrow $\psi(2S)$ charmonium resonance (i.e. $B \rightarrow H\psi(2S)$ with $H\psi(2S)\rightarrow \tau^+ \tau^-$, where $H=K,K^*, \phi,\cdots$), the relevant SM predictions are typically restricted to an invariant di-tau mass $q^2> 15$ GeV$^2$. 
Taking into account the uncertainties from the relevant form factors and CKM elements, the SM predictions for the branching ratios of the semileptonic decays into tau pairs can be determined with an accuracy between 10\% and 15\%. 
Notice that the presence of broad charmonium resonances 
(above the open charm threshold) can further lead to additional subdominant uncertainties, typically of a few percent~\cite{Beylich:2011aq}. 

\noindent 
For the $B \rightarrow K \tau^+ \tau^-$ modes, using the recent lattice $B\rightarrow K$ form factors from the Fermilab/MILC collaboration~\cite{Bailey:2015dka}, the SM predictions for the $q^2\in[15, 22]\, \text{GeV}^2$ have been reported to be~\cite{Du:2015tda},
\begin{eqnarray}
\text{BR}(B^+ \rightarrow K^+ \tau^+ \tau^-)_{\text{SM}} &=& (1.22 \pm 0.10) \times 10^{-7}\,,\nonumber\\
\text{BR}(B^0 \rightarrow K^0 \tau^+ \tau^-)_{\text{SM}} &=& (1.13 \pm 0.09) \times 10^{-7} \, .
\end{eqnarray}
Similar predictions for the $B \rightarrow K^* \tau^+ \tau^-$ modes, with $q^2\in[15, 19]\, \text{GeV}^2$,
have also been reported~\cite{Kou:2018nap,Straub:2018kue} %
\begin{eqnarray}
\text{BR}(B^+ \rightarrow K^{*+} \tau^+ \tau^-)_{\text{SM}} &=& (0.99 \pm 0.12) \times 10^{-7}\,,\nonumber\\
\text{BR}(B^0 \rightarrow K^{*0} \tau^+ \tau^-)_{\text{SM}} &=& (0.91 \pm 0.11) \times 10^{-7} \,.
\end{eqnarray}
The above results rely on the combined fit of lattice QCD and light cone sum rules (LCSR) 
results for $B\rightarrow K$ form 
factors~\cite{Straub:2015ica}.
Finally, the SM prediction for $B_s \rightarrow \phi \tau^+ \tau^-$ mode can also be obtained for the same kinematic region ($q^2 \in [15, 19]\, \text{GeV}^2$)~\cite{Capdevila:2017iqn}
\begin{eqnarray}
\text{BR}(B_s \rightarrow \phi \tau^+ \tau^-)_{\text{SM}} = (0.86 \pm 0.06) \times 10^{-7} \, .
\end{eqnarray}
As already discussed in Section~\ref{VLQmodel}, 
sizeable $b-\tau$ and $s-\tau$ couplings are necessary to explain the charged current anomalous data on $R_{D^{(*)}}$; if $R_{D^{(*)}}$ anomalies are indeed due to NP then one expects 
a significant enhancement of the rates of 
$b\to s\tau^+\tau^-$ processes, up to three orders of magnitude from the SM predictions~\cite{Alonso:2015sja,Crivellin:2017zlb,Calibbi:2017qbu,Capdevila:2017iqn}.
This renders searches for $b\to s\tau^+\tau^-$ modes extremely interesting probes of vector leptoquark models aiming at explaining anomalous LFUV data.

Although the LHCb programme includes searches 
for
$B \to {K^{(*)}}{\tau^+\tau^-}$ and $B_s \to {\phi}{\tau^+\tau^-}$ modes, being an $e^+e^-$ experiment Belle II is expected to be more efficient than the LHCb in reconstructing 
$B$ to tau-lepton decays, since many of these modes require reconstructing additional tracks originating from the final state mesons ($K$, $K^*$ or $\phi$). 
Therefore, $b\to s\tau^+\tau^-$ observables will be among the {\it ``golden modes''} aiming at probing the vector leptoquark hypothesis at Belle II. 

\begin{figure}[h!]
	\centering
	\includegraphics[width=0.8\textwidth]{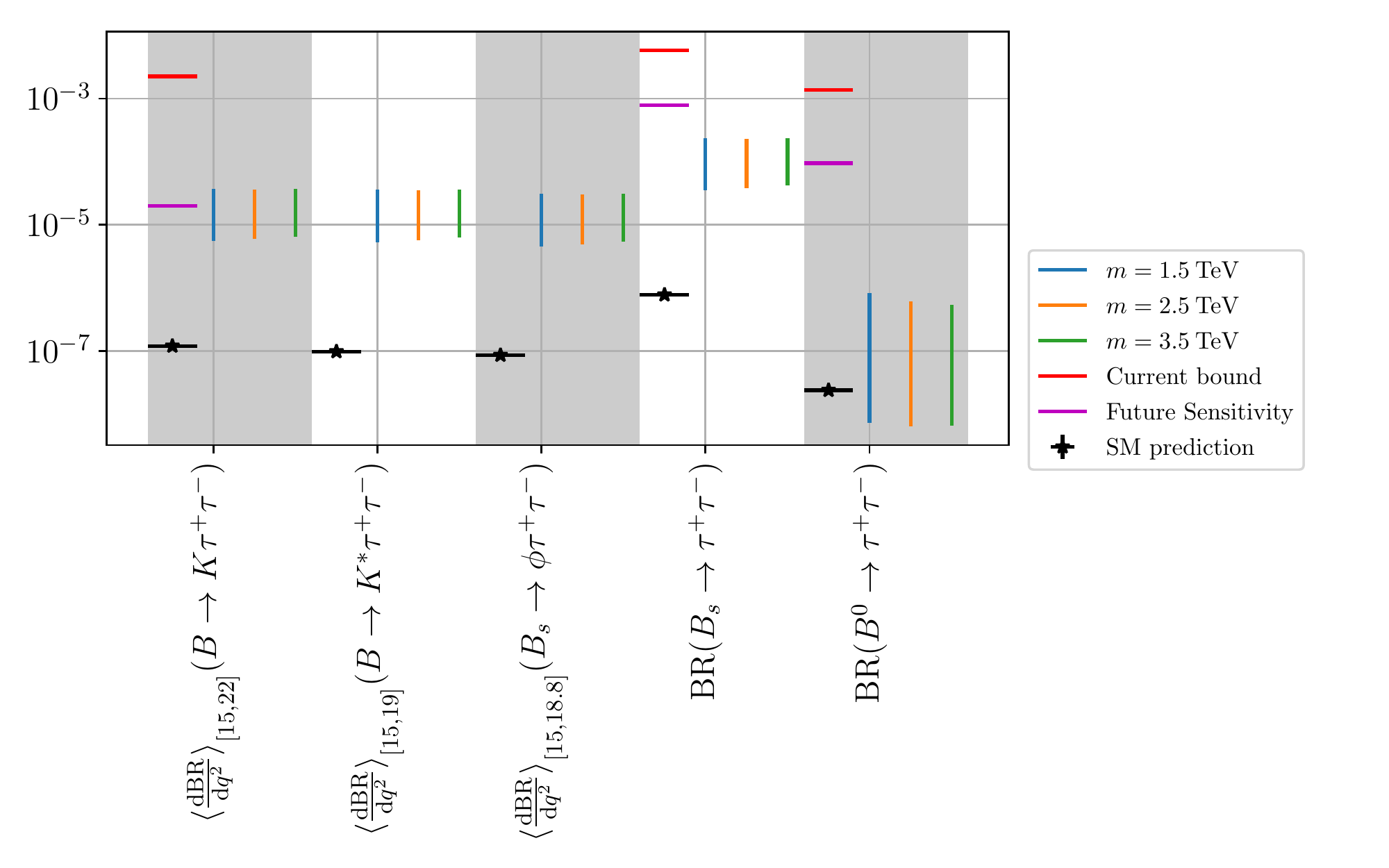}
	\caption{Predictions for several leptonic and semileptonic $B_{(s)}$ to $\tau^+ \tau^-$ decays, for three benchmark values of the vector leptoquark mass (coloured vertical bars). Also displayed, to the left of the different predictions, are the current experimental  limits and the future projected sensitivity from Belle II (horizontal lines), as well as the corresponding SM prediction (black). The ranges correspond to the interval between the $5^\text{th}$ and $95^\text{th}$ percentiles of the posterior distributions, as described in Appendix~\ref{app:stats}.}
	\label{fig:tau_sm_allowed_predictions}
\end{figure}

In Fig.~\ref{fig:tau_sm_allowed_predictions} we present
the predictions for several leptonic and semileptonic $B_{(s)}$ to $\tau^+ \tau^-$ decays, as arising in the present  
vector leptoquark scenario. 
We display the results for three benchmark leptoquark masses (coloured vertical bars, corresponding to $m_V=$1.5~TeV, 2.5~TeV and 3.5~TeV), together with the current limits and the future projected sensitivity from Belle II, and the corresponding SM predictions. 

As can be clearly observed from Fig.~\ref{fig:tau_sm_allowed_predictions}, all $b\to s\tau\tau$ branching fractions are enhanced with respect to the SM (typically by one to two orders of magnitude). 
This is a direct consequence of accommodating the charged current anomalies (i.e. $R_{D^{(*)}}$), as these call upon sizeable $b-\tau$ and $s(c)-\tau$ couplings. The decay $B^0\to\tau^+\tau^-$ is subject to a milder enhancement due to having the $d-\tau$ coupling already constrained by other observables.

\bigskip
Tau-lepton decays offer powerful probes of vector leptoquark models.
The Belle experiment has searched for 46 distinct cLFV $\tau$ decay modes, using almost its entire data sample of approximately 1000 $\mathrm{fb}^{-1}$;  no evidence for cLFV decays was found, but new 90\% C.L. upper limits on the branching fractions were set, at a level of around  $\mathcal{O}(10^{-8})$. 
At Belle II, if on the one hand the higher beam-induced background will render these searches more challenging, on the other hand its impressive luminosity will 
allow to significantly ameliorate the sensitivities to these modes. As much as 45 billion $\tau$ pairs (in the full dataset) are expected to be produced in $e^+ e^-$ collisions at Belle II, clearly providing very bright prospects for cLFV tau decay searches.
The Belle II experiment is thus expected to improve the sensitivities of the various cLFV decays by more than one order of magnitude, reaching a level of $\mathcal{O}(10^{-9}-10^{-10})$. 

\begin{figure}[h!]
	\centering
	\includegraphics[width=0.8\textwidth]{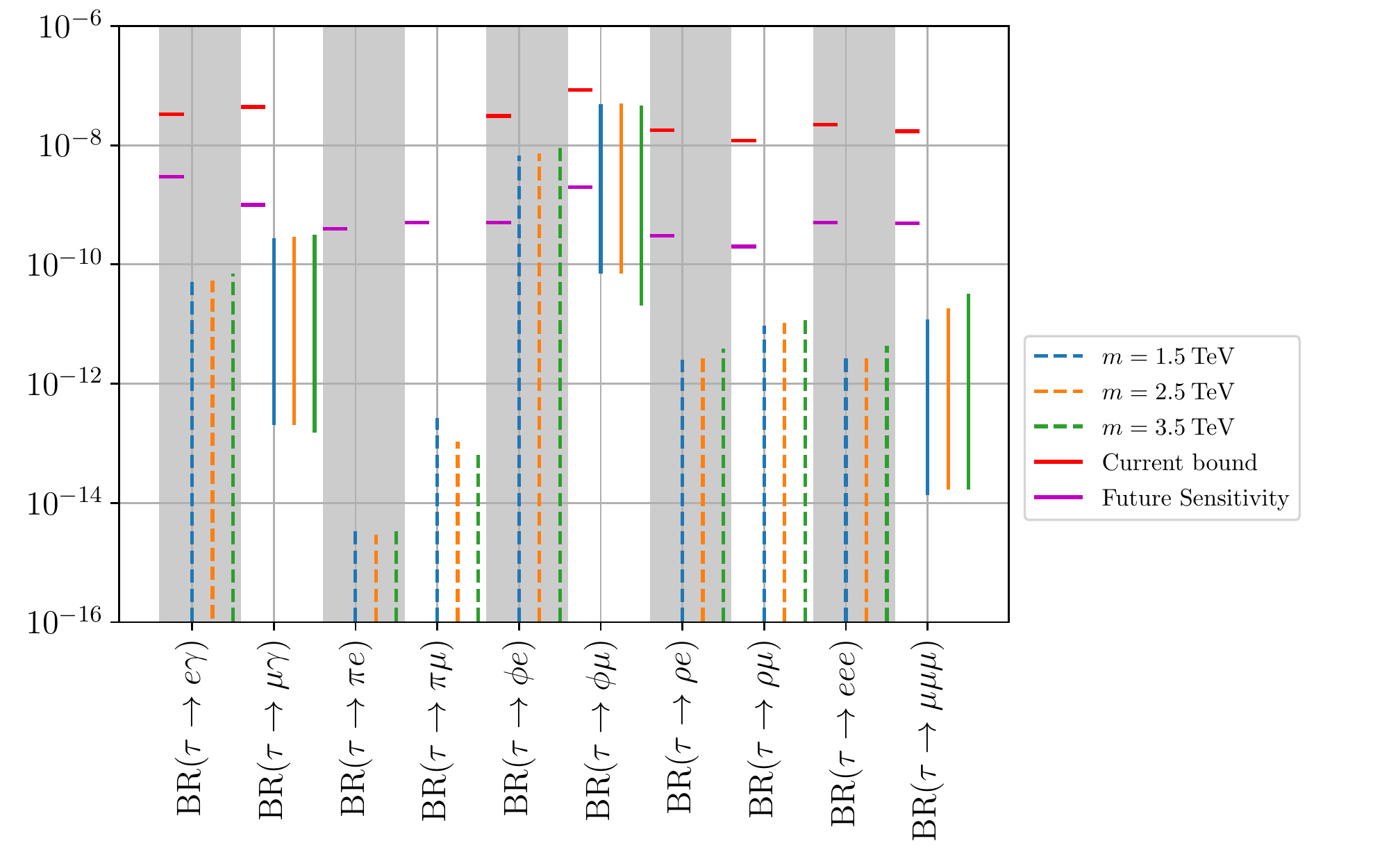}
    	\caption{Lepton flavour violating $\tau$ decay modes expected to be searched for at the Belle II experiment. The $90\%$ ranges are obtained from sampling points at the around the best-fit point. Line and colour coding as in Fig.~\ref{fig:tau_and_lfv_predictions}.}
	\label{fig:taulfv_predictions}
\end{figure}

In Fig.~\ref{fig:taulfv_predictions} we present the predictions of the vector leptoquark scenario for various cLFV tau decay modes which are programmed to be searched for at the Belle II experiment. 
It is interesting to note that among the various observables, the $\tau \to \phi\mu$ decay emerges as the most promising one to probe the vector leptoquark hypothesis  - another ``{\it golden mode}''. 

\begin{figure}[h!]
	\centering
	\includegraphics[width=0.8\textwidth]{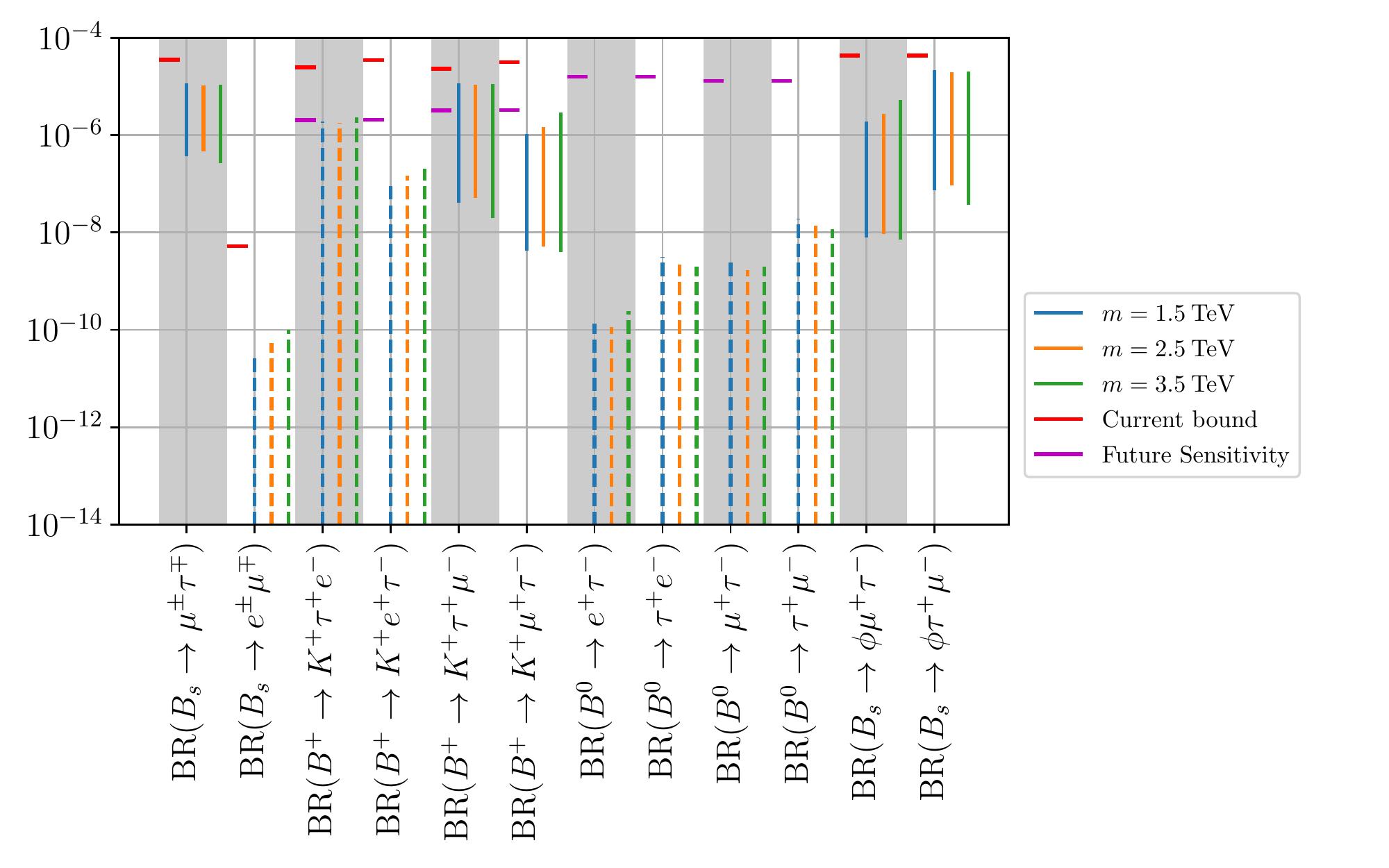}
	\caption{Lepton flavour violating decay modes of beauty-flavoured mesons to $\tau$-leptons, to be searched for at Belle II. The $90\%$  ranges are obtained from sampling points around the best-fit point. Line and colour coding as in Fig.~\ref{fig:tau_and_lfv_predictions}.}
	\label{fig:bfv_predictions}
\end{figure}

The Belle II experiment will also search for a number of cLFV leptonic and semileptonic $B$-meson decays (some into final state $\tau$s). In Fig.~\ref{fig:bfv_predictions} we present our findings for these cLFV processes.
In the context of the present vector leptoquark model, one thus expects sizeable contributions for $B_s \rightarrow \tau^+ \mu^-$,
$B^+ \rightarrow K^+ \tau^+ e^-$, $B^+ \rightarrow K^+ \tau^+ \mu^-$ and $B_s \rightarrow \phi \tau^+ \mu^-$ (for the different benchmark masses considered), close to  current bounds, and clearly within reach of future sensitivities\footnote{Notice that the rates for $B_s$ decays into $\phi \tau^- \mu^+$ are typically less enhanced than those for the (opposite charge) $\phi \tau^+ \mu^-$ mode: this is a consequence of the leptoquark couplings involved, with the combination $K_L^{22}K_L^{33}$ (entering the former) in general smaller than $K_L^{23}K_L^{32}$ (appearing in the latter), as can be inferred, for example, from Table~\ref{tab:fits_wKLCRmue}.}. Together with the decay channels identified following the results displayed in Fig.~\ref{fig:tau_sm_allowed_predictions}, these cLFV 
modes appear particularly promising to observe a signal of a vector leptoquark NP scenario explaining the $B$-meson decay anomalies.

\subsection{Impact of future negative searches}
\label{sec:future_Belle}
A final point to be addressed concerns the impact of 
future null results from Belle II and other experiments searching for cLFV: if no cLFV signal is found, and no enhancement of $B$-meson decay rates 
is observed, to which extent will this affect the prospects of a vector leptoquark hypothesis as a viable explanation of the $B$-meson decay anomalies?  
To assess the implication of such a scenario we re-conduct the fit whose results were summarised in  Table~\ref{tab:fits_wKLCRmue}, now including the projected future sensitivities from Belle II  and cLFV-dedicated experiments (COMET, Mu2e, MEG II and Mu3e).
Recall that the Belle II observables taken into account in this fit are listed in Appendix~\ref{app:Obs}  (Table~\ref{tab:belleii}), with the future sensitivities always corresponding to the assumption of the full anticipated luminosity of $50\:\mathrm{ab}^{-1}$; the future sensitivities for the cLFV dedicated experiments have been summarised in the first part of Table~\ref{tab:important_LFV}.

The results of this new fit (corresponding to null results in the several ``golden modes'' previously discussed) are presented in Table~\ref{tab:fits_wCOMET_Belle}. 
A comparison of these results with those of  Table~\ref{tab:fits_wKLCRmue} suggests that all leptoquark couplings would be well constrained (with the exception of the $d-\tau$ one). We again notice here that the vector leptoquark coupling to the first generation SM fermions remain consistent with zero.

\begin{table}[h!]
	\hspace{-1.1cm}
    \begin{tabular}{|c|c|c|c|}
	\hline
	{$m_{V_1}$} & $K_L$ {best-fit} & $K_L$ $90\%$ & {$\text{pull}_\text{SM}$}\\
	\hline
	\hline
	{\footnotesize$1.5\:\mathrm{TeV}$} &
	{\footnotesize$\begin{pmatrix}
		-1.9\times 10^{-6} & -9.5\times10^{-3} & -0.011\\
		6.4\times10^{-6} & -0.021 & 0.31\\
		-3.2\times10^{-6} & 0.061 & 0.49
	\end{pmatrix}$}&
	{\footnotesize$\begin{pmatrix}
		(-6.6\to 8.6)\times10^{-4} & (-2.3\to 8.8)\times10^{-3} & -0.056\to0.008\\
		-0.012\to 0.011 & -0.037 \to -0.009  & 0.13 \to 0.59\\
		(-3.1 \to 2.5)\times 10^{-3} & 0.030 \to 0.12  & 0.19\to1.02
	\end{pmatrix}$}&
	{\footnotesize$5.52$}\\
	\hline
	\hline
	{\footnotesize$2.5\:\mathrm{TeV}$}&
	{\footnotesize$\begin{pmatrix}
		3.8\times10^{-6} & -8.7\times10^{-3} & -0.031\\
		3.9\times10^{-5} & -0.032 & 0.53\\
		2.7\times10^{-5} & 0.11 & 0.81
	\end{pmatrix}$}&
	{\footnotesize$\begin{pmatrix}
		(-8.5\to9.0)\times10^{-4} & (-2.9\to13.9)\times10^{-3} & -0.062\to 0.013\\
		-0.017\to 0.017 & -0.077 \to -0.018 & 0.13 \to 0.92\\
		(-3.3\to 5.8)\times 10^{-3} & 0.041 \to 0.18 & 0.23\to 1.79
	\end{pmatrix}$}&
	{\footnotesize$5.58$}\\
	\hline
	\hline
	{\footnotesize$3.5\:\mathrm{TeV}$}&
	{\footnotesize$\begin{pmatrix}
		-1.2\times10^{-5} & 0.012 & -0.012\\
		3.1\times10^{-4} & -0.044 & 0.71\\
		-4.0\times10^{-5} & 0.16 & 1.19
	\end{pmatrix}$}&
	{\footnotesize$\begin{pmatrix}
		(-1.4\to 1.4)\times10^{-3} & (-6.5\to14.6)\times 10^{-3} & -0.10\to 0.011\\
		-0.025\to0.024 & -0.10\to -0.02 & 0.23\to1.39 \\
		(-7.9\to4.8)\times 10^{-3} & 0.063 \to 0.36 & 0.32\to2.41
	\end{pmatrix}$}&
	{\footnotesize$5.61$}\\
	\hline
	\end{tabular}
	\caption{Best-fit points, symmetric $90\%$ ranges (see Appendix~\ref{app:stats} for details) and SM pulls of the fits containing the envisaged sensitivities of the Belle II, COMET, Mu2e, Mu3e and MEG II experiments where the non-observation of all included cLFV observables is assumed.}
	\label{tab:fits_wCOMET_Belle}
\end{table}

One can now re-project
the new fit results onto the plane of the anomalous $B$-meson decay observables, by randomly sampling around the best fit points presented in Table~\ref{tab:fits_wCOMET_Belle}.  
For the $V_1$ scenario under consideration, the strongest impact of a non-observation of cLFV processes and non-enhanced rates for $B$-meson decays to $\tau^+ \tau^-$ final states occurs for the fit of the charged current anomalies $R_{D}$ and $R_{D^{*}}$. 
This is a consequence of having significantly stronger constraints on the vector leptoquark couplings to $\tau$-leptons following the negative search results from Belle II and future cLFV experiments, and will render $V_1$ less efficient in contributing to both $R_{D^{(*)}}$.

\begin{figure}[h!] 
	\hspace*{30mm}
	\includegraphics[width= 0.8 \textwidth]{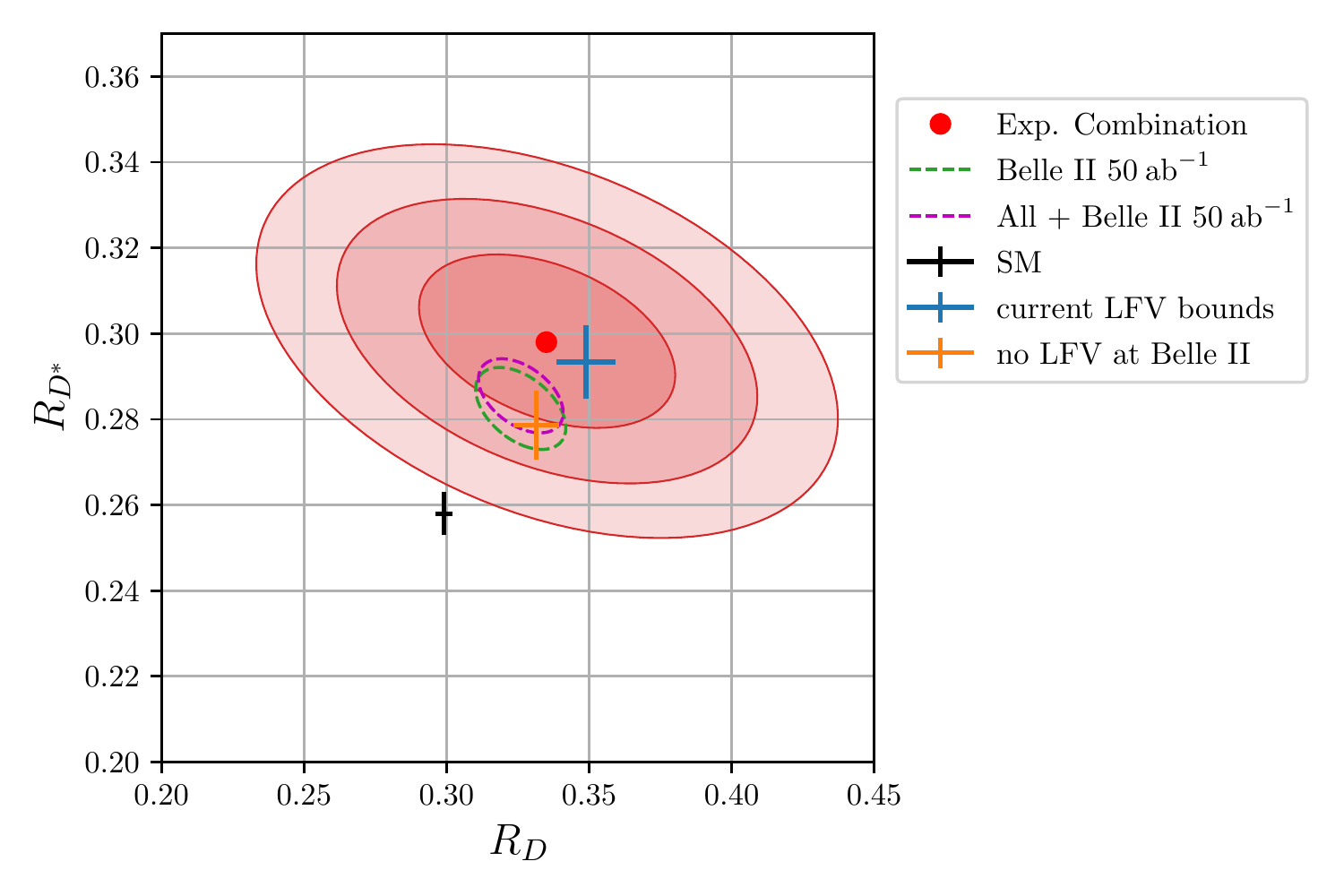}
	\caption{Likelihood contours and vector leptoquark predictions for $R_D$ and $R_{D^\ast}$. Red regions correspond to different likelihood contours obtained from a na\"ive combination of the experimental likelihoods. The blue cross denotes the predictions at the best-fit point to current LFV data. The orange cross denotes the predictions at the best-fit point with assumed null results of LFV processes at Belle II, Mu2e and COMET.
	The black cross denotes the SM prediction~\cite{Amhis:2019ckw}.
	The green dashed contour line describes the na\"ive extrapolation of the current combination of Belle data~\cite{Huschle:2015rga,Hirose:2016wfn,Abdesselam:2019dgh} to the anticipated future precision of the Belle II experiment, while the purple dashed contour line is a na\"ive combination of the Belle II projection with the current data.
	}\label{fig:rdrds_after_belleii}
\end{figure}

We present in Fig.~\ref{fig:rdrds_after_belleii} the different likelihood contours and leptoquark predictions, for different benchmark masses\footnote{The central values and uncertainties of the predictions at the best-fit points are almost identical for all mass benchmark points.} and fit set-ups, as well as best-fit points for the distinct experimental scenarios. 
The  impact for the $b\to c\ell\nu$ fit can be observed in the $R_{D}-R_{D^{*}}$ plane depicted in Fig.~\ref{fig:rdrds_after_belleii}, as the preferred ``region'' (orange cross) is pulled towards the SM prediction, and away from the current experimental best fit point (red circle).

Notice however that potential negative results from Belle II and future cLFV experiments do not significantly affect the fit to anomalous $b\rightarrow s \ell\ell$ observables. 

\bigskip
The above discussion clearly emphasises the key r\^ole played by Belle II and future cLFV experiments in probing the vector leptoquark scenario as a unified explanation to the $B$-decay anomalies, especially in view of a new determination of $R_{D^{(*)}}$ (central value and associated uncertainties).
Scenarios can be envisaged in which future experimental data 
corroborates current $R_{D^{(*)}}$ values (no change in the central value, corresponding to the red ``dot'' in Fig.~\ref{fig:rdrds_after_belleii}), but accompanied by a reduction of the associated errors (implying tighter likelihood contours): this could then potentially contribute to disfavour $V_1$ as a viable explanation to the charged current $B$-meson decay anomalies.
However, if future Belle II data (dashed contours in Fig.~\ref{fig:rdrds_after_belleii}) evolves along current Belle data, vector leptoquarks would still remain exceptional candidates to explain the $B$-meson decay anomalies, while avoiding detection in cLFV processes in the future.

\section{Concluding remarks}

Being a well-motivated new physics candidate, leptoquark extensions of the SM have been increasingly investigated, in view of their potential for a simple, minimalistic scenario to explain the current hints of LFUV arising from $B$-meson decay data. Vector leptoquarks transforming as $(\mathbf{3},\mathbf{1},2/3)$ are particularly appealing, as they offer a simultaneous explanation for both charged and neutral current $B$-meson  decay anomalies, parametrised by the $R_{K^{(\ast)}}$ and $R_{D^{(\ast)}}$ observables.

In our work, we have thus investigated how minimal constructions, containing the vector leptoquark $V_1$, successfully account for 
the anomalies in both  $R_{K^{(\ast)}}$ and $R_{D^{(\ast)}}$. Leading to our study, and relying on an EFT approach, we first presented results of global fits, which allowed to assess the impact of the  most recent LHCb data in identifying the most favoured generic classes of NP realisations (in terms of new contributions to the relevant Wilson coefficients). Our findings suggest that scenarios in which a universal contribution to $C_9^{bs\ell\ell}$ is present - in addition to the $(V-A)$ contribution -  become increasingly preferred (at a $\sim3\sigma$ level). 

In our study we have taken a  phenomenological approach for the couplings of the vector leptoquark to SM quarks and leptons: we emphasise that starting from a completely general simplified-model parametrisation, 
we presented a fit for the full $3\times 3$ 
matrix ($K_L^{ij}$) encoding the $V_1 \ell q$ couplings,  
taking into account various relevant flavour observables and the anomalous LFUV data.
In addition to providing a better guidance towards possible UV completions of vector leptoquark scenarios 
capable of addressing the LFUV anomalies, this approach can also reveal interesting prospects for observables which can be potentially used to probe the underlying  vector leptoquark hypothesis (we notice that many of the latter observables can be missed in analyses with a priori vanishing couplings to the first generation of SM fermions.
Relying on this alternative formalism for the phenomenological fitting of the vector leptoquark couplings, we thoroughly investigated the impact of such a NP scenario: 
we considered the prospects for an extensive array of observables, including (in addition to the anomalous $B$-meson decay observables) leptonic cLFV transitions, several $B$ decay modes to final states including $\tau^+ \tau^-$ pairs, flavour violating $\tau$ decays as well as cLFV (semi)leptonic decays of $B$-mesons. In view of the excellent experimental prospects, we have investigated several very promising ``{\it golden modes}'' to (indirectly) test the $V_1$ scenario. 
Among these channels one finds $\tau\to\phi\mu$ decays, $b\to s\tau\tau$ and $b\to s\tau\mu$ transitions, as well as $\mu - e$ conversion in nuclei.  
These modes, searched for at Belle II and coming cLFV experiments, will play a crucial r\^ole in testing the vector leptoquark hypothesis as a single explanation to the $R_{K^{(\ast)}}$ and $R_{D^{(\ast)}}$ anomalies.

As we have discussed, the confirmation of LFUV in $B$-meson decays,
(strongly) enhanced rates for $B$-meson decays to $\tau^+ \tau^-$ final states, as well as an observation of cLFV transitions in certain channels (by itself a massive discovery!), would all contribute to substantiate a vector leptoquark NP scenario - although some of the latter signals could indeed arise from other BSM constructions. Conversely, the non-observation of such signals at Belle II and future cLFV experiments has the potential to falsify the vector leptoquark scenario as a solution to the anomalous $R_{D^{(*)}}$ data, if the latter anomaly persists in future measurements with reduced uncertainty (without significant changes in the central values). Should this be the case, and although NP models containing vector leptoquarks could still address the neutral current $B$-decay anomalies (i.e. $R_{K^{(*)}}$), 
a common explanation of both sets of anomalies would be certainly more challenging. 

The coming years clearly offer rich and promising experimental prospects to test one of simplest - yet successful - new physics constructions that allows explaining {\it both} the LFUV $B$-meson decay anomalies.


\section*{Acknowledgements}
JK is grateful to the organisers of ICHEP 2020 where part of this work was presented. The authors are happy to thank Stephane Monteil for many fruitful discussions. The authors are also grateful to Peter Stangl and Martino Borsato for useful correspondences. CH acknowledges support from the DFG Emmy Noether Grant No. HA 8555/1-1. JK, JO, and AMT acknowledge support within the framework of the European Unions Horizon 2020 research and innovation programme under the Marie Sklodowska-Curie grant agreement No 860881-HIDDeN and from the IN2P3 (CNRS) Master Project, ``Flavour probes: lepton sector and beyond'' (16-PH-169).
\begin{appendix}

\renewcommand{\theequation}{\thesection.\arabic{equation}}
\renewcommand{\thetable}{\thesection.\arabic{table}}
\renewcommand{\thefigure}{\thesection.\arabic{figure}}

\setcounter{equation}{0}
\setcounter{table}{0}
\setcounter{figure}{0}

\section{Statistical treatment and fits}
\label{app:stats}

A proper statistical treatment of the experimental data and of the theoretical uncertainties is imperative for a precision analysis of flavour observables.
In general, the goal is to find a set of theoretical predictions for the observables of interest ($\mathcal O_i^\text{th}$) which agrees best with the experimental data on the observables ($\vec{\mathcal O}_i^\text{exp}$).
In order to determine the agreement with data, one builds a likelihood comprising the probability distributions of experimental data, evaluated at the theoretical predictions.
Schematically, we multiply the probability distribution functions (pdf) provided by the experimental data
\begin{equation}
	\mathcal L = \prod_i \mathrm{pdf}_i\left(O_i^\text{exp}, O_i^\text{th}(\vec p)\right)\,,
\end{equation}
in which the theoretical predictions depend on a set of given input parameters $\vec p$, all associated with additional sources of uncertainty.
Maximising this likelihood function then leads to the maximum likelihood estimator -- i.e. ``best-fit point'' -- as the point of highest probability.
In practice, one is only interested in a subset of the theoretical input parameters, or fit parameters ($\vec\theta$), leaving the remaining input parameters as nuisance parameters ($\vec\xi$) to be ``integrated out''. 
To do this, one in general follows either the {\it Bayesian} or the {\it Frequentist} approach, both computationally very expensive.

Another much faster approach which is used throughout this work is a {\it gaussian approximation of the likelihood}, which can be written as
\begin{equation}
	- 2\Delta \mathrm{log} \mathcal L(\vec \theta) \approx \chi^2 = (\vec{\mathcal O}_\text{th}(\vec \theta) - \vec{\mathcal O}_\text{exp})^T \times \left(\mathcal C_\text{th} + \mathcal C_\text{exp}\right)^{-1}\times\left(\vec{\mathcal O}_\text{th}(\vec \theta)- \mathcal{\vec O}_\text{exp}\right)\,\text.
	\label{eqn:loglikelihood}
\end{equation}
In the above, $\vec{\mathcal O}_\text{exp}$ are the central values of the observables as measured by experiments, $\vec{\mathcal O}_\text{th}(\vec \theta)$ the central values of the theoretical predictions with respect to the nuisance parameters (but dependent on the fit parameters $\theta_i$), $\mathcal C_\text{exp}$ the covariance matrix of the measurements of {\it all} included observables and $\mathcal C_\text{th}$ the covariance matrix of the predictions of {\it all} included observables.
The theoretical covariance matrix now contains all theoretical uncertainties of the observables (and their correlations) and is obtained by randomly sampling the nuisance parameters according to their probability distributions.
Note that in this way the nuisance parameters $\vec \xi$ are ``effectively integrated out'' and the likelihood function to be optimised only depends on the parameters of interest, $\vec\theta$.
This approach was first employed in~\cite{Altmannshofer:2014rta}.

The experimental covariance matrix is estimated by first sampling  all experimental probability distributions (with a sample size of $10^6$ random values), including the effects of correlations among them.
In a second step, the mean values and the combined covariance matrix are estimated from the random samples. 
This however leads to an incorrect inclusion of strict upper limits, for instance a half-normal distribution, since mean values of samples drawn from a half-normal distribution (or related distributions) do not correspond to the true central values, which are $0$.
To circumvent this problem, all observables that only have experimental upper bounds are not included in Eq.~\eqref{eqn:loglikelihood}. Their likelihood is evaluated using their specific probability distributions (as provided by the experiments), at the expense of neglecting theoretical uncertainties. The probability distributions are then subsequently added to the global likelihood.

To take into account the theoretical uncertainties and correlations we use a similar Monte-Carlo method - all input parameters are randomly sampled ($N_\text{MC SM} = 10^4$) according to their probability distributions.
Then all observables are computed for each sample, to estimate the theoretical covariance matrix, which then also includes the theoretical correlations between observables.

The resulting approximate log-likelihood (or $\chi^2$) is then minimised using the {\tt MIGRAD} algorithm implemented in the {\tt minuit}~\cite{James:1975dr} library. For the fits of the Wilson coefficients we compute the asymmetric errors with the {\tt MINOS} algorithm. For leptoquark fits this however requires excessively large computation times. 
Therefore, we sample the likelihoods depending on the leptoquark couplings employing MCMC-simulations using the {\tt emcee} python package~\cite{Foreman-Mackey:2012any}. This results in posterior distributions of the couplings and observables of interest.
The quoted $90\%$ ranges are derived from the histograms of the posterior distributions. Here we take symmetric intervals between the $5^\text{th}$ and $95^\text{th}$ percentiles, while predicted upper limits (denoted as dashed lines) correspond to the $90^\text{th}$ percentile.


\section{\bf Observables and data taken into account leading to the fits}
\label{app:Obs}

In this appendix we list the observables taken into account in the different fit set-ups, as well as the datasets used for the fits. The observables (and datasets) are sorted according to the different hadronic and leptonic systems.
Relevant expressions for the computation of the observables can be found in Appendix~\ref{app:observables}.

\subsection{Observables from $b\to s\ell\ell$ transitions}
\label{app:bsll}
Leading to the fits of Sections~\ref{sec:bsll} and~\ref{sec:lqfit}, we include a large number of different binned and unbinned observables into the respective likelihoods. These play a crucial r\^ole in efficiently constraining the $b\to s$ transition FCNC operators and subsequently the leptoquark couplings involved.

\paragraph{Binned observables in $b\to s\ell\ell$}
We take into account all available data for the angular observables in the optimised basis \cite{Descotes-Genon:2013vna}. Depending on the experiment providing the data, the (sub)sets of observables and bins vary. The datasets for the angular observables taken into account is summarised in Table \ref{tab:ang_data}, whereas the data on the differential branching fractions is shown in Table \ref{tab:br_bsll}.
We notice that in all cases we neglect the bin between $6$ and $8\:\mathrm{GeV}^2$ as, due to the $c\bar c$ resonances,
QCD factorisation is no longer a good approximation in this region~\cite{Beneke:2001at}.
Furthermore, we do not take into account the bin $[0.1, 0.98]\:\mathrm{GeV}^2$: 
the different form factor treatments in {\tt flavio}~\cite{Straub:2018kue} and Ref.~\cite{Descotes-Genon:2013vna} lead to significant discrepancies in the associated theoretical uncertainties in this bin, while for all other bins there is a good agreement.
Moreover, in the region of large hadronic recoil, we always take into account the narrow bins, whereas at low hadronic recoil we average over the kinematic region above the resonances.

\begin{table}[h!]
\begin{center}
	\begin{tabular}{|c|c|c|}
	\hline
	Observables & $q^2$-bins in $\mathrm{GeV}^2$ & Datasets\\
	\hline
	\hline
	$\braket{\mathcal O}(B^{0}\to K^\ast\mu^+\mu^-)$ & &\\
	\hline
	$\braket{F_L}$, $\braket{P_1}$, $\braket{P_2}$, $\braket{P_3}$, & $[1.1, 2.5]$, $[2.5, 4]$, &  LHCb'15\cite{Aaij:2015oid}, LHCb'20\cite{Aaij:2020nrf}\\
	$\braket{P_4^\prime}$, $\braket{P_5^\prime}$, $\braket{P_6^\prime}$, $\braket{P_8^\prime}$  &$[4,6]$, $[15, 19]$ & \\
	\hline
	$\braket{F_L}$, $\braket{P_1}$, $\braket{P_4^\prime}$ & $[0.04, 2]$, $[2,4]$, $[4, 6]$ & ATLAS'17\cite{Aaboud:2018krd}\\
	$\braket{P_5^\prime}$, $\braket{P_6^\prime}$, $\braket{P_8^\prime}$ & & \\
	\hline
	$\braket{F_L}$, $\braket{A_{FB}}$, & $[1,2]$, $[2, 4.3]$ & CMS'17\cite{CMS:2017ivg} \\
	 $\braket{P_1}$, $\braket{P_5^\prime}$ & $[4.3, 6]$, $[16,19]$ & \\
	\hline
	$\braket{F_L}$, $\braket{A_{FB}}$ & $[0,2]$, $[2,4.3]$, $[16, 19.3]$ & CDF'12\cite{CDF:2012qwd}\\
	\hline
    \hline
	$\braket{\mathcal O}(B^{+}\to K^\ast\mu^+\mu^-)$ & &\\
	\hline
	$\braket{F_L}$, $\braket{P_1}$, $\braket{P_2}$, $\braket{P_3}$, & $[1.1, 2.5]$, $[2.5, 4]$, &  LHCb'20\cite{Aaij:2020ruw}\\
	$\braket{P_4^\prime}$, $\braket{P_5^\prime}$, $\braket{P_6^\prime}$, $\braket{P_8^\prime}$  &$[4,6]$, $[15, 19]$ & \\
	\hline
	\hline
	$\braket{\mathcal O}(B^0\to K^\ast e^+ e^-)$ & &\\
	\hline
	$\braket{F_L}$, $\braket{P_1}$, $\braket{P_2}$, $\braket{\mathrm{Im}(A_T)}$ & $[0.002, 1.12]$ & LHCb'15~\cite{Aaij:2015dea}\\
	& $[0.0008, 0.257]$ & LHCb'20~\cite{Aaij:2020umj}\\
	\hline
	\hline
	$\braket{\mathcal O}(B_s\to \phi \mu^+ \mu^-)$ & &\\
	\hline
	$\braket{F_L}$, $\braket{S_3}$, $\braket{S_4}$, $\braket{S_7}$ & $[0.1, 2]$, $[2, 5]$, $[15, 19]$ & LHCb'15~\cite{Aaij:2015esa}\\
	\hline
	\end{tabular}
	\caption{Datasets on angular $b\to s\mu\mu$ observables taken into account in the analysis. The 2 digits appearing after each collaborations' name denote the years of the respective publications.}
	\label{tab:ang_data}
\end{center}
\end{table}

\begin{table}[h!]
\begin{center}
	\begin{tabular}{|c|c|c|}
	\hline
	Observables & $q^2$-bins in $\mathrm{GeV}^2$ & Datasets\\
	\hline
	\hline
	$\braket{\frac{\mathrm{dBR}}{\mathrm d q^2}}(B^+\to K^+\mu^+\mu^-)$ & $[1.1, 2]$, $[2,3]$, $[3,4]$ & LHCb'14~\cite{Aaij:2014pli}\\
	& $[4, 5]$, $[5, 6]$, $[15, 22]$ & \\
	\hline
	$\braket{\frac{\mathrm{dBR}}{\mathrm d q^2}}(B^0\to K^0\mu^+\mu^-)$ & $[0.1, 2]$, $[2,4]$, $[4, 6]$, $[15, 22]$ & LHCb'14~\cite{Aaij:2014pli}\\
	\hline
	$\braket{\frac{\mathrm{dBR}}{\mathrm d q^2}}(B^+\to K^*\mu^+\mu^-)$  & 
	$[0.1, 2]$, $[2, 4]$, $[4, 6]$, $[15, 19]$ & LHCb'14~\cite{Aaij:2014pli}\\
	\hline
	$\braket{\frac{\mathrm{dBR}}{\mathrm d q^2}}(B^0\to K^*\mu^+\mu^-)$ &
	$[1.1, 2.5]$, $[2.5, 4]$, $[4, 6]$, $[15, 19]$ & LHCb'16~\cite{Aaij:2016flj}\\
	\hline
	\hline
	$\braket{\frac{\mathrm{dBR}}{\mathrm d q^2}}(B_s\to \phi\mu^+\mu^-)$ & 
	$[0.1, 2]$, $[2, 5]$, $[15, 19]$ & LHCb'15~\cite{Aaij:2015esa}\\
	& $[1,2.5], [2.5,4], [4,6]$ & LHCb'21~\cite{LHCb:2021zwz}\\
	\hline
	\end{tabular}
	\caption{Datasets on binned differential branching ratios in $B\to K^{(*)}\mu\mu$ decays taken into account in the analysis.}
	\label{tab:br_bsll}
\end{center}
\end{table}

In addition to the binned observables in $b\to s\mu\mu$, we also include the $b\to s\ell\ell$ LFUV observables  into the likelihoods.
The bins and datasets of the ratios of (differential) branching fractions $R_{K^{(\ast)}}$, as well as differences of angular observables between electrons and muons in the final state,
\begin{equation}
    Q_{4,5} \equiv P_{4,5}^{'\mu\mu} - P_{4,5}^{'ee} 
\end{equation}
are listed in Table~\ref{tab:lfuv_bsll}.

\begin{table}[h!]
\begin{center}
	\begin{tabular}{|c|c|c|}
	\hline
	Observables & $q^2$-bins in $\mathrm{GeV}^2$ & Datasets\\
	\hline
	\hline
	$\braket{R_K}$ & $[1.1, 6.0]$, $[0.1, 4.0]$, $[1.0,6.0]$, $[14.18, 19.0]$ & (LHCb'19~\cite{Aaij:2019wad}), LHCb'21~\cite{LHCb:2021trn}, Belle'19~\cite{Abdesselam:2019lab}\\
	\hline
	$\braket{R_{K^\ast}}$ & $[0.045, 1.1]$, $[1.1, 6.0]$, $[15,19]$ & LHCb'17~\cite{Aaij:2017vbb}, Belle'19~\cite{Abdesselam:2019wac}\\
	\hline
	$\braket{Q_4}$, $\braket{Q_5}$ & $[0.1, 4]$, $[1.0, 6.0]$, $[14.18, 19.0]$ & Belle'16~\cite{Wehle:2016yoi}\\
	\hline
	\end{tabular}
	\caption{Datasets of observables in $B\to K^{(*)}\ell\ell$ decays sensitive to LFU violation.}
	\label{tab:lfuv_bsll}
\end{center}
\end{table}

\paragraph{Leptonic FCNC decays}
Having sizeable new physics effects in $B\to K^{(\ast)}\mu\mu$ (as required to fit the anomalous data) opens the possibility of having new contributions to other rare $b\to s\ell\ell$ decays, which have either been found to be consistent with the SM, or are yet to be observed. 

Meson decay modes without a hadron in the final state suffer from significantly smaller hadronic uncertainties, since QCD corrections can be absorbed into a redefinition of the decay constant, and all QED and electroweak corrections remain fully perturbative.
Consequently, these decays provide very clean probes for NP effects especially in $C_{7, 10}^{(')}$, but also in $C_{S,P}^{(')}$ Wilson Coefficients.
A recent LHCb analysis~\cite{Aaij:2020nol} of $B_{(s)}\to ee$ yields upper bounds at the $\mathcal{O}(10^{-9})$ level. 
For $B_{(s)}\to \mu\mu$, the situation is more complicated, since the decays are always measured in correlation to each other.
While the decay $B_s\to \mu\mu$ has been observed and measured by several experiments~\cite{LHCb:2021qbv,LHCb:2021awg,Chatrchyan:2013bka, Aaij:2017vad, Aaboud:2018mst, Sirunyan:2019xdu}, as of today only upper limits on the decay $B^0\to\mu\mu$ are available (at the $10^{-10}$ level), due to insufficient statistics. 
In order to avoid losing important correlations in the measurements, we use the 2-dimensional likelihoods (including negative values for $\mathrm{BR}(B^0\to\mu\mu)$) and sample them to obtain a naïve combination, following the prescription of Ref.~\cite{Aebischer:2019mlg,Altmannshofer:2021qrr}. 

\paragraph{Other observables}
To constrain contributions to $C^{(')\:bs\gamma}_7$ in the dipole operator, we also include the branching fractions $\mathrm{BR}(B\to K^\ast\gamma)$~\cite{Amhis:2014hma}, $\mathrm{BR}(B\to X_s\gamma)$~\cite{Misiak:2017bgg} and $\mathrm{BR}(B_s\to\phi\gamma)$\cite{Dutta:2014sxo, Aaij:2012ita}. Notice that all these observables 
correspond to the full branching fractions, implying that they are calculated and measured over the full kinematic region.

\subsection{Charged current $B$-decays}
\label{app:BFCCC}
\paragraph{Observables in $b\to c\ell\nu$}
First and foremost we include the very relevant LFUV ratios $R_{D^{(\ast)}}^{\tau\ell}$, commonly denoted $R_{D^{(\ast)}}$, into the global likelihoods. Analogously, a ratio comparing muons and electrons in the final state ($R_{D^{\ast}}^{\mu e}$) can be defined, which shows excellent agreement with the SM~\cite{Abdesselam:2017kjf, Abdesselam:2018nnh}.
For $R_{D^\ast}^{\tau\ell}$ we use the uncorrelated measurements by LHCb~\cite{Aaij:2015yra,Aaij:2017uff} and Belle~\cite{Hirose:2016wfn}, whereas for $R_D^{\tau\ell}$ there are several measurements,  obtained by BaBar~\cite{Lees:2013uzd} and Belle~\cite{Huschle:2015rga,Hirose:2016wfn,Abdesselam:2019dgh}, always in correlation with $R_{D^\ast}^{\tau\ell}$.

Numerous other observables are taken into account in addition to the anomalous ratios $R_{D^{(\ast)}}$. The extensive array of experimental data (in binned branching fractions of the decay $B\to D^{(\ast)}\ell\nu$) used in our fits is presented in Table~\ref{tab:binned_bcellnu}.

\begin{table}[h!]
\begin{center}
	\begin{tabular}{|c|c|c|}
	\hline
	Observables & $q^2$-bins in $\mathrm{GeV}^2$ & Datasets\\
	\hline
	\hline
	$\braket{\mathrm{BR}}(B^+\to D\tau\nu)$ & $[4, 4.53]$, $[4.53, 5.07]$, $[5.07, 5.6]$, $[5.6, 6.13]$ & Belle'15~\cite{Huschle:2015rga}\\
    $\braket{\mathrm{BR}}(B^0\to D\tau\nu)$	& $[6.13, 6.67]$, $[6.67, 7.2]$, $[7.2, 7.73]$, $[7.73, 8.27]$ & \\
	& $[8.27, 8.8]$, $[8.8, 9.33]$, $[9.33, 9.86]$, $[9.86, 10.4]$ & \\
	& $[10.4, 10.93]$, $[10.93, 11.47]$, $[11.47, 12.0]$ &\\
	\hline
	\hline
	$\braket{\mathrm{BR}}(B^+\to D^\ast\tau\nu)$ & $[4, 4.53]$, $[4.53, 5.07]$, $[5.07, 5.6]$, $[5.6, 6.13]$ & Belle'15~\cite{Huschle:2015rga}\\
    $\braket{\mathrm{BR}}(B^0\to D^\ast\tau\nu)$ & $[6.13, 6.67]$, $[6.67, 7.2]$, $[7.2, 7.73]$, $[7.73, 8.27]$ & \\
	& $[8.27, 8.8]$, $[8.8, 9.33]$, $[9.33, 9.86]$, $[9.86, 10.4]$ & \\
	& $[10.4, 10.93]$ & \\
	\hline
	\hline
	$\braket{\mathrm{BR}}(B^+\to D\mu\nu)$ & $[0.0, 1.03]$, $[1.03, 2.21]$, $[2.21, 3.39]$, $[3.39, 4.57]$ & Belle'15~\cite{Glattauer:2015teq}\\
	$\braket{\mathrm{BR}}(B^+\to D e\nu)$ & $[4.57, 5.75]$, $[5.75, 6.93]$, $[6.93, 8.11]$, $[8.11, 9.3]$ &\\
	&  $[9.3, 10.48]$, $[10.48, 11.66]$ &\\
	\hline
	\hline
	$\braket{\mathrm{BR}}(B^0\to D\mu\nu)$ & $[0.0, 0.97]$, $[0.97, 2.15]$, $[2.15, 3.34]$, $[3.34, 4.52]$ & Belle'15~\cite{Glattauer:2015teq}\\
	$\braket{\mathrm{BR}}(B^0\to D e\nu)$ & $[4.52, 5.71]$, $[5.71, 6.89]$, $[6.89, 8.07]$, $[8.07, 9.26]$ &\\
	& $[9.26, 10.44]$, $[10.44, 11.63]$ &\\
	\hline
	\end{tabular}
	\caption{Datasets of binned branching fractions in $B\to D^{(\ast)}\ell\nu$.}
	\label{tab:binned_bcellnu}
\end{center}
\end{table}

\noindent
Furthermore, we include the unbinned branching fractions $\mathrm{BR}(B^+\to D^{(\ast)} \mu\nu)$, $\mathrm{BR}(B^+\to D^{(\ast)} e\nu)$~\cite{Aubert:2008yv, Aubert:2007qs} and the inclusive branching fraction $\mathrm{BR}(B\to X_c e\nu)$~\cite{Urquijo:2006wd, Aubert:2009qda}.

\paragraph{Other charged current $B$-decays}
In addition to charged current $b\to c\ell\nu$ decays, we also include certain $b\to u\ell\nu$ decays to obtain further constraints on the leptoquark couplings to the first quark generation. These can be found in Table~\ref{tab:buellnu}.
\begin{table}[h!]
\begin{center}
	\begin{tabular}{|c|c|c|}
	\hline
	Observable & SM prediction & Measurement/Limit\\
	\hline
	\hline
	$\mathrm{BR}(B^0\to\pi\tau\nu)$ & $(8.4\pm1.1)\times10^{-5}$ & $(1.52 \pm 0.72 \pm 0.13) \times 10^{-4}$ Belle'15~\cite{Hamer:2015jsa}\\
	\hline
	$\mathrm{BR}(B^+\to\tau\nu)$ & $(8.8\pm0.6)\times10^{-5}$ & $(1.09 \pm 0.24) \times 10^{-4}$ PDG~\cite{Tanabashi:2018oca}\\
	\hline
	$\mathrm{BR}(B^+\to\mu\nu)$ & $(4.0\pm0.3)\times10^{-7}$ & $<1\times10^{-6}$ HFLAV'18~\cite{Amhis:2019ckw}\\
	\hline
	\end{tabular}
	\caption{Datasets on further charged current $B$-meson decays. The SM predictions are obtained using {\tt flavio}~\cite{Straub:2018kue}.}
	\label{tab:buellnu}
\end{center}
\end{table}

\subsection{Strange, charm and $\tau$-lepton decays}
\label{app:sctau}
The above listed data mostly allows to constrain combinations of second and third generation quark leptoquark couplings (to all leptons). To achieve more precise constraints for the second and first generation quarks, we further include numerous decays of strange and charm flavoured mesons.
Since the light mesons cannot decay into $\tau$-leptons, we also use data on SM allowed $\tau$-lepton decays, as a complementary source of information.

\paragraph{Binned charm decays}
In addition to the precise measurements of the full branching fractions of several charmed meson decay modes, there are also precise measurements of the $q^2$ distributions for several charged current decay modes in semileptonic charm decays with an electron in the final state. The datasets used are presented in Table~\ref{tab:binned_charm}.
\begin{table}[h!]
\begin{center}
	\begin{tabular}{|c|c|c|}
	\hline
	Observables & $q^2$-bins in $\mathrm{GeV}^2$ & Datasets\\
	\hline
	\hline
	$\braket{\mathrm{BR}}(D^{+,\,0}\to K e\nu)$ & $[0.0,0.2]$, $[0.2,0.4]$, $[0.4,0.6]$, $[0.6,0.8]$ & CLEO~\cite{Besson:2009uv}, BESIII~\cite{Ablikim:2015ixa,Ablikim:2017lks}\\
	& $[0.8, 1.0]$, $[1.2, 1.4]$, $[1.4, 1.6]$, $[1.6, 1.88]$ & \\
	\hline
	$\braket{\mathrm{BR}}(D^{0}\to \pi e\nu)$ & $[0.0,0.2]$, $[0.2,0.4]$, $[0.4,0.6]$, $[0.6,0.8]$ &  BESIII~\cite{Ablikim:2015ixa}\\
	& $[0.8, 1.0]$, $[1.2, 1.4]$, $[1.4, 1.6]$, $[1.6, 1.8]$ & \\
	& $[1.8, 2.0]$, $[2.0, 2.2]$, $[2.2, 2.4]$, $[2.4, 2.6]$ & \\
	& $[2.6, 2.98]$ & \\
	\hline
	$\braket{\mathrm{BR}}(D^{+}\to \pi e\nu)$ & $[0.0,0.3]$, $[0.3,0.6]$, $[0.6,0.9]$, $[0.9,1.2]$ &  CLEO~\cite{Besson:2009uv}, BESIII~\cite{Ablikim:2017lks}\\
	& $[1.2, 1.5]$, $[1.5, 2.0]$, $[2.0, 2.98]$ & \\
	\hline
	\end{tabular}
	\caption{Datasets on binned branching fractions in charged current charm decays.}
	\label{tab:binned_charm}
\end{center}
\end{table}

\paragraph{Unbinned observables}
Besides the binned semileptonic charm decays, we also include the full branching fractions for charged current leptonic and semileptonic charm decays, charged and neutral current decays of strange flavoured mesons, and charged current semileptonic $\tau$-lepton decays. The charged current decays are listed in Table~\ref{tab:charged_charm_strange_tau} and the neutral current ones in Table~\ref{tab:fcnc_strange}.
\begin{table}[h!]
\begin{center}
	\begin{tabular}{|c|c|c|}
	\hline
	Observable & SM prediction & Measurement/Limit\\
	\hline
	\hline
	$\mathrm{BR}(D^0\to K\mu\nu)$ & $(3.54\pm0.25)\times10^{-2}$ & $(3.31\pm0.13)\times10^{-2}\quad$~\cite{Tanabashi:2018oca}\\
	\hline
	$\mathrm{BR}(D^0\to Ke\nu)$ & $(3.55\pm0.25)\times10^{-2}$ & $(3.53\pm0.028)\times10^{-2}\quad$~\cite{Tanabashi:2018oca}\\
	\hline
	$\mathrm{BR}(D^+\to K\mu\nu)$ & $(9.04\pm0.55)\times10^{-2}$ & $(8.74\pm0.19)\times10^{-2}\quad$~\cite{Tanabashi:2018oca}\\
	\hline
	$\mathrm{BR}(D^+\to Ke\nu)$ & $(9.08\pm0.64)\times10^{-2}$ & $(8.73\pm0.0)\times10^{-2}\quad$~\cite{Tanabashi:2018oca}\\
	\hline
	$\mathrm{BR}(D^0\to\pi\mu\nu)$ & $(2.67\pm0.16)\times10^{-3}$ & $(2.37\pm0.24)\times10^{-3}\quad$~\cite{Tanabashi:2018oca}\\
	\hline
	$\mathrm{BR}(D^0\to\pi e\nu)$ & $(2.68\pm0.15)\times10^{-3}$ & $(2.91\pm0.04)\times10^{-3}\quad$~\cite{Tanabashi:2018oca}\\
	\hline
	$\mathrm{BR}(D^+\to\pi e\nu)$ & $(3.48\pm0.22)\times10^{-3}$ & $(3.72\pm0.17)\times10^{-3}\quad$~\cite{Tanabashi:2018oca}\\
	\hline
	\hline
	$\mathrm{BR}(D^+\to \tau\nu)$ & $(1.09\pm0.01)\times10^{-3}$ & $<1.2\times10^{-3}\quad$~\cite{Tanabashi:2018oca}\\
	\hline
	$\mathrm{BR}(D^+\to \mu\nu)$ & $(4.10\pm0.05)\times10^{-4}$ & $(3.74\pm0.17)\times10^{-4}\quad$~\cite{Tanabashi:2018oca}\\
	\hline
	$\mathrm{BR}(D^+\to e\nu)$ & $(9.64\pm0.12)\times10^{-9}$ & $<8.8\times10^{-6}\quad$~\cite{Tanabashi:2018oca}\\
	\hline
	\hline
	$\mathrm{BR}(D_s\to \tau\nu)$ & $(5.32\pm0.05)\times10^{-2}$ & $(5.48\pm0.23)\times10^{-2}\quad$~\cite{Tanabashi:2018oca}\\
	\hline
	$\mathrm{BR}(D_s\to \mu\nu)$ & $(5.46\pm0.05)\times10^{-3}$ & $(5.50\pm0.23)\times10^{-3}\quad$~\cite{Tanabashi:2018oca}\\
	\hline
	$\mathrm{BR}(D_s\to e\nu)$ & $(1.28\pm0.01)\times10^{-7}$ & $<8.3\times10^{-5}\quad$~\cite{Tanabashi:2018oca}\\
	\hline
	\hline
	$\mathrm{BR}(K^+\to\pi\mu\nu)$ & $(3.39\pm0.04)\times10^{-2}$ & $(3.35\pm0.03)\times10^{-2}\quad$~\cite{Tanabashi:2018oca}\\
	\hline
	$\mathrm{BR}(K^+\to\pi e\nu)$ & $(5.13\pm0.05)\times10^{-2}$ & $(5.07\pm0.04)\times10^{-2}\quad$~\cite{Tanabashi:2018oca}\\
	\hline
	$\mathrm{BR}(K_L\to\pi\mu\nu)$ & $(27.11\pm0.26)\times10^{-2}$ & $(27.04\pm0.07)\times10^{-2}\quad$~\cite{Tanabashi:2018oca}\\
	\hline
	$\mathrm{BR}(K_L\to\pi e\nu)$ & $(40.93\pm0.46)\times10^{-2}$ & $(40.55\pm0.11)\times10^{-2}\quad$~\cite{Tanabashi:2018oca}\\
	\hline
	\hline
	$\mathrm{BR}(K^+\to\mu\nu)$ & $(63.08\pm0.83)\times10^{-2}$ & $(63.56\pm0.11)\times10^{-2}\quad$~\cite{Tanabashi:2018oca}\\
	\hline
	$\mathrm{BR}(K^+\to e\nu)$ & $(1.561\pm0.023)\times10^{-5}$ & $(1.582 \pm 0.007) \times10^{-5}\quad$~\cite{Tanabashi:2018oca}\\
	\hline
	\hline
	$\mathrm{BR}(\tau\to K\nu)$ & $(7.09\pm0.11)\times10^{-3}$ & $(6.96\pm0.10)\times10^{-3}\quad$~\cite{Tanabashi:2018oca}\\
	\hline
	$\mathrm{BR}(\tau\to \pi\nu)$ & $(10.84\pm0.14)\times10^{-2}$ & $(10.82\pm0.05)\times10^{-3}\quad$~\cite{Tanabashi:2018oca}\\
	\hline
	\end{tabular}
	\caption{Data on charged current charm and strange flavoured meson decays. The SM predictions are obtained using {\tt flavio}~\cite{Straub:2018kue}.}
	\label{tab:charged_charm_strange_tau}
\end{center}
\end{table}

\begin{table}[h!]
\begin{center}
	\begin{tabular}{|c|c|c|}
	\hline
	Observable & SM prediction & Measurement/Limit\\
	\hline
	\hline
	$\mathrm{BR}(K_L\to \mu^+\mu^-)$ & $(7.45\pm1.24)\times10^{-9}$ & $(6.84\pm0.11)\times10^{-9}\quad$~\cite{Tanabashi:2018oca}\\
	\hline
	$\text{BR}(K^+ \to \pi^+ \nu\bar \nu)$ &
	$(8.4 \pm 1.0) \times 10^{-11}
	\phantom{|}^{\phantom{|}}_{\phantom{|}}$\cite{Buras:2015qea} &
	\begin{tabular}{l}
	$17.3^{+11.5}_{-10.5} \times 10^{-11}
	\quad$\cite{Artamonov:2008qb}
	\\
	$< 1.78 \times 10^{-10}
	\quad$\cite{CortinaGil:2020vlo}
	\end{tabular}
	\\
	\hline
	$\text{BR}(K_L \to \pi^0 \nu\bar \nu)$ &
	$(3.4 \pm 0.6) \times 10^{-11}\quad$\cite{Buras:2015qea} &
	$<  2.6 \times 10^{-8}
	\quad$\cite{Ahn:2009gb}\\
	\hline
	\end{tabular}
	\caption{Data on FCNC kaon decays. The SM predictions are obtained using {\tt flavio}~\cite{Straub:2018kue} if not otherwise stated.}
	\label{tab:fcnc_strange}
\end{center}
\end{table}

\subsection{Belle II Observables}
As discussed in Section~\ref{sec:future_Belle}, we use specific fit set-ups which allow for an extrapolation of the current situation into the near future.
The future sensitivities, taken into account as data, are listed in Table~\ref{tab:belleii}; these always correspond to the full anticipated luminosity of $50\:\mathrm{ab}^{-1}$.

\begin{table}[h!]
\begin{center}
	\begin{tabular}{|c|c|c|}
	\hline
	Observable & Current bound & Belle II Sensitivity\\
	\hline
	\hline
	$\text{BR}(\tau \to e \gamma)$	&
 	\quad $<3.3\times 10^{-8}$ \quad BaBar~\cite{Aubert:2009ag}	 &
 	\quad $<3\times10^{-9}$ \quad  	 	\\
	\hline
	$\text{BR}(\tau \to \mu \gamma)$	&
	 \quad $ <4.4\times 10^{-8}$ \quad BaBar~\cite{Aubert:2009ag}	 &
 	\quad $<10^{-9}$ \quad 		\\
	\hline
	$\text{BR}(\tau \to 3 e)$	&
 	\quad $<2.7\times 10^{-8}$ \quad Belle~\cite{Hayasaka:2010np}&
 	\quad $<5\times10^{-10}$ \quad  	\\
 	\hline
	$\text{BR}(\tau \to 3 \mu )$	&
 	\quad $<3.3\times 10^{-8}$ \quad Belle~\cite{Hayasaka:2010np}	 &
 	\quad $<5\times10^{-10}$ \quad 		\\
	\hline
	\hline
	$\mathrm{BR}(\tau\to\pi e)$ & $<8\times10^{-8}$\quad Belle~\cite{Miyazaki:2007jp} & $<4\times10^{-10}$\\
	\hline
	$\mathrm{BR}(\tau\to\pi \mu)$ & $<1.1\times10^{-7}$\quad Belle~\cite{Miyazaki:2007jp} & $<5\times10^{-10}$\\
	\hline
	$\mathrm{BR}(\tau\to\phi e)$ & $<3.1\times10^{-8}$\quad Belle~\cite{Miyazaki:2011xe} & $<5\times10^{-10}$\\
	\hline
	$\mathrm{BR}(\tau\to\phi \mu)$ & $<8.4\times10^{-8}$\quad Belle~\cite{Miyazaki:2011xe} & $<2\times10^{-9}$\\
	\hline
	$\mathrm{BR}(\tau\to\rho e)$ & $<1.8\times10^{-8}$\quad Belle~\cite{Miyazaki:2011xe} & $<3\times10^{-10}$\\
	\hline
	$\mathrm{BR}(\tau\to\rho \mu)$ & $<1.2\times10^{-8}$\quad Belle~\cite{Miyazaki:2011xe} & $<2\times10^{-10}$\\
	\hline
	\hline
	$\mathrm{BR}(B^+\to K^+\tau^+e^-)$ & $<1.5\times10^{-5}$\quad BaBar~\cite{Lees:2012zz} & $<2.1\times10^{-6}$\\
	$\mathrm{BR}(B^+\to K^+\tau^-e^+)$ & $<4.3\times10^{-5}$\quad BaBar~\cite{Lees:2012zz} &\\
	\hline
	$\mathrm{BR}(B^+\to K^+\tau^+\mu^-)$ & $<2.8\times10^{-5}$\quad BaBar~\cite{Lees:2012zz} & $<3.3\times10^{-6}$\\
	$\mathrm{BR}(B^+\to K^+\tau^-\mu^+)$ & $<4.5\times10^{-5}$\quad BaBar~\cite{Lees:2012zz} &\\
	\hline
	\hline
	$\mathrm{BR}(B^0\to e^\pm\tau^{\mp})$ & $<2.8\times10^{-5}$\quad BaBar~\cite{Aubert:2008cu} & $ <1.6\times10^{-5}$\\
	\hline
	$\mathrm{BR}(B^0\to \mu^\pm\tau^{\mp})$ & $<1.4\times10^{-5}$\quad LHCb~\cite{Aaij:2019okb} & $ <1.3\times10^{-5}$\\
	\hline
	\end{tabular}

	\vspace{0.5cm}
	\begin{tabular}{|c|c|c|}
	\hline
	Observable & SM prediction & Belle II Sensitivity\\
	\hline
	\hline
	$\mathrm{BR}(B^0\to\tau\tau)$ & $(2.22 \pm 0.19)\times10^{-8}$\quad~\cite{Bobeth:2013uxa,Hermann:2013kca, Bobeth:2013tba} & $<9.6\times10^{-5}$\\
	\hline
	$\mathrm{BR}(B_s\to\tau\tau)$ & $(7.73 \pm 0.49)\times10^{-7}$\quad~\cite{Bobeth:2013uxa,Hermann:2013kca, Bobeth:2013tba} & $<8.1\times10^{-4}$\\
	\hline
	$\braket{\mathrm{BR}}(B\to K\tau^+\tau^-)_{[15, 22]}$ & $(1.20\pm0.12)\times10^{-7}$\quad~\cite{Capdevila:2017iqn} & $<2\times10^{-5}$\\
	\hline
	\end{tabular}
	\caption{Observables for which Belle II will improve on current experimental sensitivities. The SM predictions are obtained using {\tt flavio}~\cite{Straub:2018kue}, unless otherwise stated.}
	\label{tab:belleii}
\end{center}
\end{table}

\section{Vector leptoquark contributions to leptonic and mesonic flavour observables}\label{app:observables}

New physics models aiming at addressing the LFUV hints in $B$-meson decays typically give rise to new contributions to several flavour observables depending on the new flavour structure; these include contributions to various flavour conserving and flavour violating leptonic and semileptonic mesonic decay modes, as well as cLFV processes. 
In particular, the vector leptoquark scenario can already contribute to some of these observables at tree level, while others receive leading contributions at the one-loop order.
In this appendix we collect information allowing to estimate the vector leptoquark contribution to several of the above mentioned processes.

\subsection{Leptonic and semileptonic meson decays }\label{app:meson}

Here we summarise the different vector leptoquark contributions to
leptonic and semileptonic meson decays which arise at tree-level, and to modes with final state neutrinos
(whose new contributions arise at one-loop level). 
We do not include neutral meson oscillations which arise at one-loop level and typically provide much weaker 
constraints if, apart from the leptoquarks, only SM fields are considered. 
However, we notice that this may no longer hold in the presence of additional heavy fermionic states (which might be present in a UV-complete model, as for example heavy vector-like leptons);
in that case, the contributions could be sizeable so that neutral meson oscillations can then lead to important constraints, as discussed in~\cite{Hati:2019ufv,Cornella:2021sby}.

\subsubsection{$P \rightarrow \ell^- \ell^{\prime +}$ decays}
Vector leptoquarks can induce new contributions to purely leptonic decays of pseudoscalar mesons, leading to important constraints on the flavour structure of $V_1$ couplings. Here, we provide a brief summary of the formalism for the
computation of the $P \rightarrow \ell^- \ell^{\prime +}$ rates. 

\noindent
Following the standard decomposition of the hadronic matrix element~\cite{Becirevic:2016zri}
\begin{equation}
  \langle 0\,| \,\bar d_j \,\gamma_\mu\,\gamma_5 \,d_i|\,P(p)\rangle
  \,=\, i \,p_\mu \,f_{P}\,,
\end{equation}
where $f_P$ corresponds to the $P$ meson decay constant,
the branching fraction can be expressed as
\begin{align}
\text{BR}(P &\rightarrow \ell^- \,\ell^{\prime +}) \,=\,
\frac{\tau_{P}}{64 \,\pi^3}\frac{\alpha^{2}
  \,G_F^{2}}{M_P^{3}}\,f_P^{2}\,|V_{3j}\,V_{3i}^{{\ast}}|^2\,
\lambda^{\frac{1}{2}}(M_P, m_{\ell}, m_{\ell^\prime}) \times\nonumber \\
&\times\Bigg\{\left(M_P^{2} - \left(m_{\ell} + m_{\ell^\prime}
\right)^{2} \right)\Bigg|\left(C_9 - C_9^{\prime}\right)\left(m_{\ell}
- m_{\ell^\prime} \right) +\left(C_S - C_S^{\prime}
\right)\frac{M_P^{2}}{m_{d_j} + m_{d_i}}  \Bigg|^{2} + \nonumber \\
&+\left(M_P^{2} - \left(m_{\ell} - m_{\ell^\prime} \right)^{2}
\right)\Bigg|\left(C_{10} - C_{10}^{\prime}\right)\left(m_{\ell} +
m_{\ell^\prime} \right) +\left(C_P -
C_P^{\prime}\right)\frac{M_P^{2}}{m_{d_j} + m_{d_i}}   \Bigg|^{2}
\Bigg\}\,,
\end{align}
where the $\lambda(a,b,c)$ is the standard K\"all\'en-function, defined as
$\lambda(a,b,c) = \left(a^{2} - \left(b-c\right)^{2} \right)\left(a^{2} -
\left(b+c\right)^{2} \right)$. 
Note that for a lepton flavour
conserving decay mode, e.g. $B_s \rightarrow \mu\mu$, one must include
the SM contribution and the relevant RG running effects. Since the vector leptoquarks contribute to the leptonic pseudoscalar meson decays at the tree level, such processes can provide important and very stringent constraints on the vector leptoquark couplings.

\subsubsection{$P \rightarrow P^{\prime} \ell^- \ell^{\prime +}$ decays}
The semileptonic decays of pseudoscalar mesons can also be the source of significant constraints on the vector leptoquark couplings. To evaluate the differential branching fractions for these modes, we parametrise
the hadronic matrix elements following the standard convention as
\begin{align}
  &\langle \bar P^{\prime}(p')\,| \,\bar d_i \,\gamma_\mu \,
  d_j\,| \,\bar P (p)\rangle
= \left[(p+p')_\mu - \frac{M_P^{2} - M_{P^{\prime}}^{2}}{q^{2}}q_\mu
  \right]\,f_+(q^{2}) +\frac{M_P^{2} - M_{P^{\prime}}^{2}}{q^{2}}q_\mu
\,f_0(q^{2})\,,\\
&\langle \bar P^{\prime}(p')\,|\,\bar d_i  \sigma_{\mu\nu} \,d_j\,
|\,\bar P(p)\rangle = -i\,\left(p_\mu \,p'_\nu - p_\nu \,p'_\mu
\right)\frac{2}{M_{P}
  + M_{P^{\prime}}}\, f_T(q^{2},\mu)\,,
\end{align}
where the momentum transfer lies in the range
$(m_{\ell} + m_{\ell^\prime})^{2} \leq q^{2}\leq (M_P - M_{P^{\prime}})^{2}$.
For the evaluation of the form factors we closely follow the prescription of~\cite{Khodjamirian:2010vf}. The final differential branching fraction for the decay $P \rightarrow P^{\prime} \ell^- \ell^{\prime +}$  can be expressed in the form
\begin{align}
\frac{d\,\mathrm{BR} (P \rightarrow P^{\prime}
\ell^-\ell^{\prime +})}{d q^{2}} &= |\mathcal
N_{P^{\prime}}(q^{2})|^{2}\times\Big\{\varphi_7(q^{2})\,|C_7 +
C_7^{\prime}|^{2} + \varphi_9(q^{2})\,|C_9 + C_9^{\prime}|^{2}
+\varphi_{10}(q^{2})\,|C_{10} + C_{10}^{\prime}|^{2} \nonumber\\
&+  \varphi_S(q^{2})\,|C_S + C_S^{\prime}|^{2} + \varphi_P(q^{2})\,|C_P +
C_P^{\prime}|^{2}  + \varphi_{79}(q^{2})\,\mathrm{Re}\left[(C_7 +
  C_{7}^{\prime})\,(C_9 + C_{9}^{\prime})^{{\ast}} \right]\nonumber\\
&+ \varphi_{9S}(q^{2})\,\mathrm{Re}\left[(C_9 + C_{9}^{\prime})\,(C_S +
  C_{S}^{\prime})^{{\ast}} \right] +
\varphi_{10P}(q^{2})\,\mathrm{Re}\left[(C_{10} + C_{10}^{\prime})\,(C_P +
  C_{P}^{\prime})^{{\ast}} \right] \Big\}\: ,
\end{align}
where
\begin{align}
\varphi_7(q^{2}) &= \frac{2\,m_{d_j}\,|f_T(q^{2})|^{2}}{(M_P +
  M_{P^{\prime}})^{2}}\,\lambda(M_P, M_{P^{\prime}},
\sqrt{q^{2}})\,\left[1 - \frac{(m_{\ell} - m_{\ell^\prime})^{2}}{q^{2}}
  - \frac{\lambda(\sqrt{q^{2}}, m_{\ell}, m_{\ell^\prime})}{3\,q^{4}}
  \right]\:\text, \nonumber\\
\varphi_{9(10)}(q^{2}) &= \frac{1}{2}\,|f_0(q^{2})|^{2}(m_{\ell} \mp
m_{\ell^\prime})^{2}\,\frac{(M_P^{2} -
  M_{P^{\prime}}^{2})^{2}}{q^{2}}\,\left[1 - \frac{(m_{\ell} \pm
    m_{\ell^\prime})^{2}}{q^{2}} \right]\nonumber\\
&+ \frac{1}{2}\,|f_+(q^{2})|^{2}\,\lambda(M_P, M_{P^{\prime}},
\sqrt{q^{2}})\,\left[1 - \frac{(m_{\ell} \mp
    m_{\ell^\prime})^{2}}{q^{2}} - \frac{\lambda(\sqrt{q^{2}},
    m_{\ell}, m_{\ell^\prime})}{3\,q^{4}} \right]\:\text,\nonumber\\
\varphi_{79}(q^{2}) &= \frac{2\,m_{d_j}\,f_+(q^{2})\,f_T(q^{2})}{M_P +
  M_{P^{\prime}}}\,\lambda(M_P, M_{P^{\prime}}, \sqrt{q^{2}})\,\left[1 -
  \frac{(m_{\ell} - m_{\ell^\prime})^{2}}{q^{2}} -
  \frac{\lambda(\sqrt{q^{2}}, m_{\ell}, m_{\ell^\prime})}{3\,q^{4}}
  \right]\:\text,\nonumber\\
\varphi_{S(P)}(q^{2}) &= \frac{q^{2}\,|f_0(q^{2})|^{2}}{2\,(m_{d_j} -
  m_{d_i})^{2}}\,\left(M_P^{2} - M_{P^{\prime}}^{2} \right)^{2}\,\left[1 -
  \frac{(m_{\ell} \pm m_{\ell^\prime})^{2}}{q^{2}} \right]\:\text,\nonumber\\
\varphi_{10P(9S)}(q^{2}) &= \frac{|f_0(q^{2})|^{2}}{m_{d_j} -
  m_{d_i}}\,(m_{\ell} \pm m_{\ell^\prime})(M_P^{2} -
M_{P^{\prime}}^{2})^{2}\,\left[1 - \frac{(m_{\ell} \mp
    m_{\ell^\prime})^{2}}{q^{2}} \right]\,,
\end{align}
and the normalisation factor is given by
\begin{equation}
|\mathcal N_{P^{\prime}}(q^{2})|^{2} = \tau_{P}\,\frac{\alpha^{2}\,
  G_F^{2} \,|V_{3j} \,V_{3i}^{{\ast}}|^{2}}{512 \,\pi^5\,
  M_P^3}\,\frac{\lambda^{\frac{1}{2}}(\sqrt{q^{2}}, m_{\ell},
  m_{\ell^\prime})}{q^{2}}\,\lambda^{\frac{1}{2}}(\sqrt{q^{2}}, M_P,
M_{P^{\prime}})\,.
\end{equation}

\subsubsection{One loop effects in modes leading to final state neutrinos}
The vector leptoquark can also contribute to $s\to d \nu\nu$ and $b\to s \nu\nu$ transitions at one-loop level.
The $|\Delta S|=1$ rare decays $K^+\,(K_L)\to \pi^+\,(\pi^0)\,\nu_\ell \bar\nu_{\ell^\prime }$ and
$B\to  K^{(\ast)} \nu_\ell\bar \nu_{\ell^\prime}$ correspond to the quark level transition $d_j\to d_i \nu_\ell\bar \nu_{\ell^\prime}$, which can be described by the short-distance effective Hamiltonian~\cite{Buras:2014fpa,Bobeth:2017ecx,Bordone:2017lsy}
\begin{eqnarray}\label{eq:eff-H-Ktopi}
-\mathcal{H}_\text{eff} =
&\frac{4 \,G_F}{\sqrt{2}} \,V_{3i}^\ast \,V_{3j}\,
\frac{\alpha_e}{2\,\pi}\,
\left[C_{L,ij}^{\ell\ell^\prime} \,\left(\bar d_i\,\gamma_\mu \,P_L\,
  d_j\right)\, \left(\bar \nu_\ell\,\gamma^\mu\,
  \,P_L\,\nu_{\ell^\prime}\right) \right. \nonumber\\
&+\left.
C_{R,ij}^{\ell\ell^\prime} \,\left(\bar d_i\,\gamma_\mu \,P_R\,
d_j\right)\, \left(\bar \nu_\ell \,\gamma^\mu
\,P_L\nu_{\ell^\prime}\right) \right] \,+\, \text{H.c.}\, ,
\end{eqnarray}
where $i,j$ corresponds to the down-type quark content of the final and
initial state mesons, respectively.
For vector leptoquarks, the one loop contributions are a priori divergent; consequently, the corresponding would-be Goldstone modes must be consistently included to obtain the correct result. Following
the prescription of~\cite{Crivellin:2018yvo}, the coefficient $C_{L,fa}^{ij}$ for $d_a \rightarrow d_f\bar\nu_i\nu_j$, due to $V_1$ leptoquark exchange is given by
\begin{align}
C_{L,fa}^{ij} = \sum_{k,l} -\frac{M_W^{2}}{2\,e^{2}\, V_{3a}\,V_{3f}^{\ast}
 \, m_{V_1}^2}\Bigg(&6\,
K_L^{fj}\,K_L^{ai\ast}\,{\ln}\left(\frac{M_W^{2}}{m_{V_1}^2} \right) +
V_{3f}^{\ast}\,V_{3k}\,K_L^{kj}\,V_{3a}\,V_{3l}^{\ast}\,
K_L^{li\ast}\,\frac{m_t^{2}}{M_W^{2}}
\nonumber\\
&+
3\left(V_{3a}\,V_{3k}^{*}\,K_L^{ki\ast}\,K_L^{fj} +
V_{3f}^{\ast}\,V_{3k}\,K_L^{kj}\,K_L^{ai\ast}
\right)\,\frac{m_t^{2}\,{\ln}\left(\frac{m_t^{2}}{M_W^{2}}
  \right)}{m_t^{2} - M_W^{2}}
\Bigg)\:\text,
\end{align}
where $M_W$ and $m_t$ respectively correspond to the masses of the $W$ boson and top quark.
The neutral and charged kaon decay branching fractions can then be obtained by~\cite{Buras:2004qb,Buras:2015qea}
\begin{align}
&\text{BR}(K^ \pm \to \pi ^ \pm \nu \bar \nu)\, =\,
  \frac{1}{3}\left( 1 + \Delta _{EM} \right)\,\eta _\pm \times
  \sum_{f,i = 1}^3 \left\{ \left[
    \frac{\text{Im}\left(\lambda_t\,\tilde X_L^{fi} \right)}{
      \lambda^5}\right]^{2}+ \left[
    \frac{\text{Re} \left(\lambda_c \right)}{
      \lambda} \,P_c\, \delta _{fi}
    + \frac{\text{Re} (
      \lambda_t \,\tilde X_L^{fi})}{ \lambda ^5}
  \right]^{2}\right \},\nonumber\\
  &\text{BR}(K_L \to \pi \nu \bar \nu )\, = \frac{1}{3}{\eta_L}
  \sum_{f,i = 1}^3 \left[\frac{\text{Im}\left(\lambda_t\,
        \tilde X_L^{fi} \right)}{\lambda^5}\right]^{2}\,,
\end{align}
where
\begin{align}
\tilde X_L^{fi} &
  = X_{L}^{\text{SM},fi} - s_W^2\,C_{L,sd}^{fi}\,,\;\quad
P_c = 0.404 \pm 0.024\,, \nonumber\\
\eta_\pm  &=
\left( 5.173 \pm 0.025 \right)\times 10^{-11}
\left[\frac{\lambda}{0.225} \right]^8\,,\nonumber\\
  \eta_L &=\left( 2.231 \pm 0.013 \right)\times 10^{-10}
 \left[\frac{\lambda}{0.225} \right]^8\,,\nonumber\\
\Delta_{EM} &=  - 0.003\,,\; \quad
X_{L}^{\text{SM},fi} =
\left(1.481 \pm 0.005 \pm 0.008\right)\,\delta _{fi}\,.
\end{align}
Here, $\lambda$ corresponds to the standard Wolfenstein parametrisation (i.e. the Cabibbo angle), $\lambda_c = V_{cs}^\ast V_{cd}$ and $\lambda_t = V_{ts}^\ast V_{td}$. The decay width for $B\to K^{(*)}\nu\bar{\nu}$ has been derived
in~\cite{Buras:2014fpa}, leading to
$C_{L,sb}^{\text{SM},fi} \approx -1.47/s_W^2\delta_{fi}$, which can be used to normalise the branching ratios as
\begin{equation}
R_{K^{(*)}}^{\nu\bar{\nu}} \,= \,
\frac{1}{3}\sum_{f,i=1}^3
\frac{ \big|C_{L,sb}^{fi} \big|^2}{
  \big|C_{L,sb}^{\text{SM},fi}\big|^2} \,\text.
\end{equation}
%


\subsection{Charged lepton flavour violating decays}{\label{app:cLFV}}
Charged lepton flavour violating observables, such as radiative decays $\ell_i \rightarrow \ell_j \gamma$, three-body decays $\ell_i \rightarrow 3\ell_j$, as well as neutrinoless $\mu - e$ conversion in nuclei, can lead to important constraints on the vector leptoquark couplings, due to the non-universal couplings to different flavours of SM charged leptons. 
We recall that while $\ell_i \rightarrow \ell_j \gamma$ and $\ell_i \rightarrow 3\ell_j$ decays can be induced at one-loop level by the vector leptoquark, $\mu - e$ conversion in nuclei can occur at tree-level. Here also, the one-loop dipole and anapole contributions from the exchange of a vector leptoquark are a priori divergent and to obtain a finite result the would-be Goldstone boson degree of freedom (degenerate in mass with vector leptoquark) must be included. After symmetry breaking, the latter degree of freedom is subsequently absorbed by the massive vector leptoquark.

\subsubsection{Radiative lepton decays $\ell_i \rightarrow \ell_j \gamma$}
Vector leptoquark exchange can induce cLFV $\ell_i \rightarrow \ell_j \gamma$ decays at one-loop level through dipole operators. We parametrise the effective Lagrangian for
radiative leptonic decays $\ell_i \rightarrow \ell_j \gamma$ as
\begin{equation}\label{eqn:reff}
\mathcal{L}^{\ell_i \to \ell_j \gamma}_\text{eff} \,=\,
-\frac{4G_F}{\sqrt{2}}\,\bar\ell_j\,
\sigma^{\mu\nu}\,F_{\mu\nu}\,
\left(C_L^{\ell_i\ell_j} \,P_L\,  +\,
C_R^{\ell_i\ell_j}\,P_R\right)\,\ell_i \,+\, \text{H.c.}\,,
\end{equation}
where $F_{\mu\nu}$ is the standard electromagnetic field strength tensor. The $\ell_i \rightarrow \ell_j \gamma$ decay width is then given by
\begin{equation}
\Gamma(\ell_i \rightarrow \ell_j\gamma) \,= \,\frac{2 G_F^2\,(m_{\ell_i}^{2}
  - m_{\ell_j}^{2})^{3}}{\pi \,m_{\ell_i}^{3}} \,
\left(|C_L^{\ell_i\ell_j}|^{2} + |C_R^{\ell_i\ell_j}|^2\right)\,.
\end{equation}
The relevant Wilson coefficients $C_{L,R}$ can be obtained in terms of the vector leptoquark couplings\footnote{As discussed in Section~\ref{sec:lqfit}, we recall that in the current study we work under the assumption that $K_R^{ij}\simeq0$.}, cf. Eq.~(\ref{eq:modelind:L:massbasis}), and are given 
by~\cite{Lavoura:2003xp}
\begin{align}
C_L^{\ell_i\ell_j} \,= \,-\frac{i\,
  N_c}{16\pi^{2}\,M^2} \frac{e}{4\sqrt{2} G_F}\sum_{k}\Bigg\{&\frac{2}{3}
\Big[\left(K_R^{kj{\ast}}\,K_R^{ki}\,m_{\ell_i}
  +K_L^{kj{\ast}}\,K_L^{ki}\,m_{\ell_j} \right) g(t_k) +
  K_R^{kj{\ast}}\,K_L^{ki}\,m_{d_k}\, y(t_k)\Big] \nonumber \\
-&\frac{1}{3}\Big[\left(K_R^{kj{\ast}}\,K_R^{ki}\,m_{\ell_i}
  +K_L^{kj{\ast}}\,K_L^{ki}\,m_{\ell_j} \right)\, f(t_k) +
  K_R^{kj{\ast}}\,K_L^{ki}\,m_{d_k}\, h(t_k)\Big] \Bigg\}\,,
  \label{eqn:radeff1}
\\
C_R^{\ell_i\ell_j} \,= \,-\frac{i \,N_c}{16\pi^{2}\,M^2} \frac{e}{4\sqrt{2} G_F}
\sum_{k}\Bigg\{&\frac{2}{3}
\Big[\left(K_L^{kj{\ast}}\,K_L^{ki}\,m_{\ell_i} +
  K_R^{kj{\ast}}\,K_R^{ki}\,m_{\ell_j} \right) \,g(t_k) +
  K_L^{kj{\ast}}\,K_R^{ki}\,m_{d_k}\, y(t_k)\Big] \nonumber\\
-&\frac{1}{3}\Big[\left(K_L^{kj{\ast}}\,K_L^{ki}\,m_{\ell_i}
 +K_R^{kj{\ast}}\,K_R^{ki}\,m_{\ell_j} \right) \,f(t_k) +
 K_L^{kj{\ast}}\,K_R^{ki}\,m_{d_k} \,h(t_k)\Big] \Bigg\}\,.
 \label{eqn:radeff2}
\end{align}
Here, $t_k = m_{d_k}^{2}/m_{V_1}^2$ and $N_c$ is the number of colours for the internal fermion in the loop. The relevant loop functions are 
\begin{align}
f(t) &= \frac{-5\,t^{3} + 9\,t^{2} - 30\,t + 8}{12\,(t-1)^{3}} +
\frac{3\,t^{2}\,{\ln}(t)}{2\,(t-1)^{4}}\,, \nonumber\\
g(t) &= \frac{-4\,t^{3} + 45\,t^{2} - 33\,t + 10}{12\,(t-1)^{3}} -
\frac{3\,t^{3}\,{\ln}(t)}{2\,(t-1)^{4}}\,, \nonumber\\
h(t) &= \frac{t^{2} + t + 4}{2\,(t-1)^{2}} -
\frac{3t\,{\ln}(t)}{(t-1)^{3}}\,, \nonumber \\
y(t) &= \frac{t^{2} - 11\,t + 4}{2\,(t-1)^2} +
\frac{3\,t^{2}\,{\ln}(t)}{(t-1)^{3}}\,.
\end{align}
\subsubsection{Three body decays $\ell \to \ell' \ell' \ell'$}
Vector leptoquarks can induce three-body cLFV decays $\ell \to \ell' \ell'
\ell'$ at the loop level, through photon penguins (dipole and off-shell ``anapole''), $Z$ penguins and box diagrams. The effective Lagrangian relevant for these  decays can be expressed as~\cite{Okada:1999zk,Kuno:1999jp}
\begin{eqnarray}\label{eq:lto3l}
\mathcal{L}^{\ell \to \ell' \ell' \ell'}_\text{eff} &=&
\mathcal{L}^{\ell_i \to \ell_j \gamma}_\text{eff} -\frac{4\,G_F}{\sqrt{2}} \left[
g_1 \,(\bar {\ell'}\,P_L\, \ell) (\bar {\ell'}
\,P_L\, \ell')\,+\,g_2 \,(\bar {\ell'} \,P_R\, \ell) (\bar {\ell'}\,
P_R \,\ell') \right. \,+\nonumber\\
&& \left.\,+\,g_3 \,(\bar {\ell'} \,\gamma^\mu \,P_R \,\ell) (\bar {\ell'}
\, \gamma_\mu \,P_R \,\ell')\,+\,
g_4\, (\bar {\ell'} \,\gamma^\mu \,P_L \,\ell) (\bar {\ell'}
\,\gamma_\mu \,P_L\, \ell') \,+\,\right. \nonumber\\
&& \left. \,+\, g_5\, (\bar {\ell'} \,\gamma^\mu \,P_R\, \ell) (\bar {\ell'}
\, \gamma_\mu \,P_L \,\ell')
\,+\, g_6 \,(\bar {\ell'} \,\gamma^\mu \,P_L \,\ell) (\bar {\ell'}
\,  \gamma_\mu \,P_R\, \ell') \right]\, +\,
\text{H.c.}\,,
\end{eqnarray}
where the photonic dipole part, cf. Eq.~(\ref{eqn:reff}), with the corresponding Wilson coefficients $C_{L(R)}^{\ell_i\ell_j}$ have already been discussed in the previous subsection; the off-shell anapole photon penguins, $Z$
penguins and box diagrams contribute to $g_3$, $g_4$, $g_5$ and $g_6$
coefficients. For our numerical analysis we only include the
log-enhanced photonic anapole contributions\footnote{This is in contrast to the $Z$-penguins and box diagrams, which (na\"ively)
scale as $ \propto |K^{i\ell}_L|^2 m_q^2/M_{V_1}^4$ and  $\propto |K^{i\ell}_L|^4 m_q^2/M_{V_1}^4$,
respectively; the off-shell anapole photon-penguin diagrams scale
as $ \propto |K^{i\ell}_L|^2 \ln(m_q^2/M^2)/M^2$~\cite{Gabrielli:2000te.}.} in addition to the dipole
ones. In the absence of right-handed couplings of
the vector leptoquark, the only non-vanishing coefficients are $g_4 = g_6$ given by
\begin{equation}
  g_4 \,= \,g_6\, = \,
  -\frac{\sqrt{2}}{4 \,G_F}\,\frac{\alpha}{4\,\pi}
  \,Q_f \,F^{\gamma \ell\ell^\prime}_L\,,
\end{equation}
where $Q_f=Q_{\ell'}$ denotes
the charge (in units of $e$) of the fermion pair attached to the end of the off-shell photon and
\begin{equation}
  F^{\gamma \ell\ell^\prime}_L \,=\,
  \frac{N_c}{m^2_V}\sum_i K_L^{i\ell}\,K_L^{i\ell^\prime \ast} \,
  f_a(x_i)\,,
\end{equation}
with the loop function $f_a(x)$ given by
\begin{equation}
  f_a(x) \,= \,\frac{4 - 26 \,x + 15 \,x^2 + x^3}{12\,(1-x)^3}
  + \frac{4 - 16\,x - 15\,x^2 + 20\,x^3 - 2\,x^4}{18\,(1-x)^4}\ln(x)\,.
\end{equation}
In the above, $N_c$ denotes the number of colours of the internal fermion and  $x_i = m_{d_i}^2/m_{V_1}^2$.
As an example, in the case of $\mu \to 3 e$ decays, the branching ratio can be written as~\cite{Okada:1999zk,Kuno:1999jp}
\begin{eqnarray}
\text{BR}(\mu \to e e e)&=&
2\,\left(|g_3|^2\,+\,|g_4|^2\right)
\,+\,|g_5|^2+|g_6|^2\,+\nonumber\\
&&+ 8\,e\, \text{Re}\left[C^{\mu e}_R\,
\left(2g_4^*\,+\,g_6^*\right)\,+\,C^{\mu e}_L \,
\left(2g_3^*\,+\,g_5^*\right)\right]\,+\nonumber\\
&&+ \frac{32\,e^2}{m_{\mu}^2}\,
\left\{\ln\frac{m_\mu^2}{m_e^2}\,-\,
\frac{11}{4}\right\}(\left|C_{R}^{\mu e}\right|^2\,+\,
\left|C_{L}^{\mu e}\right|^2)\,;
 \end{eqnarray}
similar expressions for the other cLFV 3-body decay modes can be obtained in a straightforward manner.

\subsubsection{Neutrinoless $\mu-e$ conversion}
Neutrinoless $\mu-e$ conversion can be induced by the vector leptoquark $V_1$ at tree level, in addition to the one-loop contributions through dipole and anapole photon penguins. Therefore, $\mu-e$ conversion provides very stringent limits on the vector leptoquark couplings to the first two generations of SM charged leptons. The general contribution to the neutrinoless $\mu-e$ conversion
due to vector leptoquark can be written as~\cite{Dorsner:2016wpm}
\begin{align}
  \Gamma (\mu -e, \text{N}) \,= \,2\, G_F^{2}
  \,\Big(\,\Big|&\frac{C_R^{\mu e
            \ast}}{m_\mu}\,D + \left(2 \,g_{LV}^{(u)} +
        g_{LV}^{(d)}\right)V^{(p)} + \left(g_{LV}^{(u)} +
        2\,g_{LV}^{(d)} \right)V^{(n)} \nonumber\\
	&+ (G_S^{(u,p)}\,g_{LS}^{(u)} +G_S^{(d,p)}\,g_{LS}^{(d)} +
        G_S^{(s,p)}\,g_{LS}^{(s)})\,S^{(p)} \nonumber \\
	&+ (G_S^{(u,n)}\,g_{LS}^{(u)} +G_S^{(d,n)}\,g_{LS}^{(d)} +
        G_S^{(s,n)}\,g_{LS}^{(s)})\,S^{(n)} \Big|^{2} + (L\leftrightarrow
        R)\Big)\,,
\end{align}
where the photonic dipole Wilson coefficients
$C_{L(R)}^{\ell_i\ell_j}$ can be found in Eq.~\eqref{eqn:radeff1} and~\eqref{eqn:radeff2}; the other non-vanishing Wilson coefficients, relevant for vector leptoquark exchange, are given by
\begin{align}
g_{LV}^{(d)} &= \frac{\sqrt{2}}{G_F}\left(\frac{1}{\,m_V^2}  \,K_L^{d e} \,
K_L^{d\mu\ast} + \frac{\alpha}{4 \,\pi}  \,Q_d  \,
F^{\gamma \mu e}_L \right)\,, \nonumber\\
g_{LV}^{(u)} &= \frac{\sqrt{2}}{G_F}\left(\frac{\alpha}{4 \,\pi} \, Q_u \, F^{\gamma \mu e}_L
\right)\,,\nonumber\\
g_{RV}^{(d)} &= \frac{\sqrt{2}}{G_F}\left(\frac{\alpha}{4 \,\pi} \, Q_d  \,F^{\gamma \mu e}_L
\right)\,,
\nonumber\\
g_{RV}^{(u)} &= \frac{\sqrt{2}}{G_F}\left(\frac{\alpha}{4 \,\pi}  \,Q_u  \,
F^{\gamma \mu e}_L \right)\,.
\end{align}
Here, $Q_d = -\frac{1}{3}$, $Q_u = \frac{2}{3}$, and the values for the overlap integrals ($D, V, S$) can be found for instance in~\cite{Kitano:2002mt}. The relevant scalar coefficients $G_S^{(d_i,N)}$ are given  in~\cite{Kosmas:2001mv}.

\end{appendix}



\begin{thebibliography}{99}
{\small

\bibitem{Tanabashi:2018oca}
M.~Tanabashi \textit{et al.} [Particle Data Group],
Phys. Rev. D \textbf{98} (2018) no.3, 030001.

\bibitem{ALEPH:2005ab}
S.~Schael \textit{et al.} [ALEPH, DELPHI, L3, OPAL, SLD, LEP
Electroweak Working Group, SLD Electroweak Group and SLD Heavy Flavour
Group],   
Phys. Rept. \textbf{427} (2006), 257-454
[arXiv:hep-ex/0509008 [hep-ex]].

\bibitem{Lees:2012xj}
J.~P.~Lees \textit{et al.} [BaBar],
Phys. Rev. Lett. \textbf{109} (2012), 101802
[arXiv:1205.5442 [hep-ex]].

\bibitem{Lees:2013uzd}
J.~P.~Lees \textit{et al.} [BaBar],
Phys. Rev. D \textbf{88} (2013) no.7, 072012
[arXiv:1303.0571 [hep-ex]].

\bibitem{Amhis:2019ckw}
Y.~S.~Amhis \textit{et al.} [HFLAV],
``Averages of $b$-hadron, $c$-hadron, and $\tau$-lepton properties as of 2018,''
arXiv:1909.12524 [hep-ex].

\bibitem{Huschle:2015rga}
M.~Huschle \textit{et al.} [Belle],
Phys. Rev. D \textbf{92} (2015) no.7, 072014
[arXiv:1507.03233 [hep-ex]].

\bibitem{Adachi:2009qg}
I.~Adachi \textit{et al.} [Belle],
``Measurement of B ---\ensuremath{>} D(*) tau nu using full reconstruction tags,'' 
arXiv:0910.4301 [hep-ex].

\bibitem{Bozek:2010xy}
A.~Bozek \textit{et al.} [Belle],
Phys. Rev. D \textbf{82} (2010), 072005
[arXiv:1005.2302 [hep-ex]].

\bibitem{Aaij:2015yra}
R.~Aaij \textit{et al.} [LHCb],
Phys. Rev. Lett. \textbf{115} (2015) no.11, 111803
[erratum: Phys. Rev. Lett. \textbf{115} (2015) no.15, 159901]
[arXiv:1506.08614 [hep-ex]].

\bibitem{Hirose:2016wfn}
S.~Hirose \textit{et al.} [Belle],
Phys. Rev. Lett. \textbf{118} (2017) no.21, 211801
[arXiv:1612.00529 [hep-ex]].

\bibitem{Abdesselam:2019dgh}
A.~Abdesselam \textit{et al.} [Belle],
arXiv:1904.08794 [hep-ex].

\bibitem{Aaij:2019wad}
R.~Aaij \textit{et al.} [LHCb],
Phys. Rev. Lett. \textbf{122} (2019) no.19, 191801
[arXiv:1903.09252 [hep-ex]].

\bibitem{Aaij:2017vbb}
R.~Aaij \textit{et al.} [LHCb],
JHEP \textbf{08} (2017), 055
[arXiv:1705.05802 [hep-ex]].

\bibitem{Abdesselam:2019wac}
A.~Abdesselam \textit{et al.} [Belle],
``Test of lepton flavor universality in ${B\to K^\ast\ell^+\ell^-}$ decays at Belle,'' 
arXiv:1904.02440 [hep-ex].

\bibitem{Aaij:2015esa}
R.~Aaij \textit{et al.} [LHCb],
JHEP \textbf{09} (2015), 179
[arXiv:1506.08777 [hep-ex]].

\bibitem{Wehle:2016yoi}
S.~Wehle \textit{et al.} [Belle],
Phys. Rev. Lett. \textbf{118} (2017) no.11, 111801
[arXiv:1612.05014 [hep-ex]].

\bibitem{Ligeti:2016npd}
Z.~Ligeti, M.~Papucci and D.~J.~Robinson,
JHEP \textbf{01} (2017), 083
[arXiv:1610.02045 [hep-ph]].

\bibitem{Crivellin:2016ejn}
A.~Crivellin, J.~Fuentes-Martin, A.~Greljo and G.~Isidori,
Phys. Lett. B \textbf{766} (2017), 77-85
[arXiv:1611.02703 [hep-ph]].

\bibitem{Bigi:2016mdz}
D.~Bigi and P.~Gambino,
Phys. Rev. D \textbf{94} (2016) no.9, 094008
[arXiv:1606.08030 [hep-ph]].

\bibitem{Bigi:2017jbd}
D.~Bigi, P.~Gambino and S.~Schacht,
JHEP \textbf{11} (2017), 061
[arXiv:1707.09509 [hep-ph]].

\bibitem{Bordone:2016gaq}
M.~Bordone, G.~Isidori and A.~Pattori,
Eur. Phys. J. C \textbf{76} (2016) no.8, 440
[arXiv:1605.07633 [hep-ph]].

\bibitem{Capdevila:2017bsm}
B.~Capdevila, A.~Crivellin, S.~Descotes-Genon, J.~Matias and J.~Virto,
JHEP \textbf{01} (2018), 093
[arXiv:1704.05340 [hep-ph]].

\bibitem{Iguro:2020cpg}
S.~Iguro and R.~Watanabe,
JHEP \textbf{08} (2020) no.08, 006
[arXiv:2004.10208 [hep-ph]].


\bibitem{Aaij:2014ora}
R.~Aaij \textit{et al.} [LHCb],
Phys. Rev. Lett. \textbf{113} (2014), 151601
[arXiv:1406.6482 [hep-ex]].

\bibitem{LHCb:2021trn}
R.~Aaij \textit{et al.} [LHCb],
``Test of lepton universality in beauty-quark decays,''
arXiv:2103.11769 [hep-ex].

\bibitem{Aaij:2020nrf}
R.~Aaij \textit{et al.} [LHCb],
Phys. Rev. Lett. \textbf{125} (2020) no.1, 011802
[arXiv:2003.04831 [hep-ex]].

\bibitem{Aaij:2020ruw}
R.~Aaij \textit{et al.} [LHCb],
``Angular analysis of the $B^{+}\rightarrow K^{\ast+}\mu^{+}\mu^{-}$ decay,''
arXiv:2012.13241 [hep-ex].

\bibitem{LHCb:2021zwz}
R.~Aaij \textit{et al.} [LHCb],
``Branching fraction measurements of the rare $B^0_s\rightarrow\phi\mu^+\mu^-$ and $B^0_s\rightarrow f_2^\prime(1525)\mu^+\mu^-$ decays,''
arXiv:2105.14007 [hep-ex].

\bibitem{LHCb:2021qbv}
R.~Aaij \textit{et al.} [LHCb],
Phys. Rev. D \textbf{104} (2021) no.3, 032005
[arXiv:2103.06810 [hep-ex]].

\bibitem{LHCb:2021awg}
R.~Aaij \textit{et al.} [LHCb],
``Measurement of the $B^0_s\to\mu^+\mu^-$ decay properties and search for the $B^0\to\mu^+\mu^-$ and $B^0_s\to\mu^+\mu^-\gamma$ decays,''
arXiv:2108.09283 [hep-ex].

\bibitem{Alguero:2019ptt}
M.~Alguer\'o, B.~Capdevila, A.~Crivellin, S.~Descotes-Genon,
P.~Masjuan, J.~Matias, M.~Novoa Brunet and J.~Virto, 
Eur. Phys. J. C \textbf{79} (2019) no.8, 714
[arXiv:1903.09578 [hep-ph]].

\bibitem{Aebischer:2019mlg}
J.~Aebischer, W.~Altmannshofer, D.~Guadagnoli, M.~Reboud, P.~Stangl
and D.~M.~Straub, 
Eur. Phys. J. C \textbf{80} (2020) no.3, 252
[arXiv:1903.10434 [hep-ph]].

\bibitem{Ciuchini:2019usw}
M.~Ciuchini, A.~M.~Coutinho, M.~Fedele, E.~Franco, A.~Paul,
L.~Silvestrini and M.~Valli, 
Eur. Phys. J. C \textbf{79} (2019) no.8, 719
[arXiv:1903.09632 [hep-ph]].

\bibitem{Datta:2019zca}
A.~Datta, J.~Kumar and D.~London,
Phys. Lett. B \textbf{797} (2019), 134858
[arXiv:1903.10086 [hep-ph]].

\bibitem{Arbey:2019duh}
A.~Arbey, T.~Hurth, F.~Mahmoudi, D.~M.~Santos and S.~Neshatpour,
Phys. Rev. D \textbf{100} (2019) no.1, 015045
[arXiv:1904.08399 [hep-ph]].


\bibitem{Shi:2019gxi}
R.~X.~Shi, L.~S.~Geng, B.~Grinstein, S.~J\"ager and J.~Martin Camalich,
JHEP \textbf{12} (2019), 065
[arXiv:1905.08498 [hep-ph]].

\bibitem{Bardhan:2019ljo}
D.~Bardhan and D.~Ghosh,
Phys. Rev. D \textbf{100} (2019) no.1, 011701
[arXiv:1904.10432 [hep-ph]].

\bibitem{Alok:2019ufo}
A.~K.~Alok, A.~Dighe, S.~Gangal and D.~Kumar,
JHEP \textbf{06} (2019), 089
[arXiv:1903.09617 [hep-ph]].

\bibitem{Bhattacharya:2019eji}
S.~Bhattacharya, A.~Biswas, Z.~Calcuttawala and S.~K.~Patra,
arXiv:1902.02796 [hep-ph].

\bibitem{Alok:2017qsi}
A.~K.~Alok, D.~Kumar, J.~Kumar, S.~Kumbhakar and S.~U.~Sankar,
JHEP \textbf{09} (2018), 152
[arXiv:1710.04127 [hep-ph]].

\bibitem{Ghosh:2014awa}
D.~Ghosh, M.~Nardecchia and S.~A.~Renner,
JHEP \textbf{12} (2014), 131
[arXiv:1408.4097 [hep-ph]].

\bibitem{Glashow:2014iga}
S.~L.~Glashow, D.~Guadagnoli and K.~Lane,
Phys. Rev. Lett. \textbf{114} (2015), 091801
[arXiv:1411.0565 [hep-ph]].

\bibitem{Bhattacharya:2014wla}
B.~Bhattacharya, A.~Datta, D.~London and S.~Shivashankara,
Phys. Lett. B \textbf{742} (2015), 370-374
[arXiv:1412.7164 [hep-ph]].

\bibitem{Freytsis:2015qca}
M.~Freytsis, Z.~Ligeti and J.~T.~Ruderman,
Phys. Rev. D \textbf{92} (2015) no.5, 054018
[arXiv:1506.08896 [hep-ph]].

\bibitem{Ciuchini:2017mik}
M.~Ciuchini, A.~M.~Coutinho, M.~Fedele, E.~Franco, A.~Paul,
L.~Silvestrini and M.~Valli, 
Eur. Phys. J. C \textbf{77} (2017) no.10, 688
[arXiv:1704.05447 [hep-ph]].

\bibitem{Jaiswal:2017rve}
S.~Jaiswal, S.~Nandi and S.~K.~Patra,
JHEP \textbf{12} (2017), 060
[arXiv:1707.09977 [hep-ph]].

\bibitem{Jaiswal:2020wer}
S.~Jaiswal, S.~Nandi and S.~K.~Patra,
JHEP \textbf{06} (2020), 165
[arXiv:2002.05726 [hep-ph]].

\bibitem{Bhattacharya:2019dot}
S.~Bhattacharya, A.~Biswas, S.~Nandi and S.~K.~Patra,
Phys. Rev. D \textbf{101} (2020) no.5, 055025
doi:10.1103/PhysRevD.101.055025
[arXiv:1908.04835 [hep-ph]].


\bibitem{Biswas:2020uaq}
A.~Biswas, S.~Nandi, I.~Ray and S.~K.~Patra,
``New physics in $b\to s \ell\ell$ decays with complex Wilson coefficients,''
arXiv:2004.14687 [hep-ph].


\bibitem{Bhattacharya:2018kig}
S.~Bhattacharya, S.~Nandi and S.~Kumar Patra,
Eur. Phys. J. C \textbf{79} (2019) no.3, 268
[arXiv:1805.08222 [hep-ph]].




\bibitem{Altmannshofer:2014cfa}
W.~Altmannshofer, S.~Gori, M.~Pospelov and I.~Yavin,
Phys. Rev. D \textbf{89} (2014), 095033
[arXiv:1403.1269 [hep-ph]].

\bibitem{Crivellin:2015mga}
A.~Crivellin, G.~D'Ambrosio and J.~Heeck,
Phys. Rev. Lett. \textbf{114} (2015), 151801 
[arXiv:1501.00993 [hep-ph]].

\bibitem{Crivellin:2015lwa}
A.~Crivellin, G.~D'Ambrosio and J.~Heeck,
Phys. Rev. D \textbf{91} (2015) no.7, 075006
[arXiv:1503.03477 [hep-ph]].

\bibitem{Sierra:2015fma}
D.~Aristizabal Sierra, F.~Staub and A.~Vicente,
Phys. Rev. D \textbf{92} (2015) no.1, 015001
[arXiv:1503.06077 [hep-ph]].

\bibitem{Crivellin:2015era}
A.~Crivellin, L.~Hofer, J.~Matias, U.~Nierste, S.~Pokorski and J.~Rosiek,
Phys. Rev. D \textbf{92} (2015) no.5, 054013
[arXiv:1504.07928 [hep-ph]].

\bibitem{Celis:2015ara}
A.~Celis, J.~Fuentes-Martin, M.~Jung and H.~Serodio,
Phys. Rev. D \textbf{92} (2015) no.1, 015007
[arXiv:1505.03079 [hep-ph]].

\bibitem{Bhatia:2017tgo}
D.~Bhatia, S.~Chakraborty and A.~Dighe,
JHEP \textbf{03} (2017), 117
[arXiv:1701.05825 [hep-ph]].

\bibitem{Kamenik:2017tnu}
J.~F.~Kamenik, Y.~Soreq and J.~Zupan,
Phys. Rev. D \textbf{97} (2018) no.3, 035002
[arXiv:1704.06005 [hep-ph]].

\bibitem{Chen:2017usq}
C.~H.~Chen and T.~Nomura,
Phys. Lett. B \textbf{777} (2018), 420-427
[arXiv:1707.03249 [hep-ph]].

\bibitem{Camargo-Molina:2018cwu}
J.~E.~Camargo-Molina, A.~Celis and D.~A.~Faroughy,
Phys. Lett. B \textbf{784} (2018), 284-293
[arXiv:1805.04917 [hep-ph]].

\bibitem{Darme:2018hqg}
L.~Darm\'e, K.~Kowalska, L.~Roszkowski and E.~M.~Sessolo,
JHEP \textbf{10} (2018), 052
[arXiv:1806.06036 [hep-ph]].

\bibitem{Baek:2018aru}
S.~Baek and C.~Yu,
JHEP \textbf{11} (2018), 054
[arXiv:1806.05967 [hep-ph]].

\bibitem{Biswas:2019twf}
A.~Biswas and A.~Shaw,
JHEP \textbf{05} (2019), 165
[arXiv:1903.08745 [hep-ph]].

\bibitem{Allanach:2019iiy}
B.~C.~Allanach and J.~Davighi,
Eur. Phys. J. C \textbf{79} (2019) no.11, 908
[arXiv:1905.10327 [hep-ph]].

\bibitem{Crivellin:2020oup}
A.~Crivellin, C.~A.~Manzari, M.~Alguero and J.~Matias,
``Combined Explanation of the $Z\to b\bar b$ Forward-Backward Asymmetry, the Cabibbo Angle Anomaly, $\tau\to\mu\nu\nu$ and $b\to s\ell^+\ell^-$ Data,'' 
arXiv:2010.14504 [hep-ph].

\bibitem{Hiller:2014yaa}
G.~Hiller and M.~Schmaltz,
Phys. Rev. D \textbf{90} (2014), 054014
[arXiv:1408.1627 [hep-ph]].

\bibitem{Gripaios:2014tna}
B.~Gripaios, M.~Nardecchia and S.~A.~Renner,
JHEP \textbf{05} (2015), 006
[arXiv:1412.1791 [hep-ph]].

\bibitem{Sahoo:2015wya}
S.~Sahoo and R.~Mohanta,
Phys. Rev. D \textbf{91} (2015) no.9, 094019
[arXiv:1501.05193 [hep-ph]].

\bibitem{Varzielas:2015iva}
I.~de Medeiros Varzielas and G.~Hiller,
JHEP \textbf{06} (2015), 072
[arXiv:1503.01084 [hep-ph]].

\bibitem{Alonso:2015sja}
R.~Alonso, B.~Grinstein and J.~Martin Camalich,
JHEP \textbf{10} (2015), 184
[arXiv:1505.05164 [hep-ph]].

\bibitem{Bauer:2015knc}
M.~Bauer and M.~Neubert,
Phys. Rev. Lett. \textbf{116} (2016) no.14, 141802
[arXiv:1511.01900 [hep-ph]].

\bibitem{Hati:2015awg}
C.~Hati, G.~Kumar and N.~Mahajan,
JHEP \textbf{01} (2016), 117
[arXiv:1511.03290 [hep-ph]].

\bibitem{Fajfer:2015ycq}
S.~Fajfer and N.~Ko\v{s}nik,
Phys. Lett. B \textbf{755} (2016), 270-274
[arXiv:1511.06024 [hep-ph]].

\bibitem{Das:2016vkr}
D.~Das, C.~Hati, G.~Kumar and N.~Mahajan,
Phys. Rev. D \textbf{94} (2016), 055034
[arXiv:1605.06313 [hep-ph]].

\bibitem{Becirevic:2016yqi}
D.~Be\v{c}irevi\'c, S.~Fajfer, N.~Ko\v{s}nik and O.~Sumensari,
Phys. Rev. D \textbf{94} (2016) no.11, 115021
[arXiv:1608.08501 [hep-ph]].

\bibitem{Sahoo:2016pet}
S.~Sahoo, R.~Mohanta and A.~K.~Giri,
Phys. Rev. D \textbf{95} (2017) no.3, 035027
[arXiv:1609.04367 [hep-ph]].

\bibitem{Cox:2016epl}
P.~Cox, A.~Kusenko, O.~Sumensari and T.~T.~Yanagida,
JHEP \textbf{03} (2017), 035
[arXiv:1612.03923 [hep-ph]].

\bibitem{Crivellin:2017zlb}
A.~Crivellin, D.~M\"uller and T.~Ota,
JHEP \textbf{09} (2017), 040
[arXiv:1703.09226 [hep-ph]].

\bibitem{Becirevic:2017jtw}
D.~Be\v{c}irevi\'c and O.~Sumensari,
JHEP \textbf{08} (2017), 104
[arXiv:1704.05835 [hep-ph]].

\bibitem{Cai:2017wry}
Y.~Cai, J.~Gargalionis, M.~A.~Schmidt and R.~R.~Volkas,
JHEP \textbf{10} (2017), 047
[arXiv:1704.05849 [hep-ph]].

\bibitem{Dorsner:2017ufx}
I.~Dor\v{s}ner, S.~Fajfer, D.~A.~Faroughy and N.~Ko\v{s}nik,
JHEP \textbf{10} (2017), 188
[arXiv:1706.07779 [hep-ph]].

\bibitem{Buttazzo:2017ixm}
D.~Buttazzo, A.~Greljo, G.~Isidori and D.~Marzocca,
JHEP \textbf{11} (2017), 044
[arXiv:1706.07808 [hep-ph]].

\bibitem{Greljo:2018tuh}
A.~Greljo and B.~A.~Stefanek,
Phys. Lett. B \textbf{782} (2018), 131-138
[arXiv:1802.04274 [hep-ph]].

\bibitem{Sahoo:2018ffv}
S.~Sahoo and R.~Mohanta,
J. Phys. G \textbf{45} (2018) no.8, 085003
[arXiv:1806.01048 [hep-ph]].

\bibitem{Becirevic:2018afm}
D.~Be\v{c}irevi\'c, I.~Dor\v{s}ner, S.~Fajfer, N.~Ko\v{s}nik,
D.~A.~Faroughy and O.~Sumensari, 
Phys. Rev. D \textbf{98} (2018) no.5, 055003
[arXiv:1806.05689 [hep-ph]].

\bibitem{Hati:2018fzc}
C.~Hati, G.~Kumar, J.~Orloff and A.~M.~Teixeira,
JHEP \textbf{11} (2018), 011
[arXiv:1806.10146 [hep-ph]].

\bibitem{Fornal:2018dqn}
B.~Fornal, S.~A.~Gadam and B.~Grinstein,
Phys. Rev. D \textbf{99} (2019) no.5, 055025
[arXiv:1812.01603 [hep-ph]].

\bibitem{deMedeirosVarzielas:2018bcy}
I.~de Medeiros Varzielas and S.~F.~King,
JHEP \textbf{11} (2018), 100
[arXiv:1807.06023 [hep-ph]].

\bibitem{Aebischer:2018acj}
J.~Aebischer, A.~Crivellin and C.~Greub,
Phys. Rev. D \textbf{99} (2019) no.5, 055002
[arXiv:1811.08907 [hep-ph]].

\bibitem{Aydemir:2019ynb}
U.~Aydemir, T.~Mandal and S.~Mitra,
Phys. Rev. D \textbf{101}, no.1, 015011 (2020)
[arXiv:1902.08108 [hep-ph]].

\bibitem{Mandal:2018kau}
T.~Mandal, S.~Mitra and S.~Raz,
Phys. Rev. D \textbf{99}, no.5, 055028 (2019)
[arXiv:1811.03561 [hep-ph]].

\bibitem{deMedeirosVarzielas:2019okf}
I.~De Medeiros Varzielas and S.~F.~King,
Phys. Rev. D \textbf{99} (2019) no.9, 095029
[arXiv:1902.09266 [hep-ph]].

\bibitem{Yan:2019hpm}
H.~Yan, Y.~D.~Yang and X.~B.~Yuan,
Chin. Phys. C \textbf{43} (2019) no.8, 083105
[arXiv:1905.01795 [hep-ph]].

\bibitem{Bigaran:2019bqv}
I.~Bigaran, J.~Gargalionis and R.~R.~Volkas,
JHEP \textbf{10} (2019), 106
[arXiv:1906.01870 [hep-ph]].

\bibitem{Popov:2019tyc}
O.~Popov, M.~A.~Schmidt and G.~White,
Phys. Rev. D \textbf{100} (2019) no.3, 035028
[arXiv:1905.06339 [hep-ph]].

\bibitem{Hati:2019ufv}
C.~Hati, J.~Kriewald, J.~Orloff and A.~M.~Teixeira,
JHEP \textbf{12} (2019), 006
[arXiv:1907.05511 [hep-ph]].

\bibitem{Crivellin:2019dwb}
A.~Crivellin, D.~M\"uller and F.~Saturnino,
JHEP \textbf{06} (2020), 020
[arXiv:1912.04224 [hep-ph]].

\bibitem{Saad:2020ihm}
S.~Saad,
Phys. Rev. D \textbf{102} (2020) no.1, 015019
[arXiv:2005.04352 [hep-ph]].

\bibitem{Dev:2020qet}
P.~S.~Bhupal Dev, R.~Mohanta, S.~Patra and S.~Sahoo,
Phys. Rev. D \textbf{102} (2020) no.9, 095012 
[arXiv:2004.09464 [hep-ph]].

\bibitem{Saad:2020ucl}
S.~Saad and A.~Thapa,
Phys. Rev. D \textbf{102} (2020) no.1, 015014
[arXiv:2004.07880 [hep-ph]].

\bibitem{Balaji:2019kwe}
S.~Balaji and M.~A.~Schmidt,
Phys. Rev. D \textbf{101} (2020) no.1, 015026
[arXiv:1911.08873 [hep-ph]].

\bibitem{Cornella:2019hct}
C.~Cornella, J.~Fuentes-Martin and G.~Isidori,
JHEP \textbf{07} (2019), 168
[arXiv:1903.11517 [hep-ph]].

\bibitem{Mandal:2019gff}
R.~Mandal and A.~Pich,
JHEP \textbf{12} (2019), 089
[arXiv:1908.11155 [hep-ph]].

\bibitem{Babu:2020hun}
K.~S.~Babu, P.~S.~B.~Dev, S.~Jana and A.~Thapa,
``Unified Framework for $B$-Anomalies, Muon $g-2$, and Neutrino Masses,''
arXiv:2009.01771 [hep-ph].

\bibitem{Martynov:2020cjd}
M.~V.~Martynov and A.~D.~Smirnov,
``Chiral gauge leptoquark mass limits and branching ratios of $ K_L^0,
B^0, B_s \to l^+_i l^-_j $ decays with account of the general fermion
mixing in leptoquark currents,'' 
arXiv:2011.08240 [hep-ph].

\bibitem{Fuentes-Martin:2020bnh}
J.~Fuentes-Mart\'\i{}n and P.~Stangl,
Phys. Lett. B \textbf{811} (2020), 135953
[arXiv:2004.11376 [hep-ph]].

\bibitem{Guadagnoli:2020tlx}
D.~Guadagnoli, M.~Reboud and P.~Stangl,
JHEP \textbf{10} (2020), 084
[arXiv:2005.10117 [hep-ph]].

\bibitem{Deshpand:2016cpw}
N.~G.~Deshpande and X.~G.~He,
Eur. Phys. J. C \textbf{77} (2017) no.2, 134
[arXiv:1608.04817 [hep-ph]].

\bibitem{Altmannshofer:2017poe}
W.~Altmannshofer, P.~S.~Bhupal Dev and A.~Soni,
Phys. Rev. D \textbf{96} (2017) no.9, 095010
[arXiv:1704.06659 [hep-ph]].

\bibitem{Das:2017kfo}
D.~Das, C.~Hati, G.~Kumar and N.~Mahajan,
Phys. Rev. D \textbf{96} (2017) no.9, 095033
[arXiv:1705.09188 [hep-ph]].

\bibitem{Earl:2018snx}
K.~Earl and T.~Gr\'egoire,
JHEP \textbf{08} (2018), 201
[arXiv:1806.01343 [hep-ph]].

\bibitem{Trifinopoulos:2018rna}
S.~Trifinopoulos,
Eur. Phys. J. C \textbf{78} (2018) no.10, 803
[arXiv:1807.01638 [hep-ph]].

\bibitem{Trifinopoulos:2019lyo}
S.~Trifinopoulos,
Phys. Rev. D \textbf{100} (2019) no.11, 115022
[arXiv:1904.12940 [hep-ph]].

\bibitem{Cohen:2019cge}
J.~Cohen, S.~Bar-Shalom, G.~Eilam and A.~Soni,
Phys. Rev. D \textbf{100} (2019) no.11, 115051
[arXiv:1906.04743 [hep-ph]].

\bibitem{Earl:2019adq}
K.~Earl,
``Exploring supersymmetry and naturalness in light of new experimental
data,'' [PhD thesis]
doi:10.22215/etd/2019-13685.

\bibitem{Hu:2019ahp}
Q.~Y.~Hu and L.~L.~Huang,
Phys. Rev. D \textbf{101} (2020) no.3, 035030
[arXiv:1912.03676 [hep-ph]].

\bibitem{Hu:2020yvs}
Q.~Y.~Hu, Y.~D.~Yang and M.~D.~Zheng,
Eur. Phys. J. C \textbf{80} (2020) no.5, 365
[arXiv:2002.09875 [hep-ph]].

\bibitem{Altmannshofer:2020axr}
W.~Altmannshofer, P.~S.~B.~Dev, A.~Soni and Y.~Sui,
Phys. Rev. D \textbf{102} (2020) no.1, 015031
[arXiv:2002.12910 [hep-ph]].

\bibitem{Greljo:2015mma}
A.~Greljo, G.~Isidori and D.~Marzocca,
JHEP \textbf{07} (2015), 142
[arXiv:1506.01705 [hep-ph]].

\bibitem{Arnan:2017lxi}
P.~Arnan, D.~Be\v{c}irevi\'c, F.~Mescia and O.~Sumensari,
Eur. Phys. J. C \textbf{77} (2017) no.11, 796
[arXiv:1703.03426 [hep-ph]].

\bibitem{Geng:2017svp}
L.~S.~Geng, B.~Grinstein, S.~J\"ager, J.~Martin Camalich, X.~L.~Ren
and R.~X.~Shi, 
Phys. Rev. D \textbf{96} (2017) no.9, 093006
[arXiv:1704.05446 [hep-ph]].

\bibitem{Choudhury:2017qyt}
D.~Choudhury, A.~Kundu, R.~Mandal and R.~Sinha,
Phys. Rev. Lett. \textbf{119} (2017) no.15, 151801
[arXiv:1706.08437 [hep-ph]].

\bibitem{Choudhury:2017ijp}
D.~Choudhury, A.~Kundu, R.~Mandal and R.~Sinha,
Nucl. Phys. B \textbf{933} (2018), 433-453
[arXiv:1712.01593 [hep-ph]].

\bibitem{Grinstein:2018fgb}
B.~Grinstein, S.~Pokorski and G.~G.~Ross,
JHEP \textbf{12} (2018), 079
[arXiv:1809.01766 [hep-ph]].

\bibitem{Cerdeno:2019vpd}
D.~G.~Cerde\~no, A.~Cheek, P.~Mart\'\i{}n-Ramiro and J.~M.~Moreno,
Eur. Phys. J. C \textbf{79} (2019) no.6, 517
[arXiv:1902.01789 [hep-ph]].

\bibitem{Crivellin:2019dun}
A.~Crivellin, D.~M\"uller and C.~Wiegand,
JHEP \textbf{06} (2019), 119
[arXiv:1903.10440 [hep-ph]].

\bibitem{Arnan:2019uhr}
P.~Arnan, A.~Crivellin, M.~Fedele and F.~Mescia,
JHEP \textbf{06} (2019), 118
[arXiv:1904.05890 [hep-ph]].

\bibitem{Gomez:2019xfw}
J.~D.~G\'omez, N.~Quintero and E.~Rojas,
Phys. Rev. D \textbf{100} (2019) no.9, 093003
[arXiv:1907.08357 [hep-ph]].

\bibitem{Assad:2017iib}
N.~Assad, B.~Fornal and B.~Grinstein,
Phys. Lett. B \textbf{777} (2018), 324-331
[arXiv:1708.06350 [hep-ph]].

\bibitem{Calibbi:2017qbu}
L.~Calibbi, A.~Crivellin and T.~Li,
Phys. Rev. D \textbf{98} (2018) no.11, 115002
[arXiv:1709.00692 [hep-ph]].

\bibitem{Bordone:2017bld}
M.~Bordone, C.~Cornella, J.~Fuentes-Martin and G.~Isidori,
Phys. Lett. B \textbf{779} (2018), 317-323
[arXiv:1712.01368 [hep-ph]].

\bibitem{Blanke:2018sro}
M.~Blanke and A.~Crivellin,
Phys. Rev. Lett. \textbf{121} (2018) no.1, 011801
[arXiv:1801.07256 [hep-ph]].

\bibitem{Bordone:2018nbg}
M.~Bordone, C.~Cornella, J.~Fuentes-Mart\'\i{}n and G.~Isidori,
JHEP \textbf{10} (2018), 148
[arXiv:1805.09328 [hep-ph]].

\bibitem{Kumar:2018kmr}
J.~Kumar, D.~London and R.~Watanabe,
Phys. Rev. D \textbf{99} (2019) no.1, 015007
[arXiv:1806.07403 [hep-ph]].

\bibitem{Angelescu:2018tyl}
A.~Angelescu, D.~Be\v{c}irevi\'c, D.~A.~Faroughy and O.~Sumensari,
JHEP \textbf{10} (2018), 183
[arXiv:1808.08179 [hep-ph]].

\bibitem{Balaji:2018zna}
S.~Balaji, R.~Foot and M.~A.~Schmidt,
Phys. Rev. D \textbf{99} (2019) no.1, 015029
[arXiv:1809.07562 [hep-ph]].

\bibitem{Baker:2019sli}
M.~J.~Baker, J.~Fuentes-Mart\'\i{}n, G.~Isidori and M.~K\"onig,
Eur. Phys. J. C \textbf{79} (2019) no.4, 334
[arXiv:1901.10480 [hep-ph]].

\bibitem{DaRold:2019fiw}
L.~Da Rold and F.~Lamagna,
JHEP \textbf{12} (2019), 112
[arXiv:1906.11666 [hep-ph]].

\bibitem{Fuentes-Martin:2019ign}
J.~Fuentes-Mart\'\i{}n, G.~Isidori, M.~K\"onig and N.~Selimovi\'c,
Phys. Rev. D \textbf{101} (2020) no.3, 035024
[arXiv:1910.13474 [hep-ph]].
\bibitem{Fuentes-Martin:2020luw}
J.~Fuentes-Mart\'\i{}n, G.~Isidori, M.~K\"onig and N.~Selimovi\'c,
Phys. Rev. D \textbf{102} (2020) no.3, 035021
[arXiv:2006.16250 [hep-ph]].
\bibitem{Fuentes-Martin:2020hvc}
J.~Fuentes-Mart\'\i{}n, G.~Isidori, M.~K\"onig and N.~Selimovi\'c,
Phys. Rev. D \textbf{102} (2020), 115015
[arXiv:2009.11296 [hep-ph]].

\bibitem{Capdevila:2017iqn}
B.~Capdevila, A.~Crivellin, S.~Descotes-Genon, L.~Hofer and J.~Matias,
Phys. Rev. Lett. \textbf{120} (2018) no.18, 181802
[arXiv:1712.01919 [hep-ph]].

\bibitem{Iguro:2018vqb}
S.~Iguro, T.~Kitahara, Y.~Omura, R.~Watanabe and K.~Yamamoto,
JHEP \textbf{02} (2019), 194
[arXiv:1811.08899 [hep-ph]].

\bibitem{Sakaki:2012ft}
Y.~Sakaki and H.~Tanaka,
Phys. Rev. D \textbf{87} (2013) no.5, 054002
[arXiv:1205.4908 [hep-ph]].

\bibitem{Tanaka:2012nw}
M.~Tanaka and R.~Watanabe,
Phys. Rev. D \textbf{87} (2013) no.3, 034028
[arXiv:1212.1878 [hep-ph]].

\bibitem{Neubert:1991td}
M.~Neubert,
Phys. Lett. B \textbf{264} (1991), 455-461.

\bibitem{Hagiwara:1989cu}
K.~Hagiwara, A.~D.~Martin and M.~F.~Wade,
Nucl. Phys. B \textbf{327} (1989), 569-594.

\bibitem{Caprini:1997mu}
I.~Caprini, L.~Lellouch and M.~Neubert,
Nucl. Phys. B \textbf{530} (1998), 153-181
[arXiv:hep-ph/9712417 [hep-ph]].

\bibitem{Blanke:2019qrx}
M.~Blanke, A.~Crivellin, T.~Kitahara, M.~Moscati, U.~Nierste and
I.~Ni\v{s}and\v{z}i\'c, 
``Addendum to \textquotedblleft{}Impact of polarization observables
and $B_c\to \tau \nu$ on new physics explanations of the $b\to
c \tau \nu$ anomaly'','' 
arXiv:1905.08253 [hep-ph].

\bibitem{Aaij:2013qta}
R.~Aaij \textit{et al.} [LHCb],
Phys. Rev. Lett. \textbf{111} (2013), 191801
[arXiv:1308.1707 [hep-ex]].

\bibitem{Aaij:2015oid}
R.~Aaij \textit{et al.} [LHCb],
JHEP \textbf{02} (2016), 104
[arXiv:1512.04442 [hep-ex]].

\bibitem{Abdesselam:2016llu}
A.~Abdesselam \textit{et al.} [Belle],
``Angular analysis of $B^0 \to K^\ast(892)^0 \ell^+ \ell^-$,''
arXiv:1604.04042 [hep-ex].

\bibitem{Aaboud:2018krd}
M.~Aaboud \textit{et al.} [ATLAS],
JHEP \textbf{10} (2018), 047
[arXiv:1805.04000 [hep-ex]].

\bibitem{Sirunyan:2017dhj}
A.~M.~Sirunyan \textit{et al.} [CMS],
Phys. Lett. B \textbf{781} (2018), 517-541
[arXiv:1710.02846 [hep-ex]].

\bibitem{Aebischer:2018iyb}
J.~Aebischer, J.~Kumar, P.~Stangl and D.~M.~Straub,
Eur. Phys. J. C \textbf{79} (2019) no.6, 509
[arXiv:1810.07698 [hep-ph]].

\bibitem{Altmannshofer:2008dz}
W.~Altmannshofer, P.~Ball, A.~Bharucha, A.~J.~Buras, D.~M.~Straub and M.~Wick,
JHEP \textbf{01} (2009), 019
[arXiv:0811.1214 [hep-ph]].

\bibitem{Bobeth:2011nj}
C.~Bobeth, G.~Hiller, D.~van Dyk and C.~Wacker,
JHEP \textbf{01} (2012), 107
[arXiv:1111.2558 [hep-ph]].

\bibitem{Matias:2012xw}
J.~Matias, F.~Mescia, M.~Ramon and J.~Virto,
JHEP \textbf{04} (2012), 104
[arXiv:1202.4266 [hep-ph]].

\bibitem{DescotesGenon:2012zf}
S.~Descotes-Genon, J.~Matias, M.~Ramon and J.~Virto,
JHEP \textbf{01} (2013), 048
[arXiv:1207.2753 [hep-ph]].

\bibitem{Matias:2014jua}
J.~Matias and N.~Serra,
Phys. Rev. D \textbf{90} (2014) no.3, 034002
[arXiv:1402.6855 [hep-ph]].

\bibitem{Khodjamirian:2010vf}
A.~Khodjamirian, T.~Mannel, A.~A.~Pivovarov and Y.~M.~Wang,
JHEP \textbf{09} (2010), 089
[arXiv:1006.4945 [hep-ph]].

\bibitem{Khodjamirian:2012rm}
A.~Khodjamirian, T.~Mannel and Y.~M.~Wang,
JHEP \textbf{02} (2013), 010
[arXiv:1211.0234 [hep-ph]].

\bibitem{Lyon:2014hpa}
J.~Lyon and R.~Zwicky,
``Resonances gone topsy turvy - the charm of QCD or new physics in
$b \to s \ell^+ \ell^-$?,'' 
arXiv:1406.0566 [hep-ph].

\bibitem{Descotes-Genon:2014uoa}
S.~Descotes-Genon, L.~Hofer, J.~Matias and J.~Virto,
JHEP \textbf{12} (2014), 125
[arXiv:1407.8526 [hep-ph]].

\bibitem{Capdevila:2017ert}
B.~Capdevila, S.~Descotes-Genon, L.~Hofer and J.~Matias,
JHEP \textbf{04} (2017), 016
[arXiv:1701.08672 [hep-ph]].

\bibitem{Blake:2017fyh}
T.~Blake, U.~Egede, P.~Owen, K.~A.~Petridis and G.~Pomery,
Eur. Phys. J. C \textbf{78} (2018) no.6, 453
[arXiv:1709.03921 [hep-ph]].

\bibitem{Jager:2012uw}
S.~J\"ager and J.~Martin Camalich,
JHEP \textbf{05} (2013), 043
[arXiv:1212.2263 [hep-ph]].

\bibitem{Jager:2014rwa}
S.~J\"ager and J.~Martin Camalich,
Phys. Rev. D \textbf{93} (2016) no.1, 014028
[arXiv:1412.3183 [hep-ph]].

\bibitem{Ciuchini:2015qxb}
M.~Ciuchini, M.~Fedele, E.~Franco, S.~Mishima, A.~Paul, L.~Silvestrini
and M.~Valli, 
JHEP \textbf{06} (2016), 116
[arXiv:1512.07157 [hep-ph]].

\bibitem{Ciuchini:2016weo}
M.~Ciuchini, M.~Fedele, E.~Franco, S.~Mishima, A.~Paul, L.~Silvestrini
and M.~Valli, 
PoS \textbf{ICHEP2016} (2016), 584
[arXiv:1611.04338 [hep-ph]].

\bibitem{Bobeth:2017vxj}
C.~Bobeth, M.~Chrzaszcz, D.~van Dyk and J.~Virto,
Eur. Phys. J. C \textbf{78} (2018) no.6, 451
[arXiv:1707.07305 [hep-ph]].

\bibitem{Gubernari:2020eft}
N.~Gubernari, D.~van Dyk and J.~Virto,
[arXiv:2011.09813 [hep-ph]].

\bibitem{Buchalla:1995vs}
G.~Buchalla, A.~J.~Buras and M.~E.~Lautenbacher,
Rev. Mod. Phys. \textbf{68} (1996), 1125-1144
[arXiv:hep-ph/9512380 [hep-ph]].

\bibitem{Bobeth:1999mk}
C.~Bobeth, M.~Misiak and J.~Urban,
Nucl. Phys. B \textbf{574} (2000), 291-330
[arXiv:hep-ph/9910220 [hep-ph]].

\bibitem{Ali:2002jg}
A.~Ali, E.~Lunghi, C.~Greub and G.~Hiller,
Phys. Rev. D \textbf{66} (2002), 034002
[arXiv:hep-ph/0112300 [hep-ph]].

\bibitem{Hiller:2003js}
G.~Hiller and F.~Kruger,
Phys. Rev. D \textbf{69} (2004), 074020
[arXiv:hep-ph/0310219 [hep-ph]].

\bibitem{Bobeth:2007dw}
C.~Bobeth, G.~Hiller and G.~Piranishvili,
JHEP \textbf{12} (2007), 040
[arXiv:0709.4174 [hep-ph]].

\bibitem{Bobeth:2010wg}
C.~Bobeth, G.~Hiller and D.~van Dyk,
JHEP \textbf{07} (2010), 098
[arXiv:1006.5013 [hep-ph]].

\bibitem{Descotes-Genon:2015uva}
S.~Descotes-Genon, L.~Hofer, J.~Matias and J.~Virto,
JHEP \textbf{06} (2016), 092
[arXiv:1510.04239 [hep-ph]].

\bibitem{Altmannshofer:2017fio}
W.~Altmannshofer, C.~Niehoff, P.~Stangl and D.~M.~Straub,
Eur. Phys. J. C \textbf{77} (2017) no.6, 377
[arXiv:1703.09189 [hep-ph]].

\bibitem{Alok:2017sui}
A.~K.~Alok, B.~Bhattacharya, A.~Datta, D.~Kumar, J.~Kumar and D.~London,
Phys. Rev. D \textbf{96} (2017) no.9, 095009
[arXiv:1704.07397 [hep-ph]].

\bibitem{Altmannshofer:2017yso}
W.~Altmannshofer, P.~Stangl and D.~M.~Straub,
Phys. Rev. D \textbf{96} (2017) no.5, 055008
[arXiv:1704.05435 [hep-ph]].

\bibitem{Kowalska:2019ley}
K.~Kowalska, D.~Kumar and E.~M.~Sessolo,
Eur. Phys. J. C \textbf{79} (2019) no.10, 840
[arXiv:1903.10932 [hep-ph]].

\bibitem{Alguero:2018nvb}
M.~Alguer\'o, B.~Capdevila, S.~Descotes-Genon, P.~Masjuan and J.~Matias,
Phys. Rev. D \textbf{99} (2019) no.7, 075017
[arXiv:1809.08447 [hep-ph]].

\bibitem{Crivellin:2018yvo}
A.~Crivellin, C.~Greub, D.~M\"uller and F.~Saturnino,
Phys. Rev. Lett. \textbf{122} (2019) no.1, 011805
[arXiv:1807.02068 [hep-ph]].

\bibitem{Aaij:2020nol}
R.~Aaij \textit{et al.} [LHCb],
Phys. Rev. Lett. \textbf{124} (2020) no.21, 211802
[arXiv:2003.03999 [hep-ex]].

\bibitem{Aaij:2020umj}
R.~Aaij \textit{et al.} [LHCb],
JHEP \textbf{12} (2020), 081
[arXiv:2010.06011 [hep-ex]].

\bibitem{Barbieri:2016las}
R.~Barbieri, C.~W.~Murphy and F.~Senia,
Eur. Phys. J. C \textbf{77} (2017) no.1, 8
[arXiv:1611.04930 [hep-ph]].

\bibitem{Cline:2017aed}
J.~M.~Cline,
Phys. Rev. D \textbf{97} (2018) no.1, 015013
doi:10.1103/PhysRevD.97.015013
[arXiv:1710.02140 [hep-ph]].

\bibitem{Hung:1981pd}
P.~Q.~Hung, A.~J.~Buras and J.~D.~Bjorken,
Phys. Rev. D \textbf{25} (1982), 805.

\bibitem{Valencia:1994cj}
G.~Valencia and S.~Willenbrock,
Phys. Rev. D \textbf{50} (1994), 6843-6848
[arXiv:hep-ph/9409201 [hep-ph]].

\bibitem{Smirnov:2007hv}
A.~D.~Smirnov,
Mod. Phys. Lett. A \textbf{22} (2007), 2353-2363
[arXiv:0705.0308 [hep-ph]].

\bibitem{Carpentier:2010ue}
M.~Carpentier and S.~Davidson,
Eur. Phys. J. C \textbf{70} (2010), 1071-1090
[arXiv:1008.0280 [hep-ph]].

\bibitem{Kuznetsov:2012ai}
A.~V.~Kuznetsov, N.~V.~Mikheev and A.~V.~Serghienko,
Int. J. Mod. Phys. A \textbf{27} (2012), 1250062
[arXiv:1203.0196 [hep-ph]].

\bibitem{Smirnov:2018ske}
A.~D.~Smirnov,
Mod. Phys. Lett. A \textbf{33} (2018), 1850019
[arXiv:1801.02895 [hep-ph]].

\bibitem{Feruglio:2017rjo}
F.~Feruglio, P.~Paradisi and A.~Pattori,
JHEP \textbf{09} (2017), 061
[arXiv:1705.00929 [hep-ph]].

\bibitem{Dorsner:2016wpm}
I.~Dor\v{s}ner, S.~Fajfer, A.~Greljo, J.~F.~Kamenik and N.~Ko\v{s}nik,
Phys. Rept. \textbf{641} (2016), 1-68
[arXiv:1603.04993 [hep-ph]].

\bibitem{Aebischer:2018bkb}
J.~Aebischer, J.~Kumar and D.~M.~Straub,
Eur. Phys. J. C \textbf{78} (2018) no.12, 1026
[arXiv:1804.05033 [hep-ph]].

\bibitem{Straub:2018kue}
D.~M.~Straub,
``flavio: a Python package for flavour and precision phenomenology in
the Standard Model and beyond,'' 
arXiv:1810.08132 [hep-ph].

\bibitem{DiLuzio:2018zxy}
L.~Di Luzio, J.~Fuentes-Martin, A.~Greljo, M.~Nardecchia and S.~Renner,
JHEP \textbf{11} (2018), 081
[arXiv:1808.00942 [hep-ph]].

\bibitem{Descotes-Genon:2013vna}
S.~Descotes-Genon, T.~Hurth, J.~Matias and J.~Virto,
JHEP \textbf{05} (2013), 137
[arXiv:1303.5794 [hep-ph]].


\bibitem{Chatrchyan:2013bka}
S.~Chatrchyan \textit{et al.} [CMS],
Phys. Rev. Lett. \textbf{111} (2013), 101804
[arXiv:1307.5025 [hep-ex]].

\bibitem{Aaij:2017vad}
R.~Aaij \textit{et al.} [LHCb],
Phys. Rev. Lett. \textbf{118} (2017) no.19, 191801
[arXiv:1703.05747 [hep-ex]].

\bibitem{Aaboud:2018mst}
M.~Aaboud \textit{et al.} [ATLAS],
JHEP \textbf{04} (2019), 098
[arXiv:1812.03017 [hep-ex]].


\bibitem{Sirunyan:2019xdu}
A.~M.~Sirunyan \textit{et al.} [CMS],
JHEP \textbf{04} (2020), 188
[arXiv:1910.12127 [hep-ex]].

\bibitem{Altmannshofer:2021qrr}
W.~Altmannshofer and P.~Stangl,
``New Physics in Rare B Decays after Moriond 2021,''
arXiv:2103.13370 [hep-ph].
\bibitem{Amhis:2014hma}
Y.~Amhis \textit{et al.} [Heavy Flavor Averaging Group (HFAG)],
``Averages of $b$-hadron, $c$-hadron, and $\tau$-lepton properties as
of summer 2014,'' 
arXiv:1412.7515 [hep-ex].

\bibitem{Misiak:2017bgg}
M.~Misiak and M.~Steinhauser,
Eur. Phys. J. C \textbf{77} (2017) no.3, 201
[arXiv:1702.04571 [hep-ph]].

\bibitem{Dutta:2014sxo}
D.~Dutta \textit{et al.} [Belle],
Phys. Rev. D \textbf{91} (2015) no.1, 011101
[arXiv:1411.7771 [hep-ex]].

\bibitem{Aaij:2012ita}
R.~Aaij \textit{et al.} [LHCb],
Nucl. Phys. B \textbf{867} (2013), 1-18
[arXiv:1209.0313 [hep-ex]].

\bibitem{Abdesselam:2017kjf}
A.~Abdesselam \textit{et al.} [Belle],
``Precise determination of the CKM matrix element $\left|
V_{cb}\right|$ with $\bar B^0 \to D^{*\,+} \, \ell^- \, \bar \nu_\ell$
decays with hadronic tagging at Belle,'' 
arXiv:1702.01521 [hep-ex].

\bibitem{Abdesselam:2018nnh}
E.~Waheed \textit{et al.} [Belle],
Phys. Rev. D \textbf{100} (2019) no.5, 052007
[arXiv:1809.03290 [hep-ex]].

\bibitem{Aaij:2017uff}
R.~Aaij \textit{et al.} [LHCb],
Phys. Rev. Lett. \textbf{120} (2018) no.17, 171802
[arXiv:1708.08856 [hep-ex]].

\bibitem{Aubert:2008yv}
B.~Aubert \textit{et al.} [BaBar],
Phys. Rev. D \textbf{79} (2009), 012002
[arXiv:0809.0828 [hep-ex]].

\bibitem{Aubert:2007qs}
B.~Aubert \textit{et al.} [BaBar],
Phys. Rev. Lett. \textbf{100} (2008), 231803
[arXiv:0712.3493 [hep-ex]].



\bibitem{Urquijo:2006wd}
P.~Urquijo \textit{et al.} [Belle],
Phys. Rev. D \textbf{75} (2007), 032001
[arXiv:hep-ex/0610012].

\bibitem{Aubert:2009qda}
B.~Aubert \textit{et al.} [BaBar],
Phys. Rev. D \textbf{81} (2010), 032003
[arXiv:0908.0415 [hep-ex]].

\bibitem{Grygier:2017tzo}
J.~Grygier \textit{et al.} [Belle],
Phys. Rev. D \textbf{96} (2017) no.9, 091101
[arXiv:1702.03224 [hep-ex]].

\bibitem{Lutz:2013ftz}
O.~Lutz \textit{et al.} [Belle],
Phys. Rev. D \textbf{87} (2013) no.11, 111103
[arXiv:1303.3719 [hep-ex]].

\bibitem{Lees:2013kla}
J.~P.~Lees \textit{et al.} [BaBar],
Phys. Rev. D \textbf{87} (2013) no.11, 112005
[arXiv:1303.7465 [hep-ex]].

\bibitem{delAmoSanchez:2010bk}
P.~del Amo Sanchez \textit{et al.} [BaBar],
Phys. Rev. D \textbf{82} (2010), 112002
[arXiv:1009.1529 [hep-ex]].

\bibitem{TheMEG:2016wtm}
A.~M.~Baldini \textit{et al.} [MEG],
Eur. Phys. J. C \textbf{76} (2016) no.8, 434
[arXiv:1605.05081 [hep-ex]].

\bibitem{Baldini:2018nnn}
A.~M.~Baldini \textit{et al.} [MEG II],
Eur. Phys. J. C \textbf{78} (2018) no.5, 380
[arXiv:1801.04688 [physics.ins-det]].

\bibitem{Aubert:2009ag}
B.~Aubert \textit{et al.} [BaBar],
Phys. Rev. Lett. \textbf{104} (2010), 021802
[arXiv:0908.2381 [hep-ex]].

\bibitem{Kou:2018nap}
E.~Kou \textit{et al.} [Belle-II],
PTEP \textbf{2019} (2019) no.12, 123C01
[erratum: PTEP \textbf{2020} (2020) no.2, 029201]
[arXiv:1808.10567 [hep-ex]].

\bibitem{Bellgardt:1987du}
U.~Bellgardt \textit{et al.} [SINDRUM],
Nucl. Phys. B \textbf{299} (1988), 1-6.

\bibitem{Blondel:2013ia}
A.~Blondel, A.~Bravar, M.~Pohl, S.~Bachmann, N.~Berger, M.~Kiehn,
A.~Schoning, D.~Wiedner, B.~Windelband and P.~Eckert, \textit{et al.} 
arXiv:1301.6113 [physics.ins-det].

\bibitem{Hayasaka:2010np}
K.~Hayasaka, K.~Inami, Y.~Miyazaki, K.~Arinstein, V.~Aulchenko,
T.~Aushev, A.~M.~Bakich, A.~Bay, K.~Belous and V.~Bhardwaj, \textit{et
al.} 
Phys. Lett. B \textbf{687} (2010), 139-143
[arXiv:1001.3221 [hep-ex]].

\bibitem{Bertl:2006up}
W.~H.~Bertl \textit{et al.} [SINDRUM II],
Eur. Phys. J. C \textbf{47} (2006), 337-346.

\bibitem{Nguyen:2015vkk}
T.~M.~Nguyen [DeeMe],
PoS \textbf{FPCP2015} (2015), 060.

\bibitem{Krikler:2015msn}
B.~E.~Krikler [COMET],
``An Overview of the COMET Experiment and its Recent Progress,''
arXiv:1512.08564 [physics.ins-det].

\bibitem{Adamov:2018vin}
R.~Abramishvili \textit{et al.} [COMET],
PTEP \textbf{2020} (2020) no.3, 033C01
[arXiv:1812.09018 [physics.ins-det]].

\bibitem{KunoESPP19}
Y.~Kuno,
``Physics prospects with muons'', presentation at the Flavour Session of the CERN Council Open Symposium on the Update of the European Strategy for Particle Physics, Granada, Spain, 13-16 May 2019.


\bibitem{Bartoszek:2014mya}
L.~Bartoszek \textit{et al.} [Mu2e],
``Mu2e Technical Design Report,''
arXiv:1501.05241 [physics.ins-det].

\bibitem{Aaij:2019okb}
R.~Aaij \textit{et al.} [LHCb],
Phys. Rev. Lett. \textbf{123} (2019) no.21, 211801
[arXiv:1905.06614 [hep-ex]].

\bibitem{Lees:2012zz}
J.~P.~Lees \textit{et al.} [BaBar],
Phys. Rev. D \textbf{86} (2012), 012004
[arXiv:1204.2852 [hep-ex]].

\bibitem{Khachatryan:2014ura}
  V.~Khachatryan {\it et al.} [CMS Collaboration],
  Phys.\ Lett.\ B {\bf 739} (2014) 229
  [arXiv:1408.0806 [hep-ex]].

\bibitem{Aad:2015caa}
  G.~Aad {\it et al.} [ATLAS Collaboration],
  Eur.\ Phys.\ J.\ C {\bf 76} (2016) no.1,  5
  [arXiv:1508.04735 [hep-ex]].
 
\bibitem{Aaboud:2016qeg}
M.~Aaboud \textit{et al.} [ATLAS],
New J. Phys. \textbf{18} (2016) no.9, 093016
[arXiv:1605.06035 [hep-ex]].

\bibitem{Aaboud:2019jcc}
M.~Aaboud \textit{et al.} [ATLAS],
Eur. Phys. J. C \textbf{79} (2019) no.9, 733
[arXiv:1902.00377 [hep-ex]].

\bibitem{Aaboud:2019bye}
M.~Aaboud \textit{et al.} [ATLAS],
JHEP \textbf{06} (2019), 144
[arXiv:1902.08103 [hep-ex]].

\bibitem{Aad:2020iuy}
G.~Aad \textit{et al.} [ATLAS],
JHEP \textbf{10} (2020), 112
[arXiv:2006.05872 [hep-ex]].

\bibitem{Sirunyan:2017yrk}
  A.~M.~Sirunyan {\it et al.} [CMS Collaboration],
  JHEP {\bf 1707} (2017) 121
  [arXiv:1703.03995 [hep-ex]].

\bibitem{Sirunyan:2018vhk}
  A.~M.~Sirunyan {\it et al.} [CMS Collaboration],
  JHEP {\bf 1903} (2019) 170
  [arXiv:1811.00806 [hep-ex]].

\bibitem{Sirunyan:2018kzh}
  A.~M.~Sirunyan {\it et al.} [CMS Collaboration],
  Phys.\ Rev.\ D {\bf 98} (2018) no.3,  032005
  [arXiv:1805.10228 [hep-ex]].

\bibitem{Cornella:2021sby}
C.~Cornella, D.~A.~Faroughy, J.~Fuentes-Martin, G.~Isidori and M.~Neubert,
JHEP \textbf{08} (2021), 050
[arXiv:2103.16558 [hep-ph]].

\bibitem{DeBruyn:2016tiq}
K.~De Bruyn [LHCb],
``Search for the rare decays $B^0_{(s)}\to\tau^+\tau^-$,''
LHCb-CONF-2016-011.

\bibitem{TheBaBar:2016xwe}
J.~P.~Lees \textit{et al.} [BaBar],
Phys. Rev. Lett. \textbf{118} (2017) no.3, 031802
[arXiv:1605.09637 [hep-ex]].

\bibitem{Bobeth:2013uxa}
C.~Bobeth, M.~Gorbahn, T.~Hermann, M.~Misiak, E.~Stamou and M.~Steinhauser,
Phys. Rev. Lett. \textbf{112} (2014), 101801
[arXiv:1311.0903 [hep-ph]].

\bibitem{Hermann:2013kca}
T.~Hermann, M.~Misiak and M.~Steinhauser,
JHEP \textbf{12} (2013), 097
[arXiv:1311.1347 [hep-ph]].

\bibitem{Bobeth:2013tba}
C.~Bobeth, M.~Gorbahn and E.~Stamou,
Phys. Rev. D \textbf{89} (2014) no.3, 034023
[arXiv:1311.1348 [hep-ph]].

\bibitem{Guetta:1997fw}
D.~Guetta and E.~Nardi,
Phys. Rev. D \textbf{58} (1998), 012001
[arXiv:hep-ph/9707371].

\bibitem{Bobeth:2011st}
C.~Bobeth and U.~Haisch,
Acta Phys. Polon. B \textbf{44} (2013), 127-176
[arXiv:1109.1826 [hep-ph]].

\bibitem{Hewett:1995dk}
J.~L.~Hewett,
Phys. Rev. D \textbf{53} (1996), 4964-4969
[arXiv:hep-ph/9506289].

\bibitem{Bouchard:2013mia}
C.~Bouchard \textit{et al.} [HPQCD],
Phys. Rev. Lett. \textbf{111} (2013) no.16, 162002
[erratum: Phys. Rev. Lett. \textbf{112} (2014) no.14, 149902]
[arXiv:1306.0434 [hep-ph]].

\bibitem{Kamenik:2017ghi}
J.~F.~Kamenik, S.~Monteil, A.~Semkiv and L.~V.~Silva,
Eur. Phys. J. C \textbf{77} (2017) no.10, 701
[arXiv:1705.11106 [hep-ph]].

\bibitem{Beylich:2011aq}
M.~Beylich, G.~Buchalla and T.~Feldmann,
Eur. Phys. J. C \textbf{71} (2011), 1635
[arXiv:1101.5118 [hep-ph]].

\bibitem{Bailey:2015dka}
J.~A.~Bailey, A.~Bazavov, C.~Bernard, C.~M.~Bouchard, C.~DeTar, D.~Du,
A.~X.~El-Khadra, J.~Foley, E.~D.~Freeland and E.~G\'amiz, \textit{et
al.} 
Phys. Rev. D \textbf{93} (2016) no.2, 025026
[arXiv:1509.06235 [hep-lat]].

\bibitem{Du:2015tda}
D.~Du, A.~X.~El-Khadra, S.~Gottlieb, A.~S.~Kronfeld, J.~Laiho,
E.~Lunghi, R.~S.~Van de Water and R.~Zhou, 
Phys. Rev. D \textbf{93} (2016) no.3, 034005
[arXiv:1510.02349 [hep-ph]].

\bibitem{Straub:2015ica}
A.~Bharucha, D.~M.~Straub and R.~Zwicky,
JHEP \textbf{08} (2016), 098
[arXiv:1503.05534 [hep-ph]].



\bibitem{Foreman-Mackey:2012any}
D.~Foreman-Mackey, D.~W.~Hogg, D.~Lang and J.~Goodman,
Publ. Astron. Soc. Pac. \textbf{125} (2013), 306-312
[arXiv:1202.3665 [astro-ph.IM]].

\bibitem{Altmannshofer:2014rta}
W.~Altmannshofer and D.~M.~Straub,
Eur. Phys. J. C \textbf{75} (2015) no.8, 382
[arXiv:1411.3161 [hep-ph]].

\bibitem{James:1975dr}
F.~James and M.~Roos,
Comput. Phys. Commun. \textbf{10} (1975), 343-367.

\bibitem{Beneke:2001at}
M.~Beneke, T.~Feldmann and D.~Seidel,
Nucl. Phys. B \textbf{612} (2001), 25-58
[arXiv:hep-ph/0106067].

\bibitem{CMS:2017ivg}
 [CMS],
``Measurement of the $P_1$ and $P_5'$ angular parameters of the decay
$\mathrm{B}^0 \to \mathrm{K}^{*0} \mu^+ \mu^-$ in proton-proton
collisions at $\sqrt{s}=8~\mathrm{TeV}$,'' 
CMS-PAS-BPH-15-008.

\bibitem{CDF:2012qwd}
 [CDF],
``Precise Measurements of Exclusive b \textrightarrow{}
s\textmu{}+\textmu{} \ensuremath{-} Decay Amplitudes Using the Full
CDF Data Set,'' 
CDF-NOTE-10894.

\bibitem{Aaij:2015dea}
R.~Aaij \textit{et al.} [LHCb],
JHEP \textbf{04} (2015), 064
[arXiv:1501.03038 [hep-ex]].

\bibitem{Aaij:2014pli}
R.~Aaij \textit{et al.} [LHCb],
JHEP \textbf{06} (2014), 133
[arXiv:1403.8044 [hep-ex]].

\bibitem{Aaij:2016flj}
R.~Aaij \textit{et al.} [LHCb],
JHEP \textbf{11} (2016), 047
[erratum: JHEP \textbf{04} (2017), 142]
[arXiv:1606.04731 [hep-ex]].

\bibitem{Abdesselam:2019lab}
A.~Abdesselam \textit{et al.} [Belle],
arXiv:1908.01848 [hep-ex].

\bibitem{Glattauer:2015teq}
R.~Glattauer \textit{et al.} [Belle],
Phys. Rev. D \textbf{93} (2016) no.3, 032006
[arXiv:1510.03657 [hep-ex]].

\bibitem{Hamer:2015jsa}
P.~Hamer \textit{et al.} [Belle],
Phys. Rev. D \textbf{93} (2016) no.3, 032007
[arXiv:1509.06521 [hep-ex]].

\bibitem{Besson:2009uv}
D.~Besson \textit{et al.} [CLEO],
Phys. Rev. D \textbf{80} (2009), 032005
[arXiv:0906.2983 [hep-ex]].

\bibitem{Ablikim:2015ixa}
M.~Ablikim \textit{et al.} [BESIII],
Phys. Rev. D \textbf{92} (2015) no.7, 072012
[arXiv:1508.07560 [hep-ex]].

\bibitem{Ablikim:2017lks}
M.~Ablikim \textit{et al.} [BESIII],
Phys. Rev. D \textbf{96} (2017) no.1, 012002
[arXiv:1703.09084 [hep-ex]].

\bibitem{Buras:2015qea}
A.~J.~Buras, D.~Buttazzo, J.~Girrbach-Noe and R.~Knegjens,
JHEP \textbf{11} (2015), 033
[arXiv:1503.02693 [hep-ph]].

\bibitem{Artamonov:2008qb}
A.~V.~Artamonov \textit{et al.} [E949],
Phys. Rev. Lett. \textbf{101} (2008), 191802
[arXiv:0808.2459 [hep-ex]].

\bibitem{CortinaGil:2020vlo}
E.~Cortina Gil \textit{et al.} [NA62],
JHEP \textbf{11} (2020), 042
[arXiv:2007.08218 [hep-ex]].

\bibitem{Ahn:2009gb}
J.~K.~Ahn \textit{et al.} [E391a],
Phys. Rev. D \textbf{81} (2010), 072004
[arXiv:0911.4789 [hep-ex]].


\bibitem{Miyazaki:2007jp}
Y.~Miyazaki \textit{et al.} [Belle],
Phys. Lett. B \textbf{648} (2007), 341-350
[arXiv:hep-ex/0703009].

\bibitem{Miyazaki:2011xe}
Y.~Miyazaki \textit{et al.} [Belle],
Phys. Lett. B \textbf{699} (2011), 251-257
[arXiv:1101.0755 [hep-ex]].

\bibitem{Aubert:2008cu}
B.~Aubert \textit{et al.} [BaBar],
Phys. Rev. D \textbf{77} (2008), 091104
[arXiv:0801.0697 [hep-ex]].

\bibitem{Becirevic:2016zri}
D.~Be\v{c}irevi\'c, O.~Sumensari and R.~Zukanovich Funchal,
Eur. Phys. J. C \textbf{76} (2016) no.3, 134
[arXiv:1602.00881 [hep-ph]].

\bibitem{Buras:2014fpa}
A.~J.~Buras, J.~Girrbach-Noe, C.~Niehoff and D.~M.~Straub,
JHEP \textbf{02} (2015), 184
[arXiv:1409.4557 [hep-ph]].

\bibitem{Bobeth:2017ecx}
C.~Bobeth and A.~J.~Buras,
JHEP \textbf{02} (2018), 101
[arXiv:1712.01295 [hep-ph]].

\bibitem{Bordone:2017lsy}
M.~Bordone, D.~Buttazzo, G.~Isidori and J.~Monnard,
Eur. Phys. J. C \textbf{77} (2017) no.9, 618
[arXiv:1705.10729 [hep-ph]].

\bibitem{Buras:2004qb}
A.~J.~Buras, T.~Ewerth, S.~Jager and J.~Rosiek,
Nucl. Phys. B \textbf{714} (2005), 103-136
[arXiv:hep-ph/0408142].

\bibitem{Lavoura:2003xp}
L.~Lavoura,
Eur. Phys. J. C \textbf{29} (2003), 191-195
[arXiv:hep-ph/0302221].

\bibitem{Okada:1999zk}
Y.~Okada, K.~i.~Okumura and Y.~Shimizu,
Phys. Rev. D \textbf{61} (2000), 094001
[arXiv:hep-ph/9906446].

\bibitem{Kuno:1999jp}
Y.~Kuno and Y.~Okada,
Rev. Mod. Phys. \textbf{73} (2001), 151-202
[arXiv:hep-ph/9909265].

\bibitem{Gabrielli:2000te.}
E.~Gabrielli,
Phys. Rev. D \textbf{62} (2000), 055009
[arXiv:hep-ph/9911539].

\bibitem{Kitano:2002mt}
R.~Kitano, M.~Koike and Y.~Okada,
Phys. Rev. D \textbf{66} (2002), 096002
[erratum: Phys. Rev. D \textbf{76} (2007), 059902]
[arXiv:hep-ph/0203110].

\bibitem{Kosmas:2001mv}
T.~S.~Kosmas, S.~Kovalenko and I.~Schmidt,
Phys. Lett. B \textbf{511} (2001), 203
[arXiv:hep-ph/0102101].



}

\end{thebibliography}
\end{document}